\documentclass[pra,twocolumn,showpacs,showkeys]{revtex4}
\usepackage[colorlinks=true,citecolor=green,linkcolor=red,urlcolor=blue]{hyperref}
\usepackage{graphicx,epsf,bm,bbm,color,amsmath,ulem,amssymb,subfigure}

	\newcommand{\beq}{\begin{equation}}
	\newcommand{\be}{\begin{equation}}
	\newcommand{\beqn}{\begin{eqnarray}}
	\newcommand{\eeq}{\end{equation}}
	\newcommand{\ee}{\end{equation}}
	\newcommand{\eeqn}{\end{eqnarray}}
	\newcommand{\nn}{\nonumber}
	\newcommand{\tr}{\textrm}

\newcommand{\bem}{\begin{pmatrix}}
\newcommand{\eem}{\end{pmatrix}}

\newcommand{\f}{\frac}

\newcommand{\mf}[1]{\boldsymbol{#1}}

\newcommand{\s}{St\"uckelberg }

\begin{document}

	\today
	
		\title{Geometry of Bloch states probed by \s interferometry}
        \author{Lih-King Lim}
        \affiliation{Max-Planck-Institut f\"ur Physik komplexer Systeme, N\"othnitzer Strasse 38, D-01187 Dresden}
         \author{Jean-No\"el Fuchs}
        \affiliation{LPTMC, CNRS UMR 7600, Universit\'{e} Pierre et Marie Curie, 4 place Jussieu, F-75252 Paris}
        \affiliation{Laboratoire de Physique des Solides, CNRS UMR 8502, Univ. Paris-Sud, F-91405 Orsay}
         \author{Gilles Montambaux}
        \affiliation{Laboratoire de Physique des Solides, CNRS UMR 8502, Univ. Paris-Sud, F-91405 Orsay}

\begin{abstract}
Inspired by recent experiments with cold atoms in optical lattices, we consider a \s interferometer for a particle performing Bloch oscillations in a tight-binding model on the honeycomb lattice. The interferometer is made of two avoided crossings at the saddle points of the band structure (i.e. at M points of the reciprocal space). This problem is reminiscent of the double Dirac cone \s interferometer that was recently studied in the continuum limit [Phys. Rev. Lett. \textbf{112}, 155302 (2014)]. Although the two problems share similarities -- such as the appearance of a geometric phase shift -- lattice effects, not captured by the continuum limit, make them truly different. The particle dynamics in the presence of a force is described by the Bloch Hamiltonian $H(\mf{k})$ defined from the tight-binding Hamiltonian {\it and} the position operator. This leads to many interesting effects for the lattice \s interferometer: a twisting of the two Landau-Zener tunnelings, saturation of the inter-band transition probability in the sudden (infinite force) limit and extended periodicity or even non-periodicity beyond the first Brillouin zone. In particular, \s interferometry gives access to the overlap matrix of cell-periodic Bloch states thereby allowing to fully characterize the geometry of Bloch states, as e.g. to obtain the quantum metric tensor.
\end{abstract}
	\maketitle

\section{Introduction}
St\"uckelberg interferometry \cite{Stuck1932,SAN2010} has been shown to be a powerful tool to probe the motion and merging of the Dirac points in a graphene-like optical lattice \cite{Tarruell2012,LFM2012}. In this experiment, a Fermi sea of cold atoms is accelerated and performs Bloch oscillations, which are described as a linear motion in reciprocal space. By measuring the proportion of atoms having tunneled from the lower to the upper band in the vicinity of the Dirac points after one Bloch oscillation, it is possible to reconstruct the energy spectrum. For atoms encountering two Dirac points in succession, they may tunnel either through the first or through the second, so that a two-path interferometer is realized in energy-momentum space, along a scenario  first imagined by \s in a slightly different version \cite{Stuck1932}. The interband transition probability is $P= 4P_{LZ}(1-P_{LZ})\sin^2 \frac{\varphi_\textrm{tot}}{2}$ in terms of the Landau-Zener (LZ) probability $P_{LZ}$ to tunnel at a single avoided band crossing \cite{LZ1932} and of a phase $\varphi_\textrm{tot}$. It turns out that, in this experiment, the accumulated phase $\varphi_\textrm{tot}$ between the two LZ events is washed out and the experiment simply probes a sum of intensities $P\approx 2P_{LZ}(1-P_{LZ})$. However this phase $\varphi_\textrm{tot}$ is expected to be quite rich. It is the sum of three contributions: a phase accumulated at the tunneling events (known as the Stokes phase), a dynamical phase which depends on the energy difference between the two paths \cite{SAN2010}, and possibly a geometric phase which depends on the nature (such as the chirality) of the Dirac points and on the trajectory in reciprocal space \cite{LFM2014,LFM2015}. This geometric phase probes the coupling between bands and is captured by the Bloch Hamiltonian but not by the band energy dispersion relation. Therefore the possibility of new experiments which may probe this phase is particularly interesting.

The merging and motion of Dirac points as well as their signature on the LZ transitions have been analyzed theoretically in the framework of the so-called universal Hamiltonian for a pair of Dirac points \cite{Montambaux2009}, which is a long wavelength version of the graphene Hamiltonian in which the periodicity of the reciprocal lattice has disappeared \cite{LFM2012,FLM2012}. It has been used to reveal geometric phases in a \s interferometer formed by two Dirac cones of the energy spectrum in two-dimensional systems \cite{LFM2014,LFM2015}. However it is not suitable for any trajectory in reciprocal space, far from the Dirac points.

Other regions of the reciprocal lattice may be also interesting to investigate  and the goal of the present work is to consider new scenarios of general LZ processes. This is motivated by a recent experiment realizing a \s interferometer with a Bose-Einstein condensate (BEC) in a honeycomb optical lattice \cite{BlochSchneider}. It is an experimental feat of this work that an initial single-particle Bloch state (a BEC in this case) can be accelerated in arbitrary directions, magnitude, and time duration. However, the optimum quasimomentum paths for a BEC must avoid the vicinity of Dirac cones of the energy spectrum (e.g., to minimize heating \cite{Chen2011}). Alternative paths have been proposed and the associated \textit{lattice} \s phenomena are no longer limited to probing characteristics of Dirac cones.

The problem to be solved is the following. A particle (or a BEC) is initially prepared in the ground state, eigenstate of the Bloch Hamiltonian (at the $\Gamma$ point). Under the action of an applied force, it evolves in the reciprocal space, following a path $\mf{k}(t)$, to reach a final state. We are interested  in the probability for the particle to have tunneled into the upper band.

\begin{figure}[ht]
\begin{center}
\subfigure[]{\label{energylanda}\includegraphics[width=7cm]{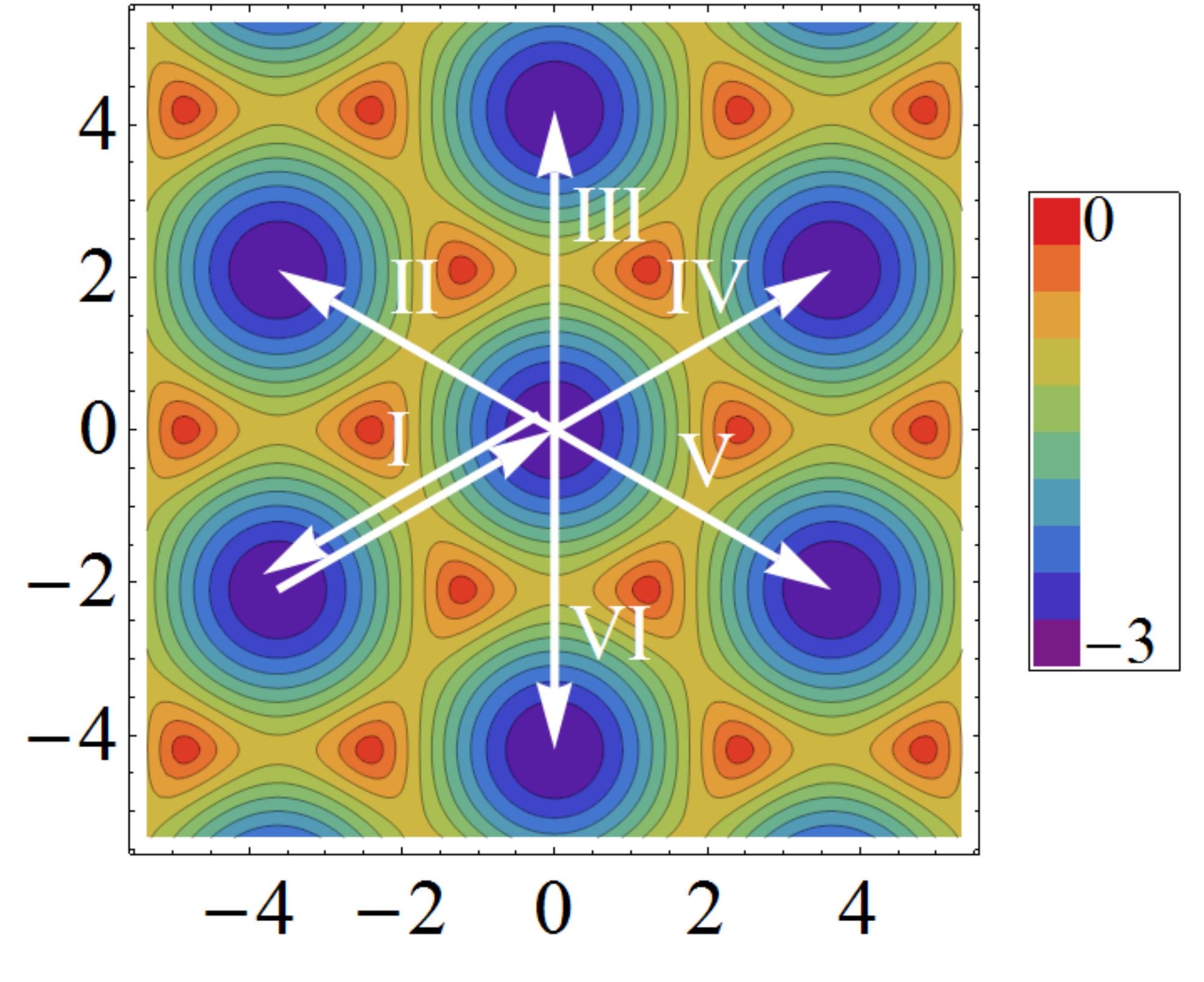}}
\subfigure[]{\label{energylandb}\includegraphics[width=6cm]{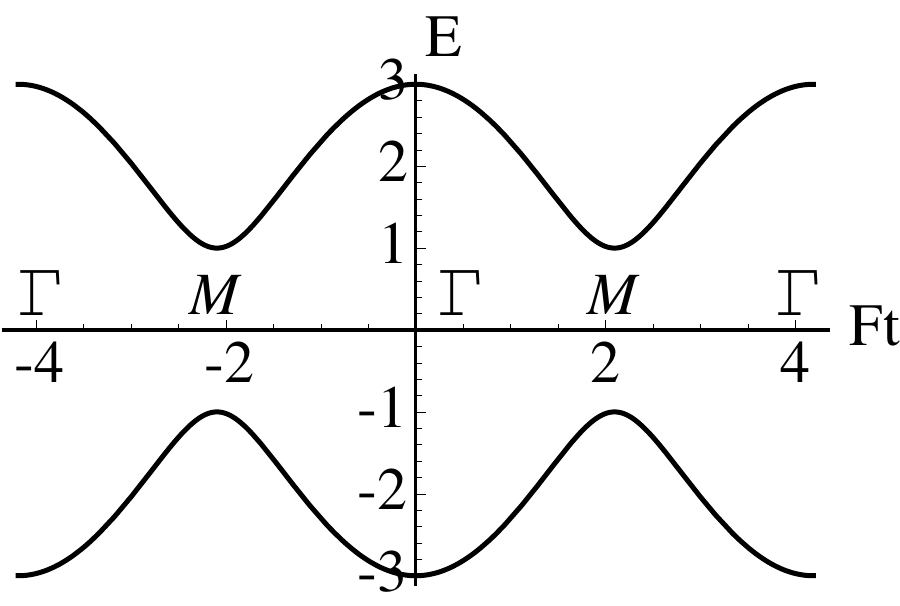}}
\end{center}
\caption{Energy band dispersion relation. (a): Iso-energy lines [energy $E$ in units of $J$] in reciprocal space [$\mf{k}$ in units of $1/a$] for the lower band. Six different trajectories (all starting at $-\mf{a}_1^*$, see Fig. \ref{kspace}(b), and ending at a $\Gamma$ point) are labeled from (I) to (VI). (b): The six trajectories lead to a single energy profile $E_\pm(t)$ [in units of $J$] as a function of $F t$ [in units of $\hbar/a$]. The avoided crossings at the M points act as the beamsplitters of the \s interferometer.}
\label{energyland}
\end{figure}
To study such phenomena, the knowledge of the Bloch Hamiltonian in the full reciprocal space is required. Following the recent experimental setup mentioned above, we consider trajectories starting from the $\Gamma$ point and moving to another $\Gamma$ point  (in the extended Brillouin zone) through a pair of  saddle points of the energy spectrum. A saddle point, which corresponds to an M point of the Brillouin zone, lies at the midpoint between two nearest Dirac points,  and constitute the ``beamsplitter" of the \s interferometer (see Fig. \ref{energyland}). We point out that the ``structure" of this beamsplitter is different from those in the vicinity of Dirac cones. By traversing distinct pairs of M points, geometric \s phases can still arise due to lattice effects. As we will show, these new features can be understood in terms of the dynamics governed by a properly defined Bloch Hamiltonian $H(\mf{k})$, that is essentially obtained via a unitary transformation of the tight-binding Hamiltonian $\hat{H}$ and crucially involving the position operator \cite{Blount,Xiao2010,FPGM2010}. As a consequence, the interferometer can be used to probe the pseudospin structure associated with the Bloch Hamiltonian, generalizing the special case of low-energy Dirac cones Hamiltonians \cite{LFM2014,LFM2015}. This is of particular interest because the Bloch Hamiltonian does not generally assume the periodicity of the Brillouin zone of the tight-binding model. Moreover, we study several related phenomena arising from this observation, including the saturation of LZ tunneling probability $P_{LZ}<1$ even in the sudden limit, and propose to observe the pseudospin structure of a Bloch Hamiltonian with no periodicity.

The outline of the paper is the following. The problem is settled in section \ref{secdoublem}. Starting from the tight-binding Hamiltonian for the honeycomb lattice, we show that the trajectories $\mf{k}(t)$ traversing two M points constitute two different \s interferometers, and we stress the importance of the external force, the parameter driving the interference pattern. In section \ref{secnum}, we present the dependence of the interband transition probability as a function of the applied force, after traversing one or two M points, the second case realizing the \s interferometer.
Section \ref{secadia} presents analytical results in the adiabatic limit, showing explicitly the role played by the geometric phase. In section \ref{secsudden}, we discuss in details the limit of infinite force, called the sudden limit. It simply measures the overlap between Bloch states. During the completion of this work, we became aware of a recent preprint \cite{Li2015} in which this interband probability is measured along a special straight trajectory joining several $\Gamma$ points, in accordance with the results obtained in this present work. The deviations from the sudden limit when the force is large but finite are also obtained. We stress that the interference pattern does not have the periodicity of the reciprocal lattice. This is a consequence of the fact that the Bloch Hamiltonian, which reflects not only the Bravais lattice but also the structure of the elementary cell, has a periodicity different from that of the reciprocal lattice. In the last section \ref{seccon}, we conclude and give perspectives. The paper also contains appendices that give details of the derivations.

\section{Double M point St\"uckelberg interferometer in the honeycomb lattice}
\label{secdoublem}
\subsection{Bloch oscillations}
The starting point is to determine the Hamiltonian governing the dynamics of the \s interferometer from a tight-binding (TB) model. The problem is that of a single particle described by a TB Hamiltonian $\hat{H}$ and subjected to a constant force $\mf{F}$ for a finite time duration.
Using the time-dependent vectorial gauge potential for the constant force (which amounts to performing a gauge transformation, see Appendix A), the resulting time-dependent Schr\"{o}dinger equation [with $\hbar=1$] is
\beq
H(\mf{F}t)|\Psi(t)\rangle  =i\f{d}{dt}|\Psi(t)\rangle
\label{tdse}
\eeq
where $H(\mf{F}t)=H(\mf{k}\to \mf{F}t)$. The latter is the so-called Bloch Hamiltonian $H(\mf{k})$ which is defined below. The replacement rule $\mf{k}$$\rightarrow$$\mf{F}t$ can also be understood as a consequence of the exact conservation law $\dot{\mf{k}}=\mf{F}$ for the quasimomentum subjected to a constant force \cite{Xiao2010}.
Next, we turn to the Bloch Hamiltonian derived for the honeycomb lattice tight-binding model and define the \s interferometer trajectories.

\begin{figure}[ht]
\begin{center}
\subfigure[]{\includegraphics[width=3.5cm]{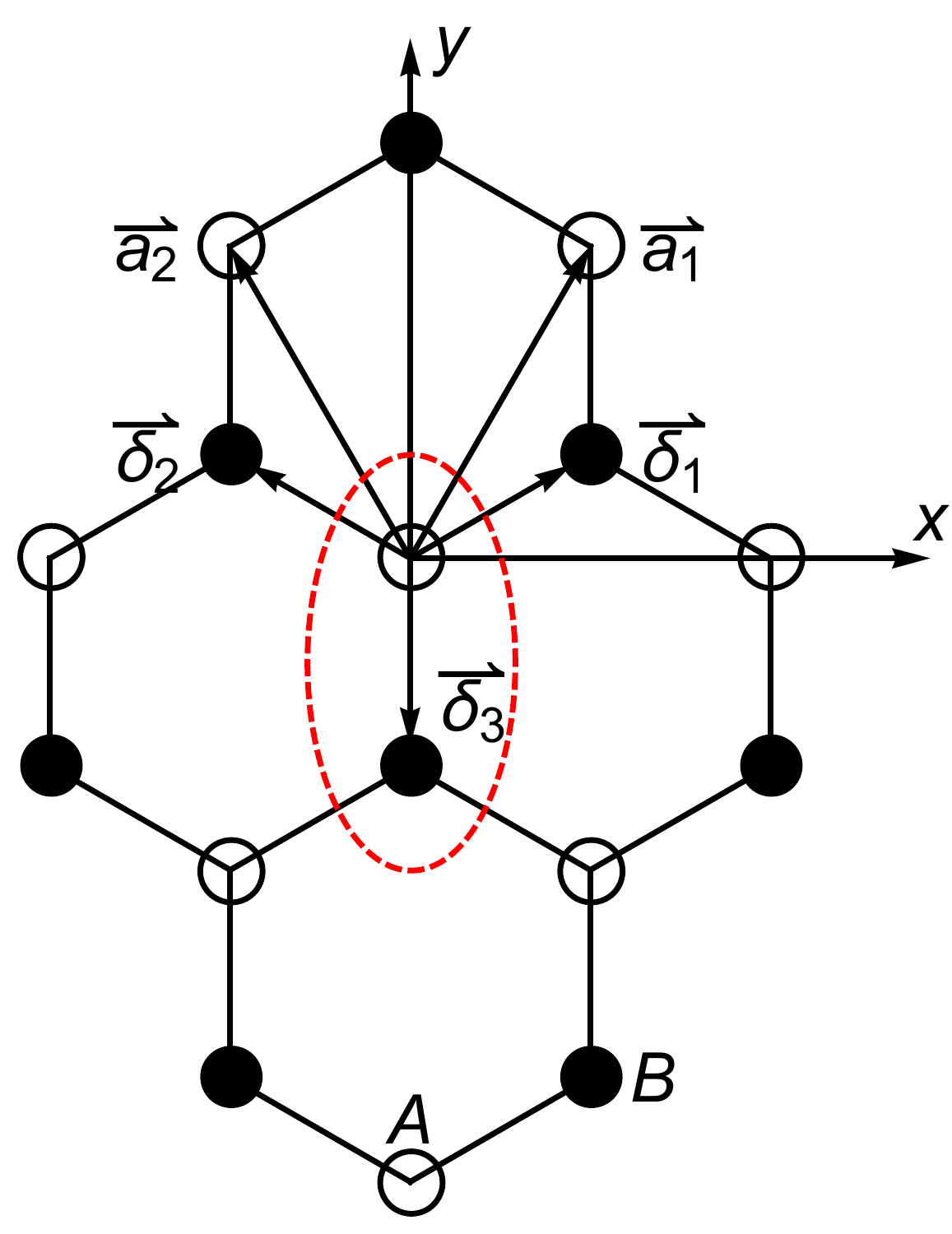}}
\subfigure[]{\includegraphics[width=5cm]{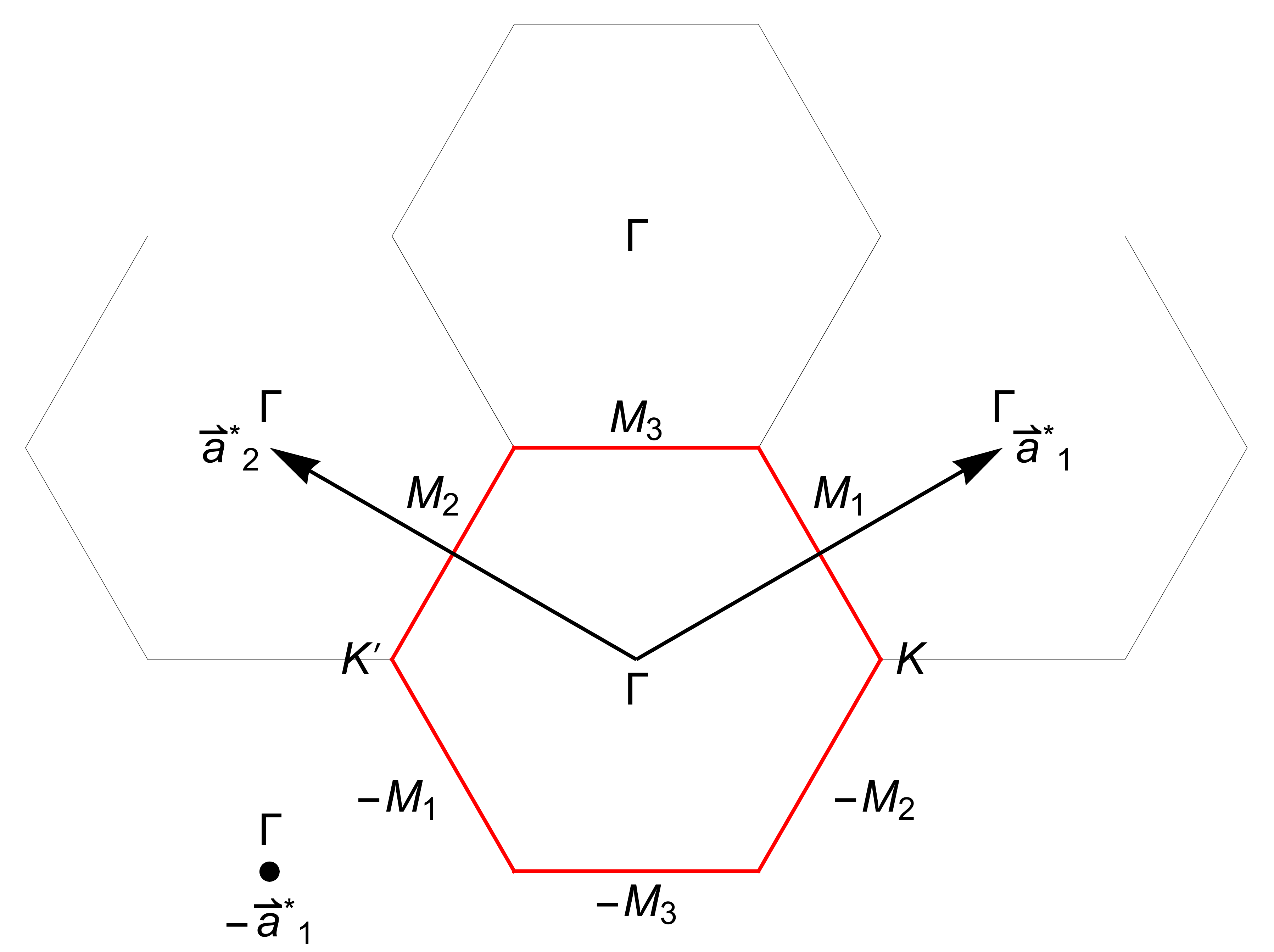}}
\end{center}
\caption{Lattice geometry. (a): Real space honeycomb lattice. The two sublattices $A$ and $B$ are indicated as empty and filled circles. The chosen basis is indicated by a dashed (red) ellipse. Definition of nearest-neighbor and Bravais lattice vectors: $\boldsymbol{\delta}_1=a (\frac{\sqrt{3}}{2},\frac{1}{2})$, $\boldsymbol{\delta}_2=a (-\frac{\sqrt{3}}{2},\frac{1}{2})$, $\boldsymbol{\delta}_3=a (0,-1)$ and $\mf{a}_1=a(\frac{\sqrt{3}}{2},\frac{3}{2})$, $\mf{a}_2=a(-\frac{\sqrt{3}}{2},\frac{3}{2})$, $\mf{a}_3=(0,0)$, where $a$ is the nearest-neighbor distance. (b): Unit cells of the reciprocal lattice.  First Brillouin zone (thick red line), definition of $\Gamma$, M and K points and reciprocal lattice vectors: $\mf{a}_1^*=\frac{4\pi}{3a}(\frac{\sqrt{3}}{2},\frac{1}{2})$, $\mf{a}_2^*=\frac{4\pi}{3a}(-\frac{\sqrt{3}}{2},\frac{1}{2})$. Our choice of initial $\Gamma$ point (at $-\mf{a}_1^*$) is indicated by a black dot.}
\label{kspace}
\end{figure}
\subsection{Tight-binding honeycomb lattice model and Bloch Hamiltonian}
The honeycomb lattice is made of a triangular Bravais lattice with a two-site basis \cite{Wallace1947,BM2009}. The two sublattices are usually called $A$ and $B$; the $A$ sublattice positions are specified by $\mf{R}=m_1 \mf{a}_1+m_2 \mf{a}_2$, with $\mf{a}_{1,2}$ being the basis vectors and $m_{1,2}$ being integers, and $\{\boldsymbol{\delta}_1,\boldsymbol{\delta}_2,\boldsymbol{\delta}_3\}$ are the nearest-neighbor vectors, see Fig. \ref{kspace}(a). The positions of the atoms $A$ and $B$ are $\mf{r}_A=\mf{R}$ and $\mf{r}_B=\mf{R}+\mf{\delta}_3$ (see the elementary basis in Fig. \ref{kspace}(a)). The reciprocal space is spanned by the vector basis $\mf{a}_{1,2}^*$ with the reciprocal lattice vectors $\mf{G}=m_1 \mf{a}^*_1+m_2 \mf{a}^*_2$ and the first Brillouin zone (BZ) is hexagonal, see Fig. \ref{kspace}(b). On the edges of the BZ, there are 3 geometrically inequivalent M points at mid distance between the K and K' corners of the BZ. We call them $\mf{M}_1=\mf{a}_1^*/2$, $\mf{M}_2=\mf{a}_2^*/2$ and $\mf{M}_3=(\mf{a}_1^*+\mf{a}_2^*)/2$. The other three M points are equivalent through a reciprocal lattice vector: $-\mf{M}_1\sim \mf{M}_1$, $-\mf{M}_2\sim \mf{M}_2$ and $-\mf{M}_3 \sim \mf{M}_3$.

The nearest-neighbor TB Hamiltonian is
\beq
\hat{H}=-J\sum_{\mf{R}}\sum_{j=1}^3 |\mf{R}+\mf{a}_j,B\rangle\langle \mf{R},A|+\tr{h.c.}
\eeq
where $\mf{a}_3=\mf{0}$ and $J>0$ is the hopping amplitude \cite{Wallace1947}. It is represented by a $2 N\times 2 N$ matrix, where $N$ is the number of unit cells and there are two basis states (two ``atomic orbitals'') per unit cell $| \mf{R},l \rangle$ with $l=A,B$ and therefore two bands.
The TB Hamiltonian contains only information about the connectivity of basis states but not about their spatial location (in particular, the relative position between $A$ and $B$ sublattices is arbitrary at this point). Using Bloch's theorem, the eigenstates (known as Bloch states) $|\psi_{n\mf{k}}\rangle$ and eigenenergies $E_{n}(\mf{k})$ satisfy
\beq
\hat{H} |\psi_{n\mf{k}}\rangle = E_{n}(\mf{k}) |\psi_{n\mf{k}}\rangle
\eeq
where $\mf{k}$ spans the BZ and $n=\pm1$ is the band index. In the following we will use units such that $J=a=\hbar=1$.

We now perform a unitary transformation -- involving the position operator $\hat{\mf{r}}=\sum_{\mf{R},l=A,B}\mf{r}_l |\mf{r}_l\rangle\langle\mf{r}_l |$ -- to obtain a parameter-dependent Hamiltonian \cite{Xiao2010}:
\beqn
\hat{H}(\mf{k})&\equiv& e^{-i \mf{k} \cdot\hat{\mf{r}}} \hat{H} e^{i \mf{k} \cdot \hat{\mf{r}}}\\
&=&-\sum_{\mf{R}}\sum_{j=1}^3 e^{-i\mf{k}\cdot \mf{\delta}_j} |\mf{R}+\mf{a}_j,B\rangle\langle \mf{R},A|+\tr{h.c.}\nn
\eeqn
with $\mf{\delta}_j=\mf{a}_j+\mf{\delta}_3$. This Hamiltonian is also represented by a $2N\times 2N$ matrix for each value of the parameter $\mf{k}$. Its periodicity is clearly different from that of the BZ as $\hat{H}(\mf{k}+\mf{G})\neq \hat{H}(\mf{k})$ in general. This is because the unitary transformation contains the position operator $\hat{\mf{r}}$, the eigenvalues of which need not be Bravais lattice vectors (i.e., the position of $B$-sublattice sites here). It is important to realize that $\hat{H}(\mf{k})$ depends on the spatial embedding of basis states \cite{Haldane2014}, i.e. on the exact position in real space of the TB basis states. In the unitary transformation, the Bloch states are transformed into cell-periodic Bloch states
$|u_{n,\mf{k}}\rangle=e^{-i \mf{k} \cdot\hat{\mf{r}}}|\psi_{n\mf{k}}\rangle$. For a given $\mf{k}$, the latter are two ($n=\pm 1$) particular eigenvectors of $\hat{H}(\mf{k})$ among $2N$ and read
\beqn
|u_{n\mf{k}}\rangle&=&\f{1}{\sqrt{2N}}\sum_{\mf{R}}\biggl(|\mf{R},A\rangle+n\, e^{i\phi(\bf{k})}|\mf{R},B\rangle\biggr)\nn \\
&=&\f{1}{\sqrt{2}}\left(|\tilde{\mf{k}}=0,A\rangle+n\, e^{i\phi(\bf{k})}|\tilde{\mf{k}}=0,B\rangle\right)
\label{unk}
\eeqn
where $f(\mf{k})=-\sum_{j=1}^3 e^{-i\mf{k}\cdot \boldsymbol{\delta}_j}$ and $\phi(\mf{k})=\textrm{Arg }f(\mf{k})$. They verify $\hat{H}(\mf{k})|u_{n\mf{k}}\rangle=E_{n}(\mf{k})|u_{n\mf{k}}\rangle$, where the band energy spectrum is $E_{n}(\mf{k})=n|f(\mf{k})|$. In the second line of Eq. (\ref{unk}), we introduced Fourier modes for the basis states $|\mf{R},A\rangle=(1/\sqrt{N})\sum_{\tilde{\mf{k}}\in BZ}e^{-i\tilde{\mf{k}}\cdot \mf{R}}|\tilde{\mf{k}},A\rangle$, $|\mf{R},B\rangle=(1/\sqrt{N})\sum_{\tilde{\mf{k}}\in BZ}e^{-i\tilde{\mf{k}}\cdot(\mf{R}+\boldsymbol{\delta}_3)}|\tilde{\mf{k}},B\rangle$. Note that Eq. (\ref{unk}) does not depend on a specific choice for the Fourier transform. More generally, the complete eigenvalue equation of $\hat{H}(\mf{k})$ reads:
\beq
\hat{H}(\mf{k}) \left[e^{i \mf{k} \cdot\hat{\mf{r}}}|\psi_{n,\mf{k'}}\rangle \right]=E_{n}(\mf{k'})\left[e^{i \mf{k} \cdot\hat{\mf{r}}}|\psi_{n,\mf{k'}}\rangle \right]
\eeq
where $\mf{k}$ is a fixed parameter and $\mf{k'}$ is a quantum number spanning the BZ. Only when $\mf{k'}=\mf{k}$ does the eigenvectors $e^{i \mf{k} \cdot\hat{\mf{r}}}|\psi_{n,\mf{k'}}\rangle$ reduce to $|u_{n,\mf{k}}\rangle$.

Due to crystal momentum conservation (see Appendix \ref{ps}), we can project the Hamiltonian on each $\mf{k}$ subspace to obtain a $2\times 2$ matrix -- called $H(\mf{k})$ -- written in the $\mf{k}$-independent $\{|\tilde{\mf{k}}=0,A\rangle,|\tilde{\mf{k}}=0,B\rangle\}$ basis \cite{FPGM2010}. This matrix only acts in band (or sublattice) subspace and has the following expression:
\beq
H(\mf{k})
=\left(\begin{array}{cc}0&f(\mf{k})^*\\f(\mf{k})&0 \end{array} \right) \textrm{ with } f(\mf{k})=-\sum_{j=1}^3 e^{-i\mf{k}\cdot \boldsymbol{\delta}_j} \, .
\label{2x2bh}
\eeq
Note that $f(\mf{k}+\mf{G})=e^{-i \mf{G}\cdot \boldsymbol{\delta}_3}f(\mf{k})\neq f(\mf{k})$ in general. In the same basis, the cell-periodic Bloch states are spinors: 
\beq
|u_{n}(\mf{k})\rangle=\frac{1}{\sqrt{2}}\left(\begin{array}{c}1\\n e^{i\phi(\mf{k})}\end{array}\right)
\eeq
where $\phi(\mf{k})=\textrm{Arg }f(\mf{k})$ is the azimuthal angle along the equator of the Bloch sphere (the polar angle being $\theta(\mf{k})=\pi/2$ here). The change of notation from $|u_{n\mf{k}}\rangle$ to $|u_{n}(\mf{k})\rangle$ is here to emphasize that the two objects are different (the first is of size $2N$, the second of size $2$) and that in the second $\mf{k}$ plays the role of a parameter and no longer that of a quantum number. These spinors are the two eigenvectors of $H(\mf{k})$ for a given $\mf{k}$:
\beq
H(\mf{k})|u_{n}(\mf{k})\rangle=E_{n}(\mf{k})|u_{n}(\mf{k})\rangle
\eeq
In the following, we will mainly work with the $2\times 2$ matrix $H(\mf{k})$ and only rarely refer back to the $2N\times 2N$ Hamiltonians $\hat{H}$ and $\hat{H}(\mf{k})$. For a discussion of the relation between the three Hamiltonians see Appendix B in Ref. \cite{FPGM2010}. $H(\mf{k})$ is now the parameter-dependent Hamiltonian that will be called ``the Bloch Hamiltonian'' in the following. It is the starting point of our study of geometrical effects of Bloch states \cite{Xiao2010}.

\subsection{Reciprocal space trajectories}
We now consider the six trajectories in the reciprocal space shown in Fig. \ref{energylanda}. Geometrically, they are simply depicted as piecewise linear trajectories that begin at a $\Gamma$ point ($-\mf{a}_1^*$ in the figure) and first uniformly accelerate to the next $\Gamma$ point via the nearest M point. Then there are six possible choices for the second segment of the trajectory that ends at another $\Gamma$ point. The equation of motion is $\dot{\mf{k}}(t)=\mf{F}(t)$ with the force $\mf{F}(t)$ being a \textit{piecewise-constant} function of time with magnitude $F$. We choose the initial condition as $\mf{k}(t_i)=-\mf{a}_1^*$ and parameterize the trajectory by the variable $Ft$ with the initial/final point $Ft_{i,f}=\mp 4\pi/3$; $t=0$ corresponds to $\mf{k}(0)=0$, i.e. to the ``central'' $\Gamma$ point. The force profiles $\mf{F}(t)/F$ that realize the six trajectories are specified by:\\
\\
$\bullet$ case (I): $\f{\mf{a}_1^*}{|\mf{a}_1^*|}\,\Theta(-t)-\f{\mf{a}_1^*}{|\mf{a}_1^*|}\,\Theta(t)$\\
$\bullet$ case (II): $\f{\mf{a}_1^*}{|\mf{a}_1^*|} \,\Theta(-t) + \f{\mf{a}_2^*}{|\mf{a}_2^*|} \,\Theta(t) $\\
$\bullet$ case (III): $\f{\mf{a}_1^*}{|\mf{a}_1^*|} \,\Theta(-t) + \frac{\mf{a}_1^*+\mf{a}_2^*}{|\mf{a}_1^*+\mf{a}_2^*|} \,\Theta(t) $\\
$\bullet$ case (IV): $\f{\mf{a}_1^*}{|\mf{a}_1^*|}$\\
$\bullet$ case (V): $\f{\mf{a}_1^*}{|\mf{a}_1^*|}\, \Theta(-t) - \f{\mf{a}_2^*}{|\mf{a}_2^*|} \,\Theta(t)$\\
$\bullet$ case (VI): $\f{\mf{a}_1^*}{|\mf{a}_1^*|}\,\Theta(-t) - \frac{\mf{a}_1^*+\mf{a}_2^*}{|\mf{a}_1^*+\mf{a}_2^*|}\, \Theta(t)$\\

The resulting time-dependent Hamiltonians $H(t)=H(\mf{k}\to \mf{F}(t)t)$ correspond to the {\it same} energy landscape $E_\pm (t)=\pm \sqrt{5+4\cos \frac{3Ft}{2}}$ with $t_i\leq t \leq t_f$ (see Fig. \ref{energylandb}). It features two avoided crossings at M points with large energy gaps $\Delta=2$ at $Ft_1=-2\pi/3$ and $Ft_2=2\pi/3$, comparable to the bandwidth $W=6$. These trajectories all realize St\"uckelberg interferometers \cite{SAN2010}. Indeed, the particle initially starts in the lower band and goes through two avoided crossings at M points, that act as beamsplitters, realizing a two-path interferometer in energy-time space.

In our previous work \cite{LFM2014,LFM2015}, we showed that the energy landscape alone is not sufficient to describe fully the \s interferometer. Additional correction to the \s phase requires information about the eigenstates, which depend on the explicit form of the time-dependent Hamiltonian. These corrections can be attributed to purely geometric properties of the Bloch Hamiltonian.

\begin{figure}[ht]
\begin{center}
\subfigure[]{\raisebox{3mm}{\label{azimuthalphasea}\includegraphics[width=6cm]{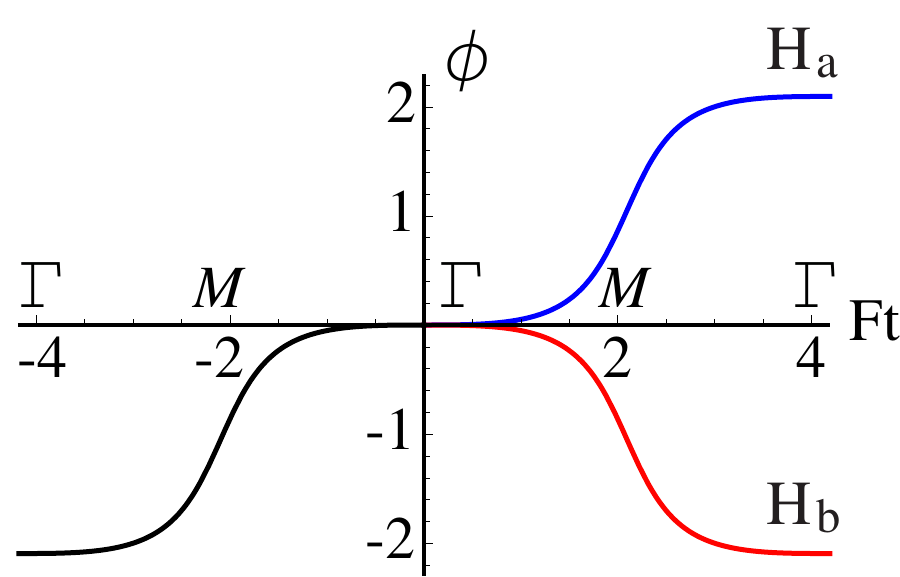}}}
\subfigure[]{\label{azimuthalphaseb}\includegraphics[width=7cm]{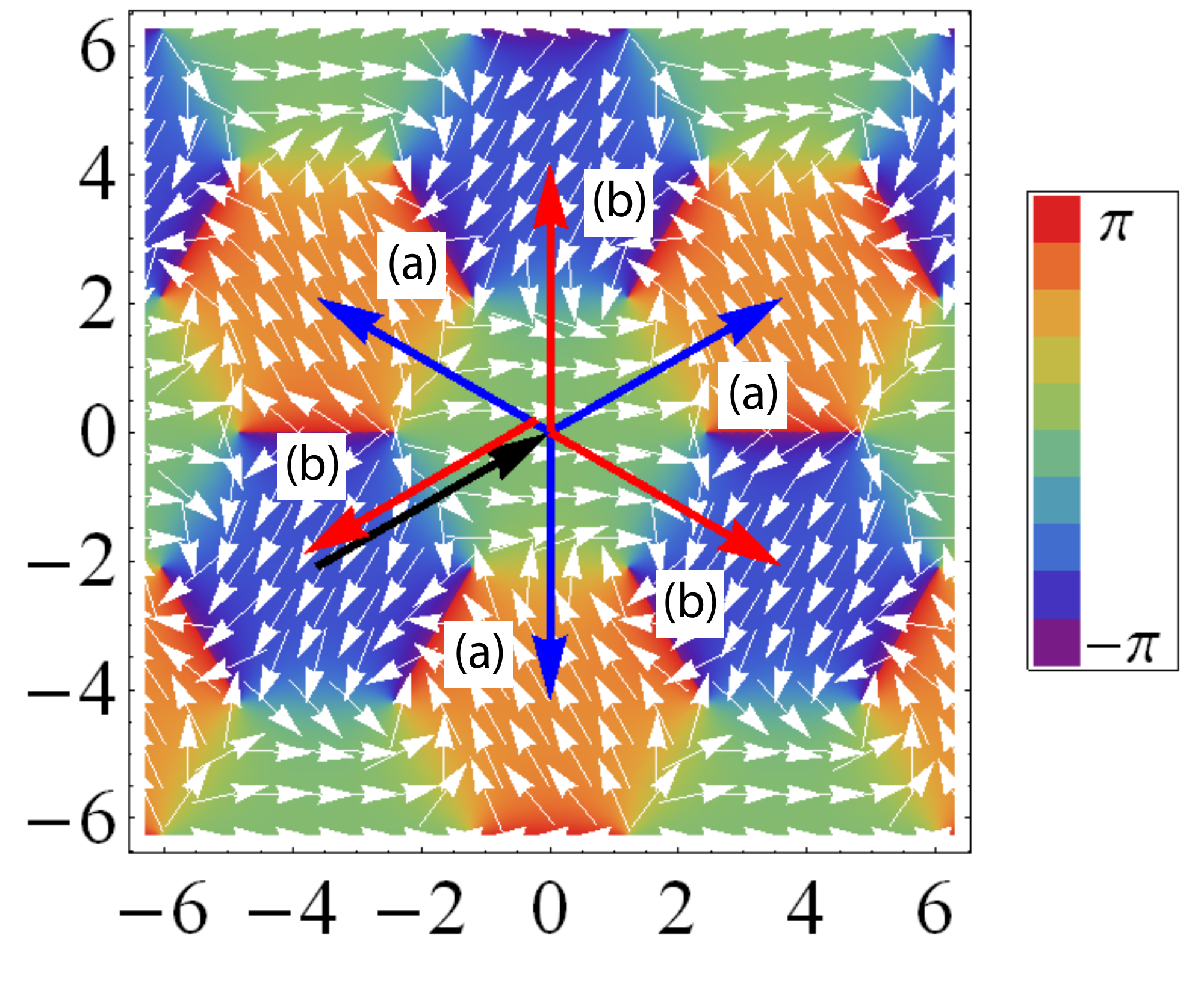}}
\end{center}
\caption{Pseudo-spin texture in reciprocal space. (a): The six trajectories in reciprocal space lead to only two different phase landscapes $\phi$ as a function of $Ft$ [in units of $\hbar/a$]. (b): Azimuthal angle $\phi(\mf{k})$ on the Bloch sphere plotted in the reciprocal space [$\mf{k}$ in units of $1/a$]. The color is proportional to the value of the angle. The first Brillouin zone is the smallest hexagonal region. This phase texture has an extended periodicity with a tripled unit cell in reciprocal space. The six trajectories which give rise to only two phase landscapes (in blue, case (a), and in red, case (b)) are represented.}
\label{azimuthalphase}
\end{figure}
\subsection{Lattice \s interferometry}
The six trajectories shown in Fig. \ref{energyland}(a) actually give rise to only {\it two} different time-dependent Hamiltonians differing by a sign$(t)$ function. Cases (II), (IV) and (VI) correspond to:
\beq
H_a(t)=-(\cos Ft +2 \cos \frac{Ft}{2})\sigma_x + (2 \sin \frac{Ft}{2}-\sin Ft)\sigma_y
\label{hat}
\eeq
and cases (I), (III) and (V) to:
\beq
H_b(t)=-(\cos Ft +2 \cos \frac{Ft}{2})\sigma_x - \textrm{sign}(t)(2 \sin \frac{Ft}{2}-\sin Ft)\sigma_y.
\label{hbt}
\eeq
Therefore there are only two different St\"uckelberg interferometers, which we refer to as case (a) and case (b). It is convenient to use the Bloch sphere parametrization $H(t)=E_+(t)\,\mf{h}(t)\cdot \boldsymbol{\sigma}$ with $h_x=\sin \theta \cos \phi$, $h_y=\sin \theta \sin \phi$, $h_z=\cos \theta$, for $H(t)=H_{a,b}(t)$ from Eqs. (\ref{hat}) and (\ref{hbt}), respectively. In the two cases, the motion of the unit vector $\mf{h}(t)$ is restricted to the equator of the Bloch sphere ($\theta=\pi/2$), and the only degree of freedom is the azimuthal angle $\phi(t)$, shown in Fig. \ref{azimuthalphasea}. The dependence $\phi(t)$ reflects the underlying pseudospin structure of the Bloch Hamiltonian in the reciprocal space ($\theta(\mf{k})=\pi/2$ and $\phi(\mf{k})$ is plotted in Fig. \ref{azimuthalphaseb}) along the trajectory. One notes that the unit cell of the pseudospin structure is three times as large as that of the first Brillouin zone (the smallest hexagonal region). The two different $H_{a,b}(t)$ reflect the extended periodicity of the pseudospin structure.

There appears to be some confusion in the literature as to the actual ``choice" -- sometimes referred to as a ``gauge choice'' -- of $\mf{k}$-dependent Hamiltonian deriving from tight-binding models (see e.g. \cite{Wu2012,Jackson2014}), which also determine the time-dependent Hamiltonians considered in this work (for a discussion, see Refs.~\cite{BM2009,FPGM2010,Fruchart2014,Milovanovic2015}). Indeed, it may seem dubious, at first sight, to see that the Bloch Hamiltonian, shown in Fig. \ref{azimuthalphaseb} for the TB model on the honeycomb lattice does not have the periodicity of the reciprocal lattice, while most physical quantities such as the energy band dispersion $E_{n}(\mf{k})$, or the Berry curvature $\Omega_{n}(\mf{k})$, do have this periodicity \cite{Xiao2010,FPGM2010,Fruchart2014}. There actually is no such choice but one unique Bloch Hamiltonian that follows from the TB model when studying the effect of an external force and which has the explicit form given in Eq. (\ref{2x2bh}) (see Appendix \ref{ps}). It is therefore interesting to ask whether the \s interferometer, which covers various Brillouin zones, can reveal the enlarged pseudospin structure in the reciprocal space.

\subsection{Summary of the problem}
The lattice \s interferometer problem is described by the time-dependent Schr\"{o}dinger equation (\ref{tdse}). A particle is initially prepared in the ground state $|\Psi(t_i)\rangle=|u_{-}(\mf{F}t_i) \rangle$, eigenstate of the Bloch Hamiltonian (at a $\Gamma$ point). Under the action of an applied force $\mf{F}(t)$, it evolves in the reciprocal space, following a path $\mf{k}(t)$, to reach a final state $| \Psi(t_f) \rangle$. We are interested  in the probability $P^{+-}(t_f)$ for the particle to have tunneled into the upper band
\be P^{+-}(t_f)= |\langle u_{+}(\mf{F}t_f)| \Psi(t_f) \rangle|^2 \ee
and, of course, the probability to stay in the lower band is $P^{--}(t_f)= 1 - P^{+-}(t_f)$.

We first write the state of the system at an arbitrary time using a general ansatz $|\Psi(t)\rangle = \sum_{n} C_n(\mf{k}(t))|u_{n}(\mf{k}(t))\rangle$ expressed in the adiabatic basis but involving all bands \cite{Houston1940} and substitute it into Eq. (\ref{tdse}) to obtain (see appendix A in Ref.~\cite{LFM2015} for a derivation)
\beqn
i\frac{d}{dt}C_{n}(\mf{k})&=&[E_n(\mf{k})-\mf{F}\cdot \boldsymbol{\mathcal{A}}_{\,n}(\mf{k})]\,C_n(\mf{k})\nn\\
&&-\sum_{n'\neq n} \mf{F}\cdot \boldsymbol{\mathcal{A}}_{\,n,n'}(\mf{k})\,C_{n'}(\mf{k}).
\eeqn
The merit is to spell out the adiabatic and non-adiabatic contributions to the amplitudes $C_n$. The adiabatic approximation would consist in retaining the two first terms ($\propto C_n$) on the right hand side and neglecting the third one ($\propto C_{n'\neq n}$). The first term depends on the band energy spectrum $E_n(\mf{k})$ and gives the dynamical phase (see below Eq. (\ref{4p})) and the second gives the line integral of the diagonal Berry connection $\boldsymbol{\mathcal{A}}_{\,n}(\mf{k})=\boldsymbol{\mathcal{A}}_{\,n,n}(\mf{k})=\langle u_{n}(\mf{k})|i\partial_{\mf{k}}|u_{n}(\mf{k})\rangle$, which results in a geometric phase \cite{Berry1984} (these two terms together contribute to the adiabatically accumulated phase in the $n^{th}$ band). The third term represents the coupling to other bands ($n'\neq n$) and depends on the off-diagonal Berry connection $\boldsymbol{\mathcal{A}}_{\,n,n'}(\mf{k})=\langle u_{n}(\mf{k})|i\partial_{\mf{k}}|u_{n'}(\mf{k})\rangle$ (giving rise to interband transitions).

In the case of the honeycomb TB model, there are two bands $n=\pm$ and by writing $C_n(t)\equiv c_n(\mf{k}(t))e^{-i\int^t dt' E_n(\mf{k}(t'))}e^{i\int^t dt' \mathcal{A}_n (t')}$ the equations simplify to
\beqn
\frac{d}{dt}c_+&=&i\mathcal{A}_{+,-}\, e^{i\int^t dt' [E_+ -E_-]}e^{i \int^t dt' [\mathcal{A}_- -\mathcal{A}_+]}\,c_- \nn\\
\frac{d}{dt}c_-&=&i\mathcal{A}_{-,+} \,e^{i\int^t dt' [E_- -E_+]}e^{i \int^t dt' [\mathcal{A}_+ -\mathcal{A}_-]}\,c_+
\label{cfode}
\eeqn
where $\mathcal{A}_{\,n,n'}(t)\equiv \langle u_{n}(\mf{k}(t))|i\partial_t|u_{n'}(\mf{k}(t))\rangle=\mf{F}\cdot \boldsymbol{\mathcal{A}}_{\,n,n'}(\mf{k}(t))$ using $d\mf{k}/dt=\mf{F}$. 
The particle is initially in the lower band $|u_{-}(\mf{F}t_i)\rangle$ (given by $c_+(t_i)=0$ and $c_-(t_i)=1$), and the quantity to be computed is the final probability for the particle to be found in the upper band $|u_{+}(\mf{F}t_f)\rangle$, which is given by $P=|c_+(t_f)|^2$.

\medskip

We emphasize now the differences with our previous work on \s interferometry with low-energy Hamiltonians for Dirac cones \cite{LFM2014,LFM2015}:

$\bullet$ Firstly, the energy landscape is bounded by a finite bandwidth $W=6$ fixed by the TB model bandstructure. Therefore there is no meaning to the question of approaching the asymptotic state of uncoupled bands \cite{SAN2010}. While the band structure is fixed by the TB model, the other parameter is the magnitude of the force, which determines how fast the sequence is executed and the non-adiabaticity in passing the avoided crossings. These remarks pose two important differences from the conventional LZ problem \cite{LZ1932}, or the double Dirac cones Hamiltonian that we studied \cite{LFM2014,LFM2015}. First, the bands are not uncoupled at the initial and final times. This may suggest that the adiabatic impulse model relying on asymptotic states could be less accurate. Second, the extend of the region $\delta k$ where LZ tunneling process take place is no longer sharply defined at the avoided crossing $\delta k/(F t_f-F t_i)\sim 1$. As we shall show, the transition probabilities are not affected much and the geometric phase is robust against the finite bandwidth effect.

$\bullet$ Secondly, the two avoided crossings at the M points in the lattice are not ``aligned" in the pseudospin space in the standard way \cite{SAN2010,LFM2014,LFM2015}. We will make this notion of alignment precise in the following. For now, it suffices to mention that the geometric expression derived from the heuristic \s theory and the adiabatic impulse model for the double Dirac cones Hamiltonian cannot be applied directly \cite{LFM2014,LFM2015}. We therefore have to first untwist the double LZ problem by applying a time-dependent unitary transformation to $H_{a,b}(t)$ \cite{Berry1990}. Only then the problem can be mapped to the framework of our previous methods.

$\bullet$ Thirdly, even the LZ process across a single M point in the sudden limit $(F\rightarrow\infty)$ already shows some interesting feature. In this limit, the time-evolution operator is trivially an \textit{identity}, i.e., the state remains as the initial state at all later times \cite{Messiah}. A measurement at time $t_f$ therefore simply amounts to a \textit{projection} of the initial state onto the instantaneous measurement (or adiabatic) basis at $t=t_f$, which is the eigenstate $|u_{n}(\mf{F}t_f)\rangle$. The overlap needs not take value of 0 or 1, since the cell-periodic Bloch functions are not orthogonal at different $\mf{k}$ points. We comment on this sudden limit in section \ref{secsudden}.

\section{Numerical solution}
\label{secnum}
Substituting $H_{a,b}(t)$ into the time-dependent Schr\"{o}dinger equation (\ref{cfode}) and after rescaling ($\tau=Ft, a_{\pm}(\tau)=c_{\pm}(t)$), we obtain the equations
\beqn
\frac{da_+}{d\tau}&=& f_F(\tau)a_-(\tau)\nn\\
\frac{da_-}{d\tau}&=& -f_F(\tau)^*a_+(\tau)
\eeqn
where $f_F^{(a)}(\tau)=-\frac{i}{2} \frac{\cos\frac{3\tau}{2}-1}{4\cos\frac{3\tau}{2}+5} e^{i \frac{8}{F}E(\frac{3\tau}{4},\frac{8}{9})}$ for $H_a(t)$ and $f_F^{(b)}(\tau)= -\textrm{sign}(\tau)\times f_F^{(a)}(\tau)$ for $H_b(t)$, with $E(\phi,m)=\int_0^\phi d\theta \sqrt{1-m\sin^2 \theta}$ the elliptic integral of the second kind. Note that $F$ is the only relevant parameter controlling the tunneling regime: adiabatic when $F\ll 1$ and sudden when $F\gg 1$. In other words, the adiabaticity parameter $\delta$, which should be dimensionless and large (resp. small) in the adiabatic (resp. sudden) limit, is expected to be proportional to $F^{-1}$. The equations are exact irrespective of the regime. We numerically solve these equations and present the results below.

\subsection{Single M point avoided crossing: saturation}
\begin{figure}[h!]
\begin{center}
\includegraphics[width=6.cm]{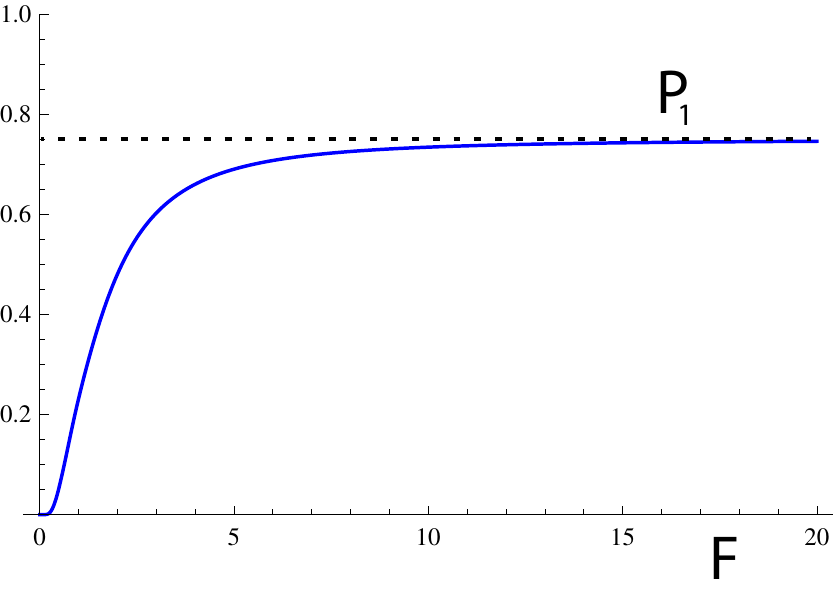}
\end{center}
\caption{Transition probability $P_1$ for the single M point crossing as a function of the force $F$ [in units of $J/a$], together with the saturation probability $3/4$.}
\label{singleMpoint}
\end{figure}
To illustrate the structure of the dynamics, we first focus on the trajectory which traverses only one single M point -- a Landau-Zener problem. We use $\tau_i=-4\pi/3$ and $\tau_f=0$, i.e., from the initial $\Gamma$ point to the next $\Gamma$ point, in the numerical integration. The result is shown in Fig. \ref{singleMpoint} with the transition probability denoted as $P_1$. As anticipated in the last section, in the sudden limit regime ($F\gg 1$), we find the saturation of the probability at $3/4$ (not 0 or 1) \cite{Blochnotes}. We will explain the saturation value in Sec. \ref{secsudden}. Note also that the phase acquired upon being reflected at the avoided crossing is no longer given by the Stokes phase, as for a conventional LZ process, but by a new phase that we call $\varphi_1$ and numerically compute in Appendix \ref{app:dpt}.

\subsection{Double M point avoided crossings}
\begin{figure}[ht]
\begin{center}
\subfigure[]{\label{numa}\includegraphics[width=6cm]{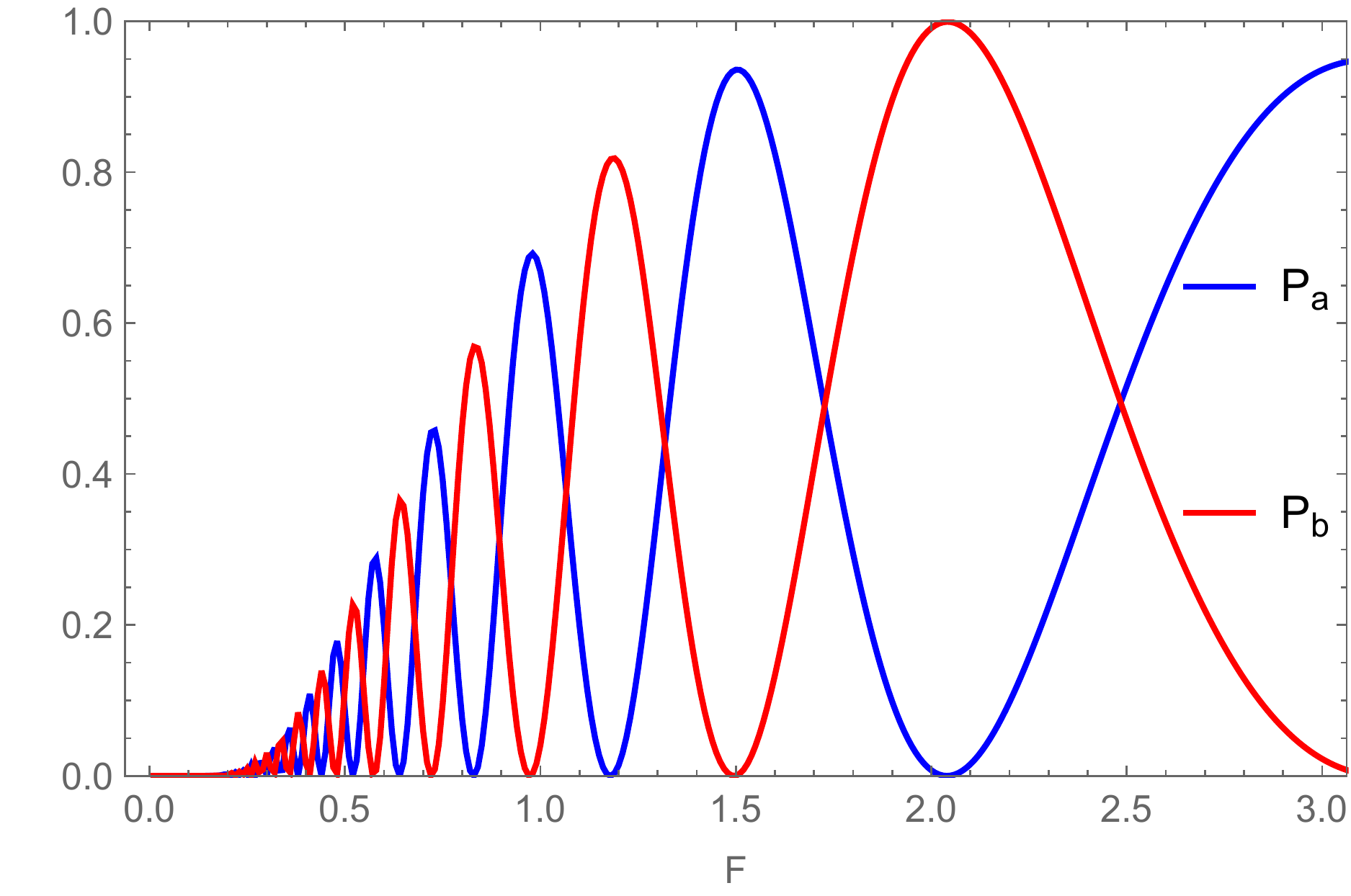}}
\subfigure[]{\label{numb}\includegraphics[width=6cm]{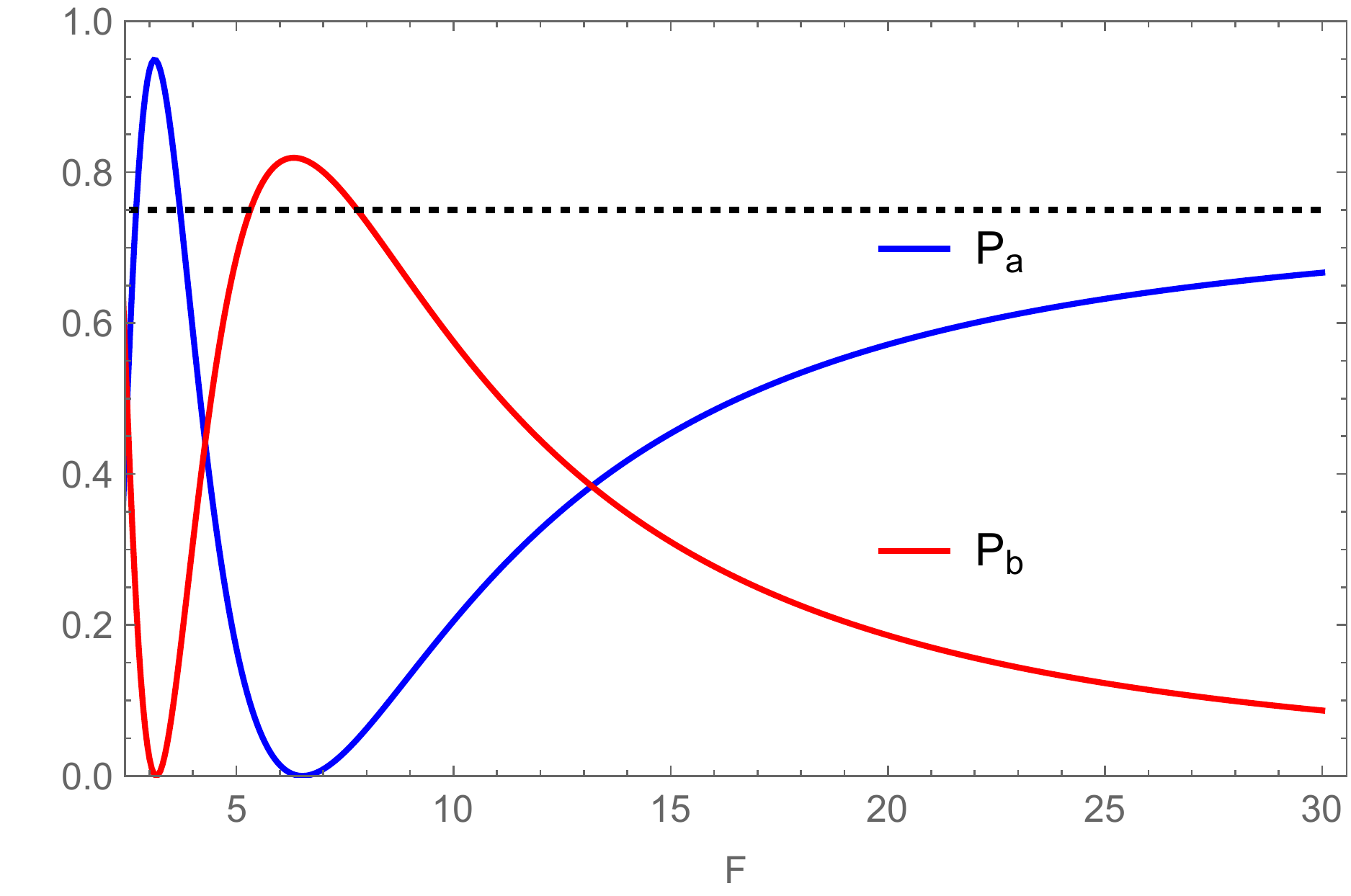}}
\end{center}
\caption{Interband transition probabilities $P_a$ (in blue) and $P_b$ (in red) computed numerically as a function of the force $F$ [in units of $J/a$] in the range: (a) between 0 and 3, (b) between 3 and 30. Note that $P_a\to 3/4$ (dashed line) and $P_b\to 0$ when $F\gg 1$. Note that $P_a$ and $P_b$ can be larger than $3/4$ for intermediate forces.}
\label{num}
\end{figure}

We now consider a trajectory which traverses two M points by setting $\tau_i=-\tau_f=-4\pi/3$. As discussed above, there are two distinct lattice \s interferometers. By calling $P_a$ and $P_b$ the associated transition probabilities for $H_{a,b}(t)$, results are shown from the adiabatic regime in Fig. \ref{num}(a), to the sudden limit in Fig. \ref{num}(b).  There is a clear $\pi$-shift between the two cases for \textit{all} $F$. We also see the saturation effect happening for $P_a$ ($P_b$) to $3/4$ (0) in the sudden limit. 

\begin{figure}[ht]
\begin{center}
\includegraphics[width=6.cm]{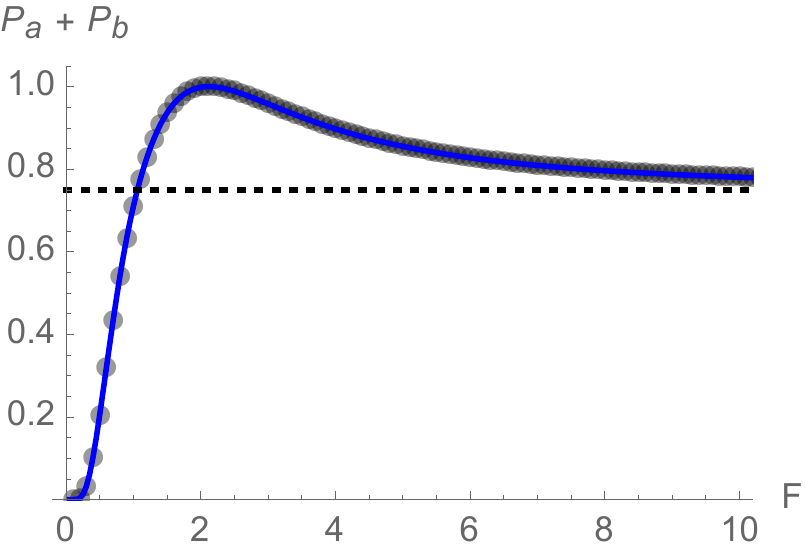}
\end{center}
\caption{$P_a+P_b$ (full blue line) plotted as a function of the force $F$ [in units of $J/a$] between 0 and 10, with $P_a$ and $P_b$ numerically computed for a double M point avoided crossing (see Fig. \ref{num}). Also shown is $4P_1(1-P_1)$ (black circles) where $P_1$ is numerically obtained for a single M point avoided crossing (see Fig. \ref{singleMpoint}). The dashed line shows the saturation at $3/4$ which occurs when $F\gg 1$. Note the maximum $P_a+P_b=1$ when $F\approx 2.1$.}
\label{papluspb}
\end{figure}

To show that these oscillations are indeed of the \s form~:  $P_{a,b}= 4 P_{LZ}(1-P_{LZ})\sin^2( {\varphi_\textrm{tot}^{a,b} \over 2})$ with an exact $\pi$ shift~: $\varphi_\textrm{tot}^{b}= \varphi_\textrm{tot}^{a} + \pi$, we have checked in Figure \ref{papluspb} that $P_a+P_b=4 P_1(1 - P_1)$, where $P_1$ is the probability after traversal of a single M point. This result shows numerically that the \textit{$\pi$-shift is geometrical in origin} -- a quantity independent of the force --  and that the interferometer actually consists in a double beamsplitter -- i.e. a two-path interferometer -- with probabilities $P_1$  and $1-P_1$ of tunneling or not at each M point. Interestingly, \s oscillations with maximum contrast are seen close to $F\approx 2.1$ (see Fig. \ref{num}(a)), corresponding to $P_1=1/2$ and $P_a+P_b=1$ (see Fig. \ref{papluspb}).

\subsection{Multiple M point avoided crossings}
\begin{figure}[ht]
\begin{center}
\subfigure[]{\includegraphics[width=6cm]{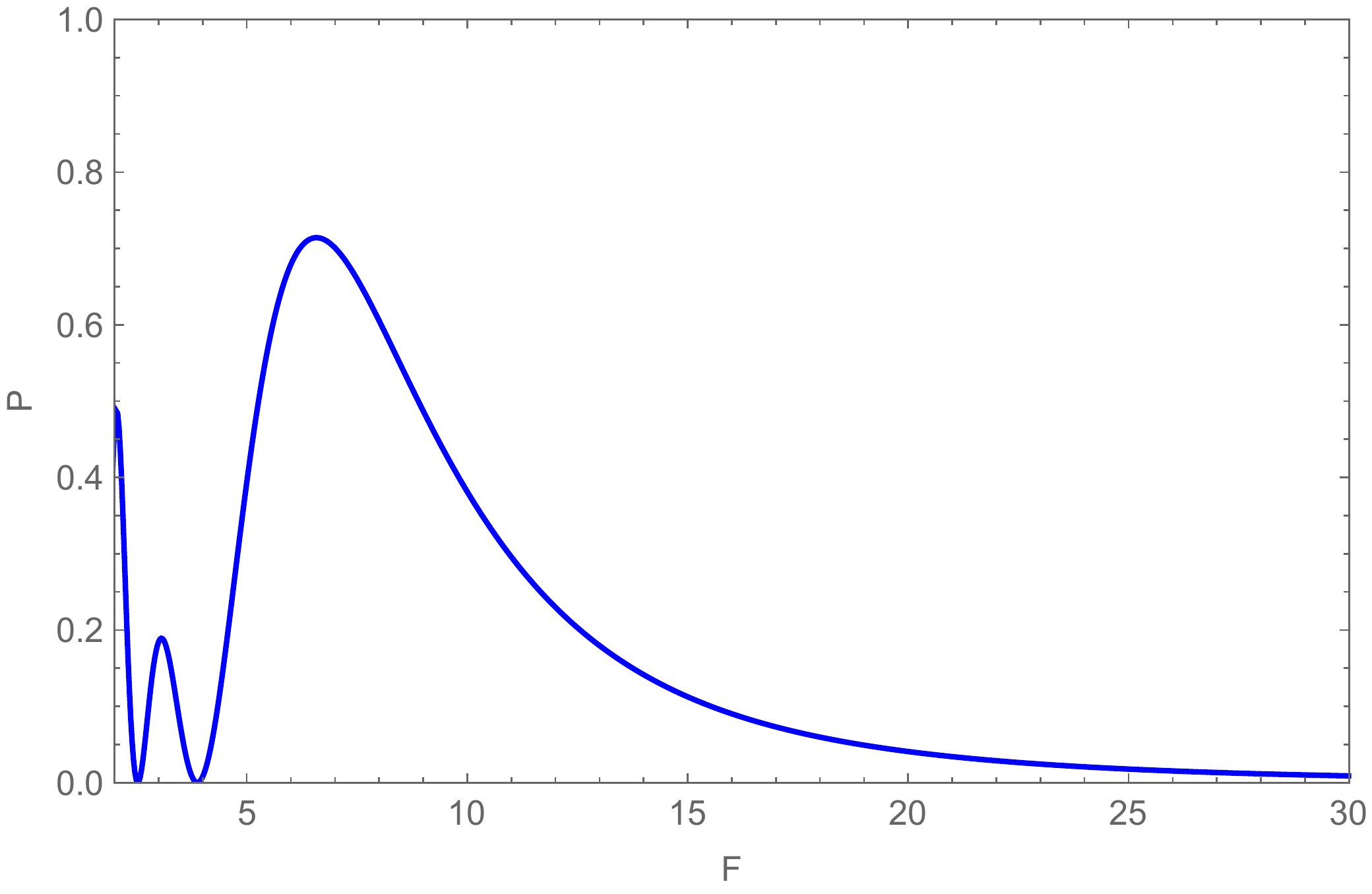}}
\subfigure[]{\includegraphics[width=6cm]{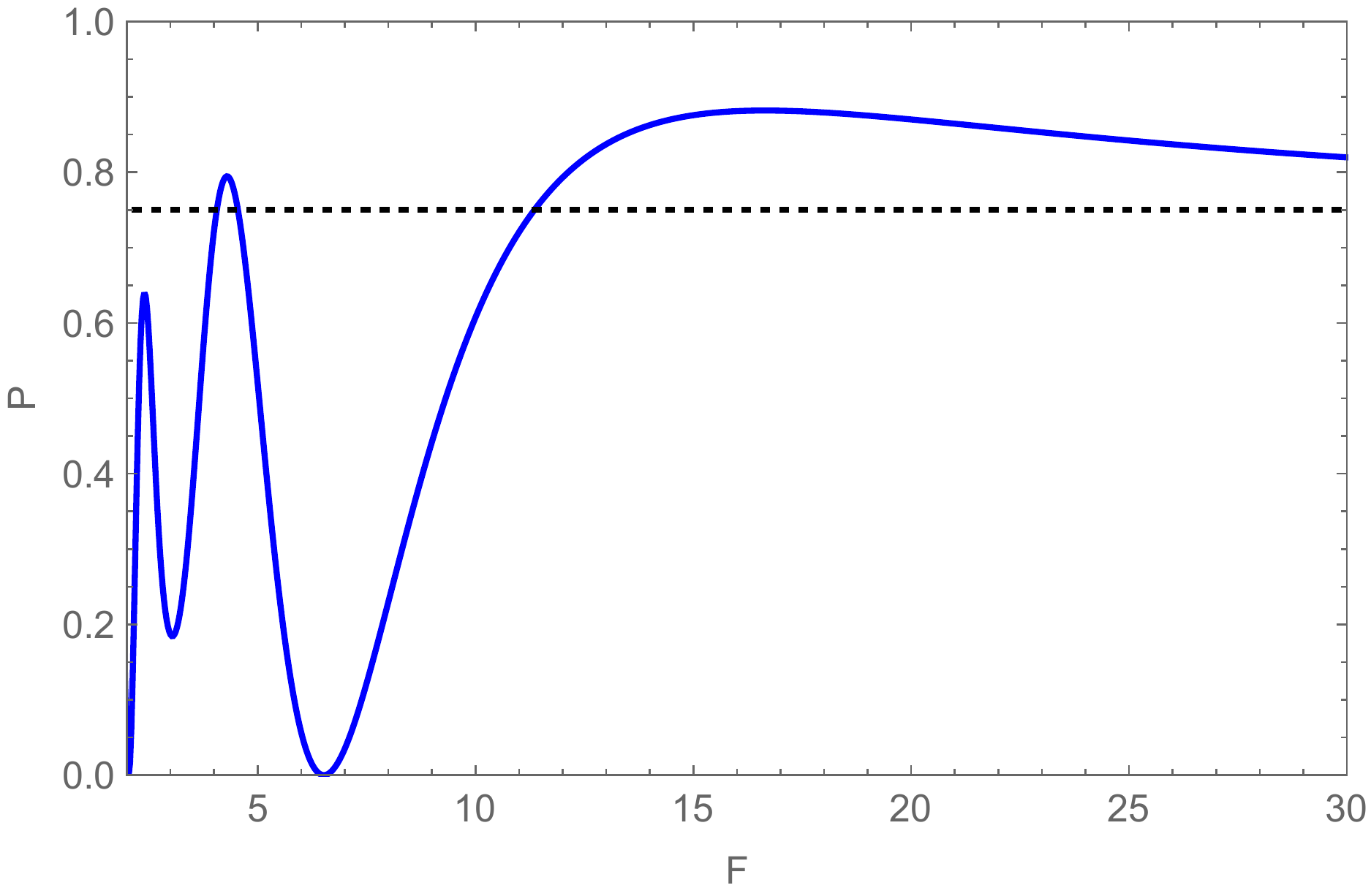}}
\end{center}
\caption{Multiple M point crossings in straight line along $\mf{a}_1^*$. Interband transition probability $P$ as a function of the force $F$ between 2 and 30 [in units of $J/a$] for (a) a triple M point and (b) a quadruple M point. The dashed line indicates a probability of $3/4$.} \label{tripleMpoint}
\end{figure}

To complete our study, we present the interband transition probability $P^{+-}(t_f)$ after three or four avoided crossings along a straight trajectory parallel to $\mf{a}_1^*$ corresponding to the continuation of a double M point interferometer of case (a) (see trajectory IV in Fig. \ref{energyland}(a)). For triple M point crossings, we find a saturation to $0$ (see Fig. \ref{tripleMpoint}(a)). For quadruple M point crossings still in straight line, we find a saturation to $3/4$ (see Fig. \ref{tripleMpoint}(b)). This saturation is now approached from above, in contrast to the cases of a single or a double M point crossing. We will discuss the saturation in section \ref{secsudden}.

\medskip

In the next section, we use several analytical approximations in order to understand the above numerical results in the case of a double M point interferometer.

\section{Adiabatic limit: geometric phase}
\label{secadia}
In the adiabatic limit $F\ll 1$, we study the generalized adiabatic impulse model \cite{SAN2010,LFM2014,LFM2015}, and the first-order adiabatic perturbation theory, following closely Refs.~\cite{FLM2012,LFM2015}, to compute the interband transition probabilities.

\subsection{Adiabatic impulse model}
The idea of the adiabatic impulse model \cite{SAN2010} is to separate the St\"uckelberg interferometer into three blocks. This identification comes from recognizing the \s interferometer as a two-path interferometer. The first block is the non-adiabatic interband transition at the first avoided crossing (M point) playing the role of a first beamsplitter. The second block is the adiabatic evolution along the two paths in energy-time space in between the two avoided crossings. The third block is the non-adiabatic process at the second M point playing the role of the second beamsplitter. For $H_{a,b}(t)$, the first and second avoided crossings take place at $Ft_1=-2\pi/3$ and $Ft_2=2\pi/3$, respectively. Such an interferometer gives rise to a final probability with the following structure
\beq
P=4P_{LZ}(1-P_{LZ})\sin^2\frac{\varphi_\textrm{tot}}{2}
\label{4p}
\eeq
where $P_{LZ}=e^{-\pi W^2/V}$ is the LZ probability for an avoided crossing locally described by a Hamiltonian $H(t)=Vt \sigma_z+W\sigma_x$ \cite{LZ1932} and $\delta=-\frac{1}{2\pi}\ln P_{LZ}$ is the adiabaticity parameter. The total phase of the interferometer $\varphi_\textrm{tot}=2\varphi_\textrm{S}+\varphi_\textrm{dyn}+\varphi_g$ is the sum of three terms: a Stokes phase $\varphi_\textrm{S}(\delta)=\frac{\pi}{4}+\delta (\ln \delta -1)+\textrm{Arg }\Gamma(1-i\delta)$ acquired at each beamsplitter, a dynamical phase $\varphi_\textrm{dyn}=\int_{t_1}^{t_2} dt [E_+(t)-E_-(t)]$ acquired during the adiabatic evolution along the two paths \cite{SAN2010} and possibly a phase shift of geometrical origin (called $\Delta \varphi$ or $\varphi_g$ in the following) \cite{LFM2015}. Next, we present two methods derived from the adiabatic impulse model that are closely related and provide a simple physical picture \cite{LFM2015}.

The first is the scattering matrix (or $N$-matrix) approach, where LZ processes are treated as scattering matrices (which may include geometric phases) and the final outcome of two successive LZ problems is treated as a total scattering matrix problem. The phase of geometric origin in the \s interferometer appears during the \textit{non-adiabatic transitions} and is called $\Delta \varphi$ below.

The second method is the heuristic St\"{u}ckelberg theory, where minimum information on the non-adiabatic transitions is used (which do not include geometric phase) and the adiabatically accumulated phase is separately computed for the upper/lower band eigenstates. In the second method, the phase of geometric origin appears during \textit{the adiabatic evolution} in between the two avoided crossing and has the appearance of a geometric phase $\varphi_g$.

At various stages in the following computations, we will perform time-independent rotations in pseudo-spin space by an angle $\theta$ around the $\sigma_i$-axis, denoted by $R(\sigma_i,\theta)=e^{i\theta \sigma_i/2}$ (we perform mostly around the $\sigma_z$-axis). These rotations of pseudo-spin neither modify the spectrum nor the overlap between states.

\subsubsection{Case (a)}
Expanding up to linear order in $t$, the Hamiltonian $H_a(t)$ around $t_1=-2\pi/(3F)$ and $t_2=2\pi/(3F)$ for case (a) becomes
\beqn
H_a(t_1+t)&\approx& \biggl( -\f{1}{2}-\sqrt{3}\,Ft \biggr)\,\sigma_x+ \biggl( -\f{\sqrt{3}}{2}+Ft \biggr)\,\sigma_y,\nn\\
H_a(t_2+t)&\approx& \biggl( -\f{1}{2}+\sqrt{3}\,Ft \biggr)\,\sigma_x+ \biggl( \f{\sqrt{3}}{2}+Ft \biggr)\,\sigma_y.\nn
\eeqn
After the rotation $R(\sigma_z,-5\pi/6)$, it becomes $\tilde{H}_a(t)=R^\dag  H_a(t)  R$:
\beqn\label{twistHam}
\tilde{H}_a(t_1+t)&\approx& 2Ft \,\sigma_x+ \sigma_y,\\
\tilde{H}_a(t_2+t)&\approx& \biggl( \f{\sqrt{3}}{2}-Ft \biggr)\,\sigma_x+ \biggl( -\f{1}{2}-\sqrt{3}\,Ft \biggr) \sigma_y.\nn
\eeqn

\begin{figure}
\begin{center}
\includegraphics[width=7cm]{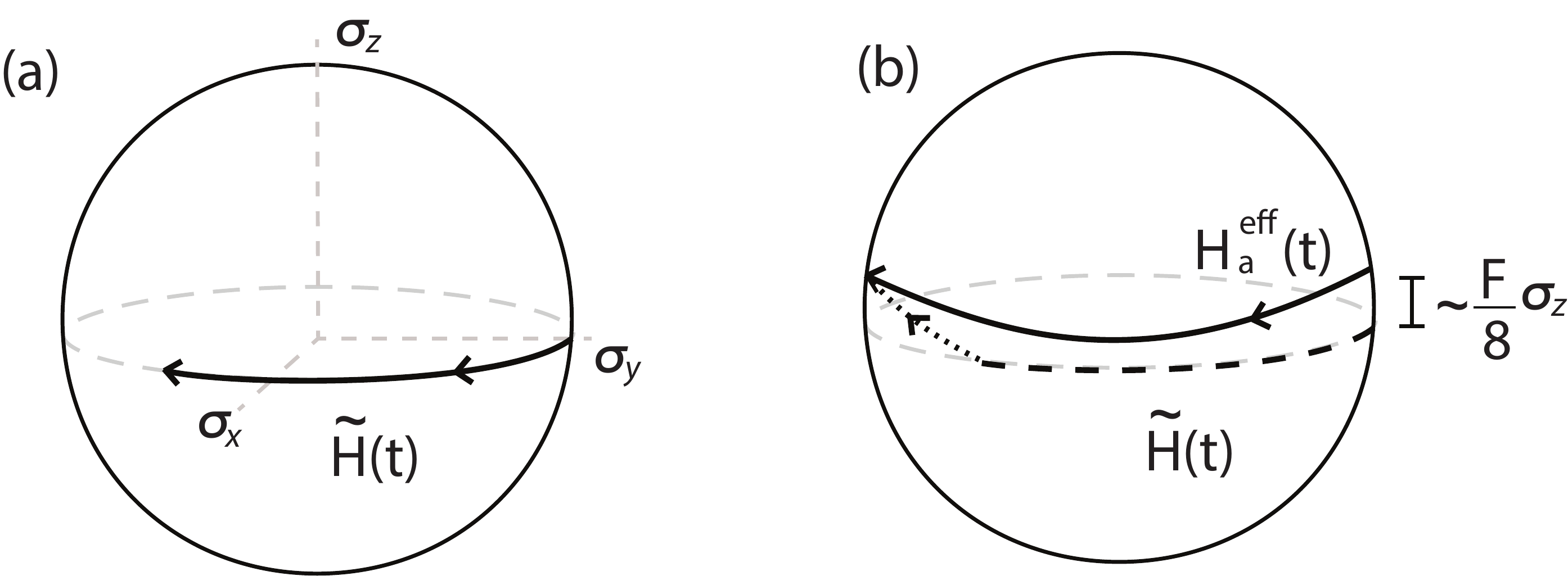}
\end{center}
\caption{(a) The original Hamiltonian curve of $\tilde{H}(t)$ where the initial and final points are on the equator and not antipodal. (b) The Hamiltonian curve for $H^\textrm{eff}(t)$, after performing the unitary transformation Eq. (\ref{hameff}), resembles that of boron nitride with a diagonal trajectory studied in \cite{LFM2015}. Note that the initial and final points belong to the same meridian so that the geodesic closure goes through the north pole. The area enclosed by the trajectory closed by a geodesic is almost a quarter of the sphere.}
\label{figlih}
\end{figure}

We note that the adiabatic outgoing states of $\tilde{H}_a(t_1+t) \approx  2F|t| \,\sigma_x$ for $Ft\gg 1$ do not join smoothly with the adiabatic incoming states of $\tilde{H}_a(t_2+t)\approx  F|t| \,\sigma_x+ \sqrt{3}\,F|t|  \,\sigma_y$ for $Ft\ll -1$. In the Bloch sphere representation, the pseudo-spin $\mf{h}_a(t_1+t)\approx \mf{e}_x$ points in another direction than $\mf{h}_b(t_2+t)\approx (\mf{e}_x+\sqrt{3}\mf{e}_y)/2$ (see Fig. \ref{figlih}(a)).
This shows that we are encountering a different kind of beamsplitters. In our previous studies \cite{LFM2014,LFM2015}, the beamsplitters realized in the vicinity of the Dirac cones were such that the outgoing states (upper and lower bands) of the first avoided crossing $t=t_1$ and the incoming states (upper and lower band) of the second avoided crossing $t=t_2$ did join smoothly. This proper ``alignement'' was due to the simple form of the linear Hamiltonians expanded around the Dirac cones.

In contrast, the adiabatic states of $\tilde{H}_a(t)$ experience an \textit{additional} twist at times $t=t_1^+$ and $t_2^-$. To tackle this problem we introduce a time-dependent unitary transformation around $\sigma_z$-axis to undo the twisting linearly in time:
\beq
U(t)=e^{\f{i}{2}\sigma_z\gamma(t)} \tr{\ \ with\ \ } \gamma(t)=-\f{\pi}{3}\f{t-t_1}{t_2-t_1}\tr{\ \ for \ \ }t_1<t<t_2.
\eeq
The idea of such an untwisting belongs to Berry, who applied it to the problem of a single avoided crossing \cite{Berry1990}.
The full time-evolution becomes
\beq\label{hameff}
\bigl[  U^\dag \tilde{H}_a U-i U^\dag (\partial_t U)   \bigr] \, |\Psi'\rangle=H_a^\textrm{eff}\, |\Psi'\rangle=i\partial_t|\Psi'\rangle
\eeq
where the part $U^\dag (t) \tilde{H}_a(t) U(t)$ is untwisted
\beqn
U^\dag (t_1) \tilde{H}_a(t_1+t) U(t_1)&\approx& 2Ft \,\sigma_x+ \sigma_y,\nn\\
U^\dag (t_2) \tilde{H}_a(t_2+t) U(t_2)&\approx& -2Ft \,\sigma_x- \sigma_y,\nn
\eeqn
at the cost of an additional constant $\sigma_z$ term (with $\gamma(t)$ linear in time)
\beq
-i U^\dag (\partial_t U)=\f{1}{2}\dot{\gamma}\,\sigma_z=\f{F}{8}\,\sigma_z.
\eeq
The full effective Hamiltonian $H_a^\textrm{eff}(t)=U^\dag \tilde{H}_a U-i U^\dag (\partial_t U)=\mf{h}(t)\cdot\boldsymbol{\sigma}$ contains the three Pauli matrices and has a modified adiabatic energy spectrum $E_{\pm}^\textrm{eff}(t)=\pm\sqrt{h_x(t)^2+h_y(t)^2+\f{F^2}{64}}$ with the minimum gap occurring at the same position as in the original problem, provided that $\dot{\gamma}$ is a constant. The untwisted local Hamiltonians now have the desirable property that the adiabatic outgoing states of $H_a^\textrm{eff}(t_1+t)\approx    2F|t|\,\sigma_x $ for $Ft\gg1$ do join smoothly with the adiabatic incoming states of $H_a^\textrm{eff}(t_2+t)\approx    2F|t|\,\sigma_x$ for $Ft\ll -1$. The adiabatic impulse model can now be applied.

\paragraph*{Scattering matrix approach.} 
In the first method, we rotate $H_a^\textrm{eff}$ into the LZ basis $(\sigma_x,\sigma_y,\sigma_z)\rightarrow
(-\sigma_z,\sigma_y,\sigma_x)$ to obtain
\beqn
H_a^{\tr{eff}}(t_1+t)&\approx& \bem -2Ft & - (-\f{F}{8}+i)\\ -
(-\f{F}{8}-i) & 2Ft \eem   \nn\\
H_a^{\tr{eff}}(t_2+t)&\approx& \bem 2Ft & - (-\f{F}{8}-i)\\ -
(-\f{F}{8}+i) & -2Ft \eem
\label{lham}
\eeqn
that have the form of standard LZ Hamiltonians $H(t)=Vt \sigma_z+\textrm{Re}W \sigma_x + \textrm{Im}W \sigma_y$ albeit with a complex gap $W$. The total phase $\varphi_\textrm{tot}$ of the \s interferometer consists of the sum of three terms: the Stokes phase $\varphi_\textrm{S}$, the dynamical phase $\varphi_\textrm{dyn}$ and the difference $\Delta \varphi_a$ in the phase of the two complex gaps - i.e., $-\f{F}{8}+i$ and $-\f{F}{8}-i$, see
Eq. (\ref{lham}) -- \cite{LFM2014, LFM2015}, which are given respectively as
\beq
\varphi_\textrm{dyn}=2\int_{t_1}^{t_2}dt\, E_{+}^\textrm{eff}(t),\tr{\ \ }\Delta \varphi_a =\pi+ 2 \arctan \biggl(\f{F}{8}\biggr).
\eeq
In the adiabatic limit $F\rightarrow 0$, $\varphi_\textrm{S}\rightarrow 0$, $\varphi_\textrm{dyn}$ becomes the dynamical phase of the original Hamiltonian (as $E_{+}^\textrm{eff}(t)\approx E_+(t)$) and $\Delta \varphi_a \approx \pi$. The final interband transition probability is given by $P_{a}=4 P_{LZ}(1-P_{LZ})\sin^2(\varphi_\textrm{S}+\varphi_\textrm{dyn}/2+\pi/2)\approx 4 P_{LZ} \sin^2(\varphi_\textrm{dyn}/2+\pi/2) $.

If one adds a staggered on-site potential $\pm M$ to the honeycomb lattice (which is a standard model for boron nitride or gapped graphene, see e.g. \cite{FPGM2010}), the Hamiltonian $H_{a}(t)$ becomes $H_{a}(t)+M\sigma_z$ and a gap opens in the band structure. One can perform the exact same derivation to obtain 
\beqn 
\Delta \varphi_a &=& \pi + 2 \arctan(\frac{F}{8}+M)\nn \\\
&\approx&-2\arctan \frac{1}{M} \, \textrm{ modulo }2\pi
\eeqn 
in the adiabatic limit. In this case, the phase shift is no longer quantized to $\pi$ but depends on the magnitude of the gap $M$. This agrees with the case \# 2 defined and studied in Ref. \cite{LFM2015}.

\paragraph*{Heuristic \s theory.} In the second method, we have the Stokes phase and the dynamical phase taking the same form as in the scattering matrix method, but the geometric phase is now acquired during the adiabatic evolution and given by \cite{LFM2014, LFM2015} 
\beqn
\varphi_g^a&=&\int_{t_1}^{t_2}dt\,\biggl(\langle u_-|i\partial_t|u_-\rangle-\langle u_+|i\partial_t|u_+\rangle\biggr)\nn\\
&+&\tr{arg} \langle u_-(t_1)| u_-(t_2) \rangle -\tr{arg} \langle u_+(t_1)| u_+(t_2) \rangle
\eeqn
with respect to an adiabatic basis of the effective Hamiltonian $H_a^\textrm{eff}(t)|u_{\pm}(t)\rangle=E^{eff}_{\pm}(t)|u_{\pm}(t)\rangle$. We can read off the value of the geometric phase without explicit computations, since we learned from our previous work \cite{LFM2015} that this quantity is equal to the area on the Bloch sphere that is enclosed by the Hamiltonian trajectory closed by a geodesic (here, because of the clockwise orientation of the trajectory, it is actually equal to minus that area). The phase is therefore given by $\varphi_g^a=-(\pi- 2 \arctan \bigl(\f{F}{8} \bigr))$, see Fig. \ref{figlih}(b). In fact, the Hamiltonian curve of the effective Hamiltonian $H_a^\textrm{eff}(t)$ is similar to the boron nitride case with a diagonal trajectory studied in Ref. \cite{LFM2015}. The term $F/8$ plays the role of the mass term in boron nitride (``gapped graphene'') and the adiabatic limit $(F\rightarrow 0)$ amounts to taking the massless limit. If one adds a staggered on-site potential $\pm M$, the Hamiltonian $H_{a}(t)$ becomes $H_{a}(t)+M\sigma_z$ and the area on the Bloch sphere becomes $\pi- 2 \arctan \bigl(\f{F}{8}+M \bigr)$ so that $\varphi_g^a=-\pi+ 2 \arctan \bigl(\f{F}{8}+M \bigr)$. In summary, in the adiabatic limit $F\to 0$, we arrive at the same result as the scattering matrix approach with $\varphi_g^a=\Delta \varphi_a=-\pi+2 \arctan M=\pi+2\arctan M=-2\arctan\frac{1}{M}$ modulo $2\pi$.

We note that the adiabatic limit $F \rightarrow 0$ is subtle here. While it renders the term $-iU^\dag (\partial_t U)=F\sigma_z/8$ vanishingly small, it does not render $U^\dag \tilde{H}_a U$ to an identity $\tilde{H}_a$.

\subsubsection{Case (b)}
Expanding up to linear order in $t$, the Hamiltonian $H_b(t)$ around $t=t_1$ and $t=t_2$ for case (b) becomes
\beqn
H_b(t_1+t)&\approx& \biggl( -\f{1}{2}-\sqrt{3}\,Ft \biggr)\,\sigma_x+ \biggl(- \f{\sqrt{3}}{2}+Ft \biggr)\,\sigma_y,\nn\\
H_b(t_2+t)&\approx& \biggl( -\f{1}{2}+\sqrt{3}\,Ft \biggr)\,\sigma_x- \biggl( \f{\sqrt{3}}{2}+Ft \biggr)\,\sigma_y.
\eeqn
By applying a constant rotation $R(\sigma_z,-\pi/3)$ they become
\beqn
R^\dag  H_b(t_1+t)  R &\approx& - \sigma_x+2Ft \,\sigma_y,\nn\\
R^\dag H_b(t_2+t)  R&\approx& - \sigma_x-2 Ft\,\sigma_y.
\eeqn
This is equivalent to the St\"{u}ckelberg interferometry problem that we have studied in \cite{FLM2012,LFM2015} without any geometric phase correction (i.e. $\Delta \varphi_b=0$). The interband transition probability is then given by $P_{b}=4 P_{LZ}(1-P_{LZ})\sin^2(\varphi_\textrm{S}+\varphi_\textrm{dyn}/2)\approx 4 P_{LZ}\sin^2(\varphi_\textrm{dyn}/2)$ for  $F\ll 1$.

\subsubsection{Discussion}
The key step in the above derivation is to untwist the original Hamiltonian to make it look like one of the universal Hamiltonian double Dirac cone problems that were studied in \cite{LFM2015}. The above considerations show that the $\pi$-shift between $P_a$ and $P_b$ can be understood either as the phase $\Delta \varphi$ of a complex gap (with $\Delta \varphi_a=\pi$ and $\Delta \varphi_b=0$) or as a geometric phase $\varphi_g$. It is due to the twisting of the two tunneling events which are no longer aligned as in the usual \s interferometer. Here the phase shift is quantized to $\pi$ unlike most cases studied in \cite{LFM2015}. When we add an on-site staggered potential $\pm M$, the phase shift becomes $\Delta \varphi_a=-2\arctan \frac{1}{M}$ and $\Delta \varphi_b=0$.

\subsection{Adiabatic perturbation theory}
\begin{figure}[h!]
\begin{center}
\subfigure[]{\includegraphics[width=6.5cm]{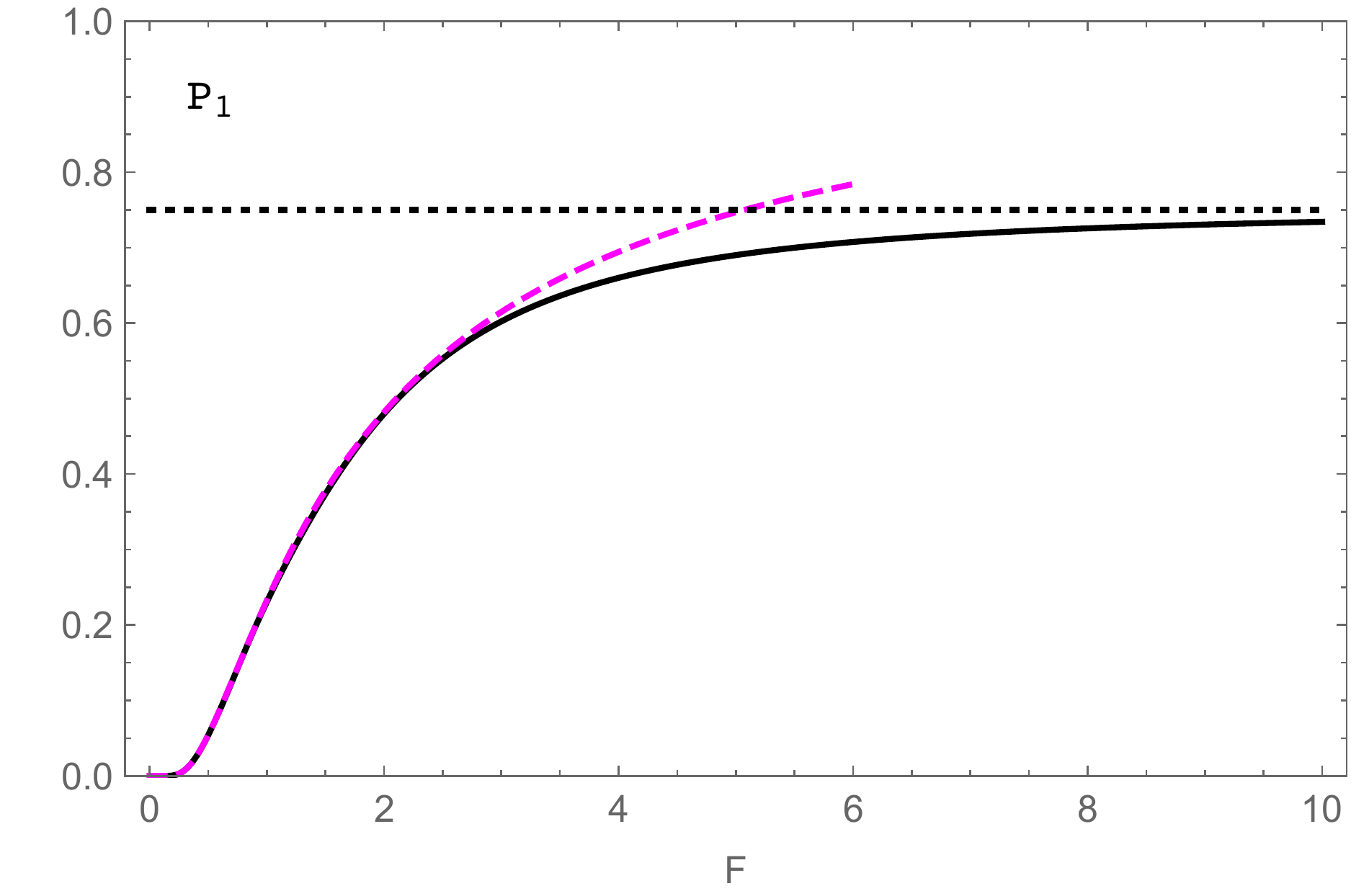}}
\subfigure[]{\includegraphics[width=6.5cm]{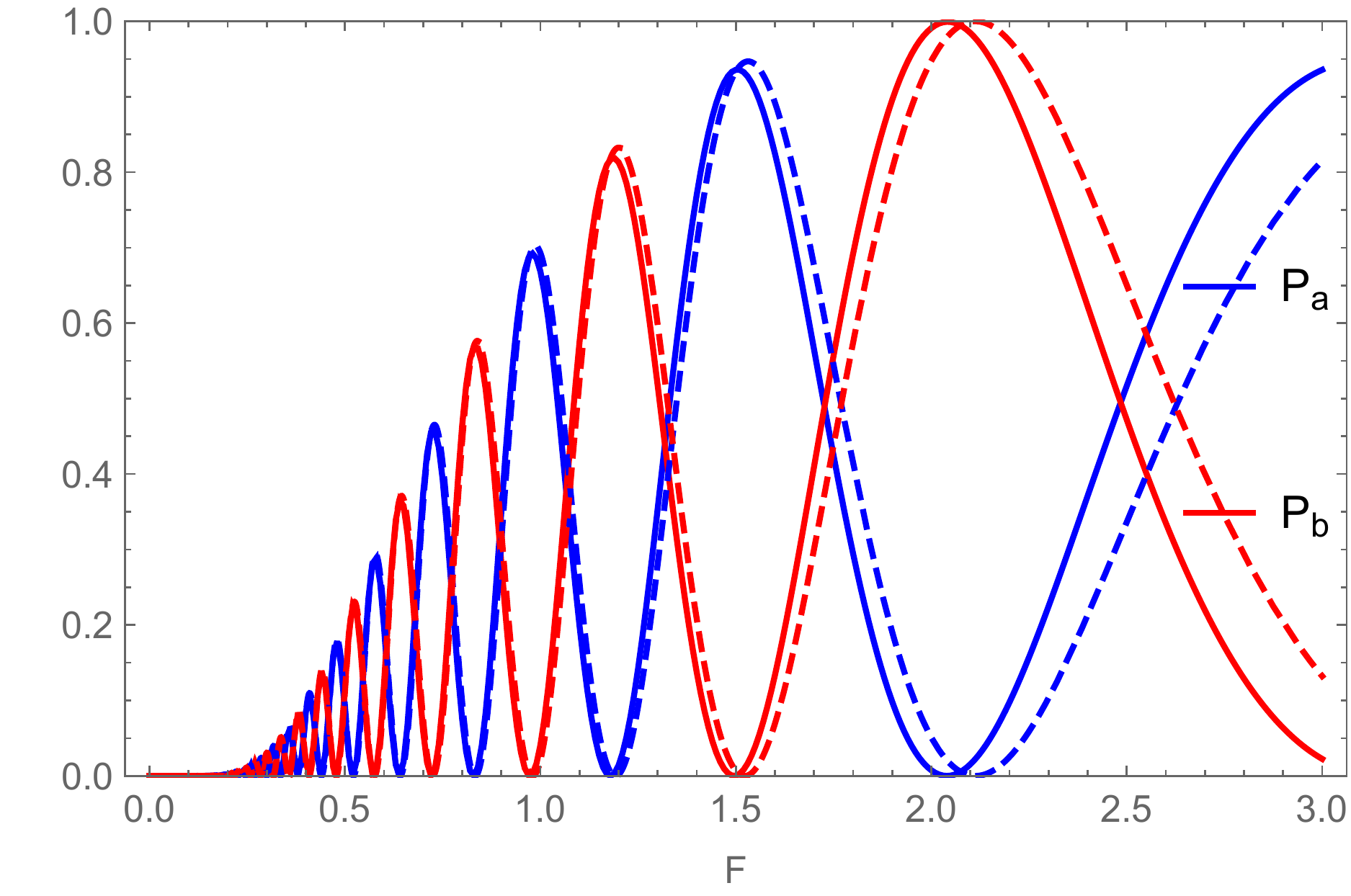}}
\end{center}
\caption{Interband transition probability $P$ as a function of the force $F$ [in units of $J/a$]. Comparison between the numerical solution (full line) and the adiabatic perturbation theory (dashed line) for: (a) the single M point avoided crossing, agreeing with a LZ-like probability $P_1\approx e^{-2\textrm{Im}\alpha_2}\approx e^{-1.459/F}$ (dashed magenta line), see Eq. (\ref{p1apt}). The dotted black line is the asymptote at $3/4$. (b) double M points interferometer with the St\"uckelberg-like formula: $P_a\approx 4e^{-2\textrm{Im}\alpha_2}(1-e^{-2\textrm{Im}\alpha_2})\cos^2(\textrm{Re}\alpha_2+\varphi_\textrm{S})$ (corresponding to $\Delta \varphi_a=\pi$, in blue), $P_b\approx 4e^{-2\textrm{Im}\alpha_2}(1-e^{-2\textrm{Im}\alpha_2})\sin^2(\textrm{Re}\alpha_2+\varphi_\textrm{S})$ (corresponding to $\Delta \varphi_b=0$, in red). Both are in good agreement in the adiabatic regime $(F\ll 1)$. The real and imaginary parts of $\alpha_2$ are given in Eqs. (\ref{B5}) and (\ref{B6}).}
\label{apt_main}
\end{figure}
The first-order adiabatic perturbation theory calculations are quite similar to those in Refs.  \cite{LFM2015,LFM2014}, except that the the initial and final times are finite (rather than at the asymptotic infinity). The details are outlined in Appendix B. Here, we briefly summarize the results in Fig. \ref{apt_main}, which contains the interband transition probability for the cases of a single M point crossing and a double M point interferometer.

\subsection{Summary of the adiabatic limit}
In summary, in the adiabatic limit, we find that the interband probabilities $P_a$ and $P_b$ are shifted by $\Delta \varphi_a = -2 \arctan \frac{1}{M}$ as $\Delta \varphi_b=0$. This phase shift is geometric in nature and does therefore not depend on the force, although it was here obtained in the $F \ll 1$ limit.

\section{Sudden limit}
\label{secsudden}
\subsection{Sudden approximation: pseudo-spin texture and extended periodicity}
When the force $F$ is very large, the initial state $|u_{-}(\mf{F}t_i)\rangle$  has no time to evolve so that the final state $|\Psi(t_f)\rangle$ is the {\it same} as this initial state: $|\Psi(t_f)\rangle=  |u_{-}(\mf{F}t_i)\rangle$. This is the so-called {\it sudden approximation} \cite{Messiah}. The probability for the particle to have tunneled in the upper band is therefore given by the overlap
\be 
P^{+-}(t_f)=  |\langle u_{+}(\mf{F} t_f )|  u_{-}(\mf{F} t_i) \rangle|^2 =\sin^2\frac{\Delta \phi}{2} 
\ee
where $\Delta \phi=\phi_f-\phi_i$, with $\phi(\mf{k})=\textrm{Arg }f(\mf{k})=\textrm{Arg} \left(-\sum_j e^{-i \mf{k}\cdot \mf{\delta}_j}\right)$ the azimuthal angle along the equator of the Bloch sphere. In this sudden limit, the interband transition probability only depends on the initial and final $\mf{k}$ points, and not on the trajectory.

\subsubsection{From a $\Gamma$ point to another $\Gamma$ point}
Consider first the special cases where the initial and the final states are localized at $\Gamma$ points. The complete evolution of the probability $P^{+-}$ when increasing the force from the adiabatic to the sudden limit has been discussed in the previous sections. We now discuss this probability for the infinite force.

For a \textit{single} M point crossing,  we find a saturation to $P^{+-}=\sin^2\frac{4\pi}{3}=3/4$ corresponding to a phase difference $\Delta \phi=2\pi/3$. The correction  to the sudden  limit is displayed in Figs. \ref{singleMpoint},\ref{singleMpointsudden} and calculated explicitly using diabatic perturbation theory in  Appendix \ref{app:dpt}. This case can be seen as a modified version of the LZ problem for a single avoided crossing, including the effect of a finite band and a finite time lapse. The main surprise here is the saturation of the interband transition probability to a finite value strictly below $1$.

For the sudden limit of a \textit{double} M point crossing, two different cases must be considered corresponding to the Hamiltonians  $H_a(t)$  (trajectories II, IV, VI in Fig. \ref{energyland}) and $H_b(t)$ (trajectories  I, III, V). As seen on Fig. \ref{azimuthalphase}, the angular changes are respectively $\Delta \phi_a=4\pi/3$ and $\Delta \phi_b=0$ and the interband probabilities are $P^{+-}_a=3/4$ and $P^{+-}_b=0$. The correction  to this sudden limit reveals the \s oscillations. They are displayed in Figs. \ref{num}, \ref{doubleMpointsudden} and calculated explicitly using diabatic perturbation theory in Appendix \ref{app:dpt}.

For triple and quadrupole M point crossings in straight line, we  respectively have $P^{+-}_a=3/4$ and $0$ as displayed in Fig. \ref{tripleMpoint}.

Therefore, going from a $\Gamma$ point to another $\Gamma$ point in straight line and in the $F\to \infty$ limit, there is a sequence of probabilities $P^{+-}=3/4,3/4,0,3/4, 3/4,0,$ etc with a tripled periodicity. Indeed the interband transition probability has the periodicity of the azimuthal angle $\phi(\mf{k})$  which is three times that of the reciprocal lattice (see Fig. \ref{azimuthalphase}). This is reminiscent of the X-ray diffraction pattern of the honeycomb lattice in which the position of the Bragg peaks has the periodicity of the reciprocal lattice and their intensity reveals the geometric form factor $S(\mf{G})= 1+e^{i \mf{G}\cdot \boldsymbol{\delta}_3}$ \cite{AshcroftMermin} which has a larger periodicity.
The diffraction pattern is spanned by the two elementary vectors  $\mf{A}_1^*=2\mf{a}_1^*+\mf{a}_2^*$ and $\mf{A}_2^*=\mf{a}_1^*+2\mf{a}_2^*$ and its periodicity is tripled.
\begin{figure}[ht]
\begin{center}
\includegraphics[width=7.5cm]{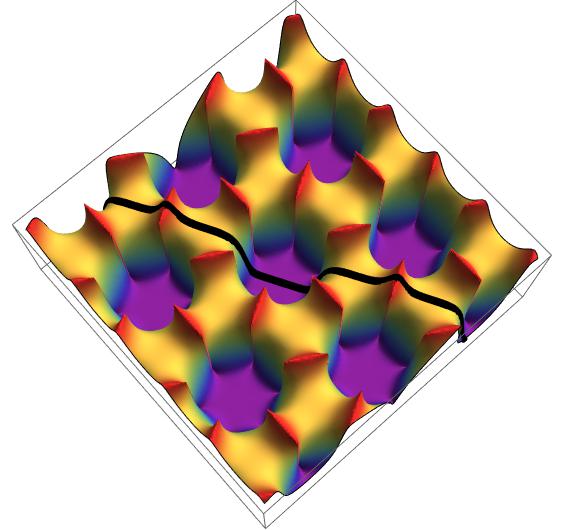}
\end{center}
\caption{Interband probability $P(\mf{k}_f)\approx |S_{+,-}(\mf{k}_f,\mf{k}_i=0)|^2$ obtained for a fixed initial state at $\mf{k}_i=0$ in the lower band $n_i=-$ and as a function of the varying final state $\mf{k}_f$ in the upper band $n_f=+$. The final $\mf{k}_f$ [in units of $1/a$] spans a few BZ. The black curve represents the probability \cite{pmmnote} measured along a straight line parallel to $\mf{a}^*_1$ in Ref. \cite{Li2015}.} \label{gilles}
\end{figure}

\subsubsection{From a $\Gamma$ point to an arbitrary final point}
More generally, interband transitions in the sudden limit give access to the modulus of the full overlap matrix $S_{n,n'}(\mf{k},\mf{k'})=\langle u_{n}(\mf{k})|u_{n'}(\mf{k}')\rangle$ as defined by Blount \cite{Blount}. It can therefore be used to obtain the complete characterization of the geometry of Bloch states. Recent experiments have used interferometric methods restricted to a single band (i.e. without interband transitions) in order to obtain Berry phases \cite{Duca2015} and maps of the Berry curvature in the BZ \cite{Esslinger2014, Sengstock2015}. Interband transitions in the sudden limit can be used to map the overlap matrix. For example, Fig. \ref{gilles} shows the interband probability $P^{+-}(\mf{k}_f)=|S_{+,-}(\mf{k}_f,\mf{k}_i=0)|^2$ obtained for a fixed initial state at $\mf{k}_i=0$ in the lower band $n_i=-$ and as a function of the varying final state $\mf{k}_f$ in the upper band $n_f=+$. 

This quantity has been recently measured in a \s interferometer build with a BEC moving in a honeycomb optical lattice \cite{Li2015}. By applying a sudden force on the BEC and starting from a $\Gamma$ point, the probability to stay in the lower band $P^{--}$ has been measured. The latter has been interpreted as the manifestation of a Wilson loop, but is merely the modulus square of the overlap between initial and final cell-periodic Bloch states. The ${\mf k}$ dependence of this probability \cite{pmmnote} measured along a straight line is reproduced as a special trajectory in Fig. \ref{gilles}. 

We emphasize that when the initial and final points are close, $P^{+-}$ gives access to the so-called quantum metric tensor $g^n_{ij}(\mf{k})$, whose physical importance has been put forward by Berry \cite{Berry1989}, as
\beq
P^{+-}=|\langle u_+(\mf{k}+\delta\mf{k})|u_-(\mf{k})\rangle|^2=1-P^{--}=g^{-}_{ij}dk_i dk_j
\eeq
which defines a distance between cell-periodic Bloch states within a given band. The quantum metric tensor $g^n_{ij}(\mf{k})$ and the Berry curvature $-\Omega_n(\mf{k})/2$ are the real and imaginary parts of the quantum geometric tensor $T^n_{ij}=\langle \partial_{k_i}u_n |(\mathbb{I}-|u_n \rangle\langle u_n|)|\partial_{k_j} u_n \rangle$ \cite{Berry1989}.

\subsection{Approaching the sudden limit}
\begin{figure}[ht]
\begin{center}
\includegraphics[width=7.5cm]{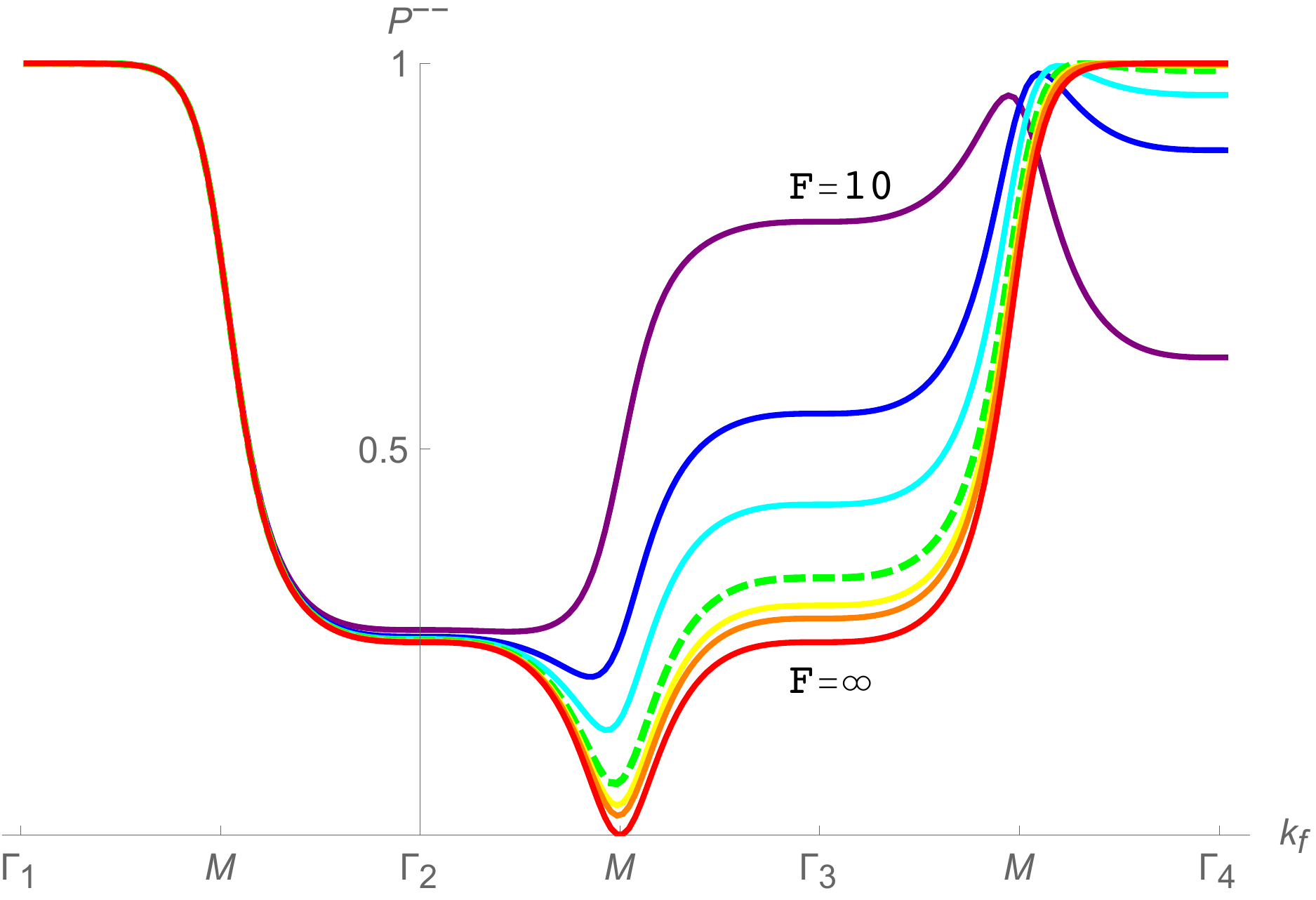}
\end{center}
\caption{Probability to stay in the lower band (i.e. $P^{--}=1-P^{+-}$ where $P^{+-}$ is the interband transition probability) when starting from $\Gamma_1$ ($k_i=Ft_i=-4\pi/3$) and as a function of the final momentum $k_f=F t_f$ [in units of $1/a$] between $\Gamma_1$ ($Ft_f=-4\pi/3$) and $\Gamma+3\mf{a}_1^*=\Gamma_4$ ($Ft_f=8\pi/3$) for several values of the force $F=\infty,50,40,30,20,15,10$ [in units of $J/a$] from bottom to top (colors span that of the rainbow). The thick dashed (green) line is for $F=30$ corresponding to the experimental value. Compare with Fig. 3(b) and S2 in \cite{Li2015}.} \label{compexp}
\end{figure}

\begin{figure}[ht]
\begin{center}
\includegraphics[width=7.5cm]{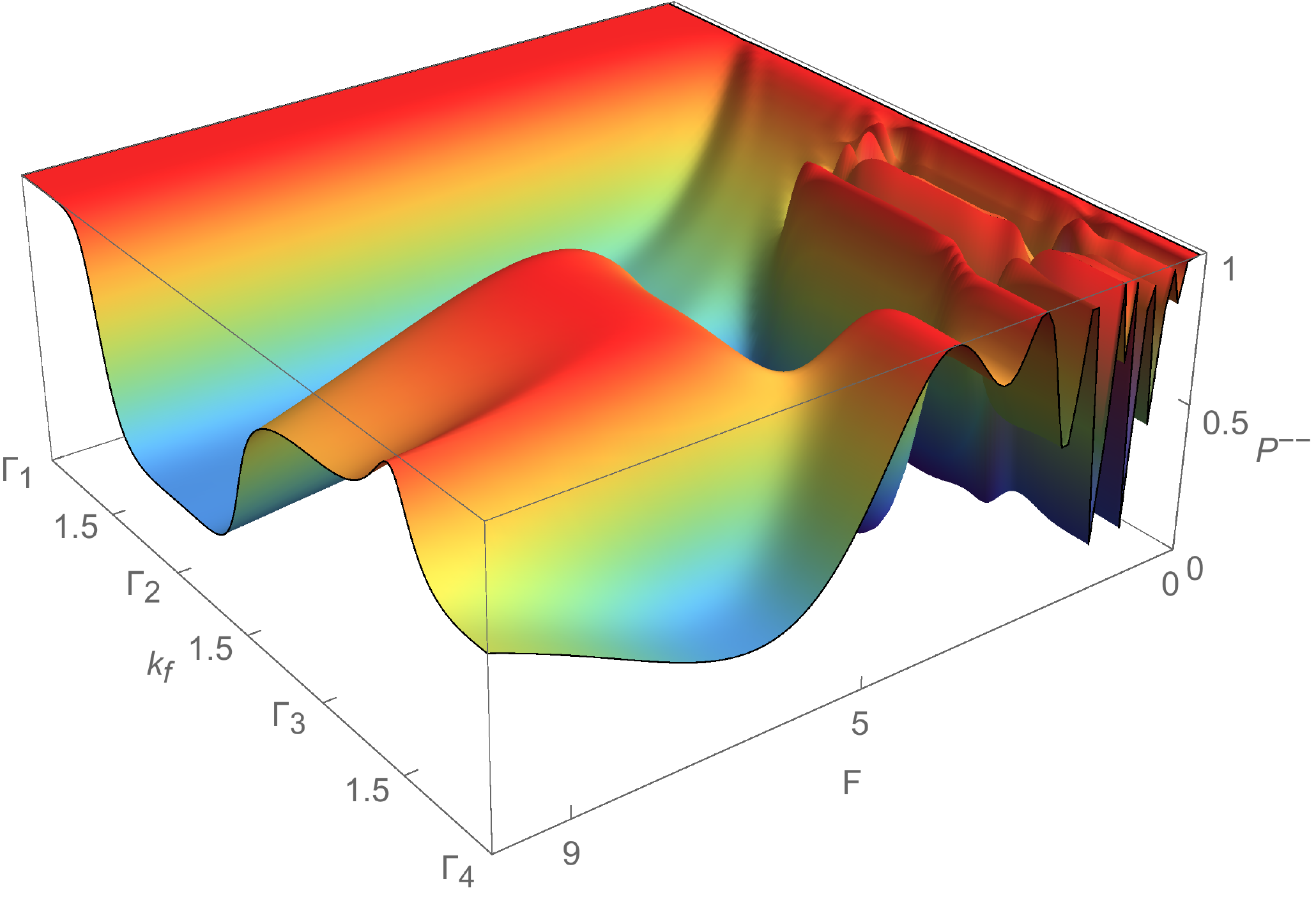}
\end{center}
\caption{Probability to stay in the lower band (i.e. $P^{--}=1-P^{+-}$ where $P^{+-}$ is the interband transition probability) when starting from $\Gamma_1$ ($k_i=Ft_i=-4\pi/3$) and as a function of the final momentum $k_f=F t_f$ [in units of $1/a$] between $\Gamma_1$ ($Ft_f=-4\pi/3$) and $\Gamma+3\mf{a}_1^*=\Gamma_4$ ($Ft_f=8\pi/3$). The force $F$ varies between 10 and 0.1 [in units of $J/a$]. The value $F=10$ corresponds to the smallest force shown in Fig. \ref{compexp}. \s oscillations are clearly visible for the double and triple M point crossings when $F\sim 1$. Notice that \s oscillations can also occur at fixed force by varying the final point $\mf{k}_f$.} \label{approachsudden}
\end{figure}

Figure \ref{compexp} presents the variation of the probability $P^{--}= 1 - P^{+-}$ measured along the direction ${\mf a}_1^*$. In the sudden limit, it is symmetric with respect to a M point. When the force ${\mf F}$ becomes finite, it becomes asymmetric. The value $F=30$ corresponds to the experimental value in \cite{Li2015}. The corresponding shape of the probability fits closely the experimental results. The shape asymmetry has been attributed to higher band effects \cite{Li2015}. We show here that the effect of a finite force is also important and has to be taken into account in a quantitative description of the experimental result. Fig. \ref{approachsudden} shows the evolution of the probability $P^{--}$ when going from the sudden limit (large ${\mf F}$) to the adiabatic regime (${\mf F} \rightarrow 0$) with the appearance of the \s oscillations.

\subsection{Toy-model: changing the intra-cell positions}
\subsubsection{Incommensurate intra-cell position and non-periodicity}
\begin{figure}[ht]
\begin{center}
\includegraphics[width=7.5cm]{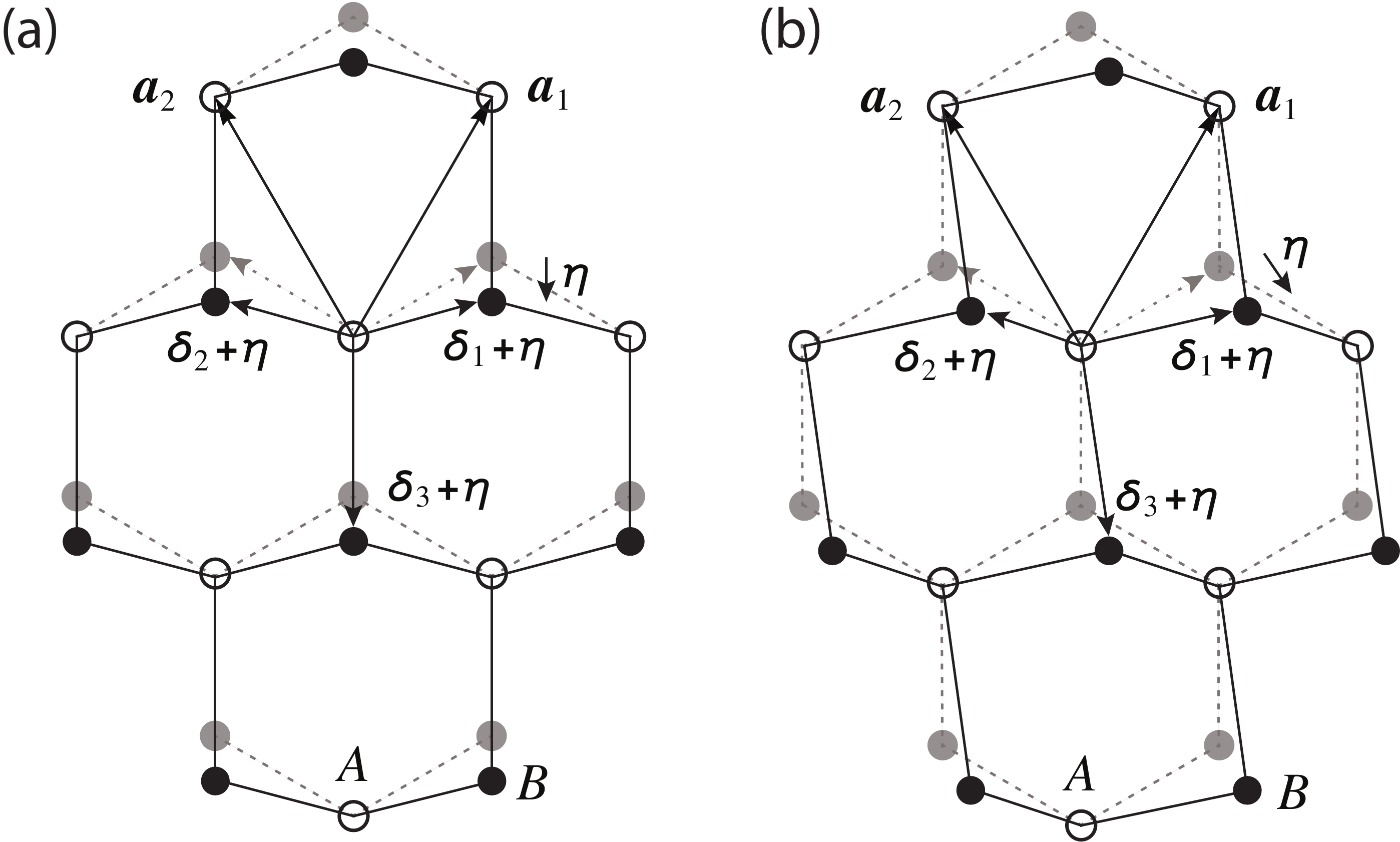}
\end{center}
\caption{Intra-cell displacement induced on the honeycomb lattice. The $B$-sublattice is rigidly slided with respect to the $A$-sublattice by a displacement vector $\boldsymbol{\eta}$. The Bravais lattices $\mf{a}_{1,2}$ are kept constant. For simplicity, we consider two types of displacement: (a) along the ``principal axis" $\boldsymbol{\eta}\propto \boldsymbol{\delta}_3$; (b) deviation from $\boldsymbol{\delta}_3$ with $\boldsymbol{\eta}\propto (1,-1)$.} \label{stretch1}
\end{figure}
In the honeycomb lattice, the periodicity of the interband transition probability is three times that of the reciprocal lattice. We generalize this observation to other atomic basis structure by considering, as a simple toy model, a ``stretched" honeycomb lattice TB model, see Fig. \ref{stretch1}. Our purpose is to show the importance of spatial embedding, i.e. of the real space position of basis states and not just their connectivity. The $B$-sublattice is rigidly displaced with respect to the $A$-sublattice by a displacement vector $\boldsymbol{\eta}$. We assume that the hopping amplitudes are unchanged (so that the TB Hamiltonian is unaffected) and that only the real space positions are modified (so that the position operator is changed). This model can be seen as complementary to the models describing uni-axially deformed graphene in which the positions of sites are left unchanged and only the hopping amplitudes are modified (such as the $t-t'$ model on the honeycomb or on the brick-wall lattices describing the merging of Dirac cones \cite{Montambaux2009,LFM2012}).

\begin{figure}[ht]
\begin{center}
\subfigure[]{\includegraphics[width=6.8cm]{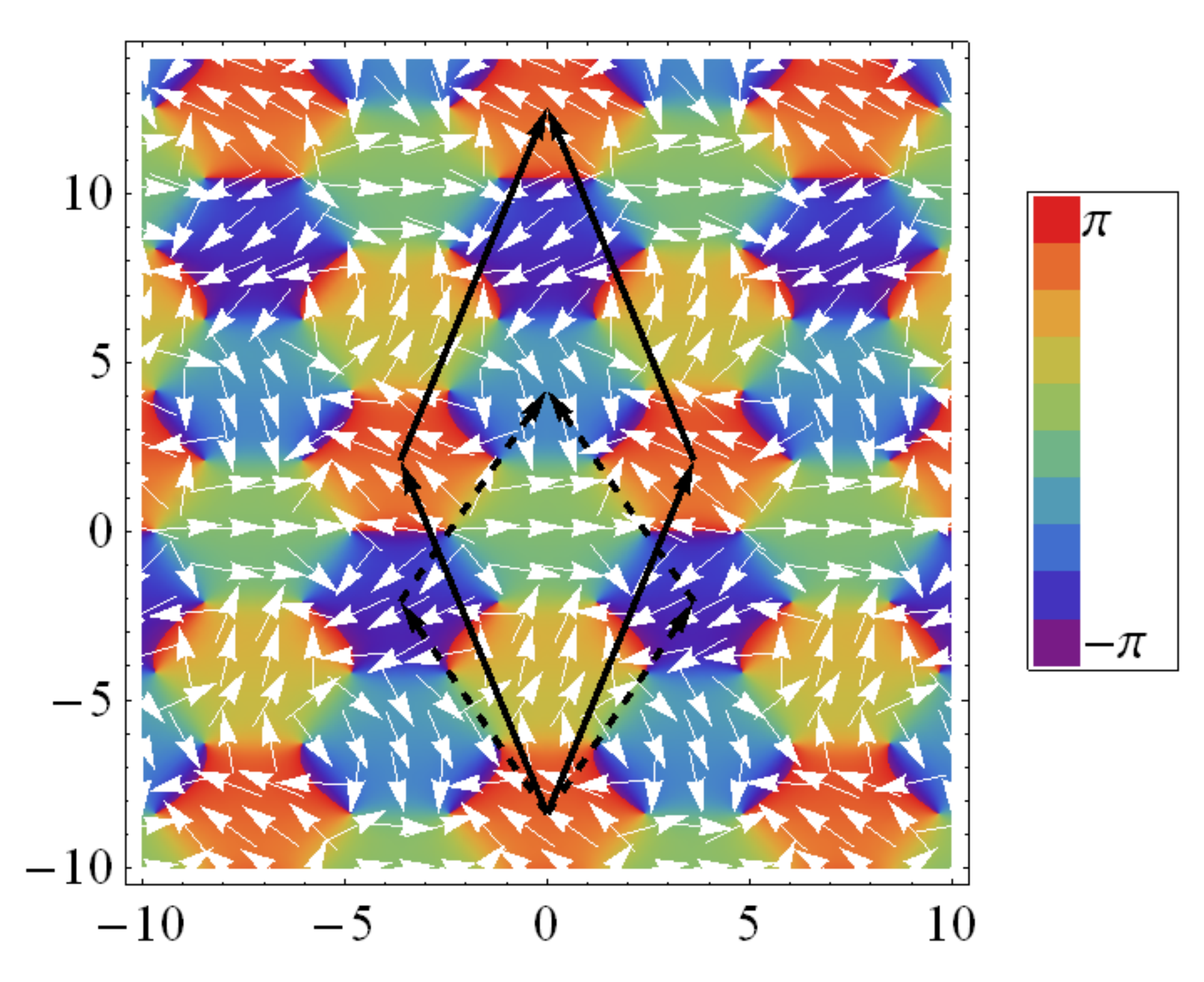}}
\subfigure[]{\includegraphics[width=6.8cm]{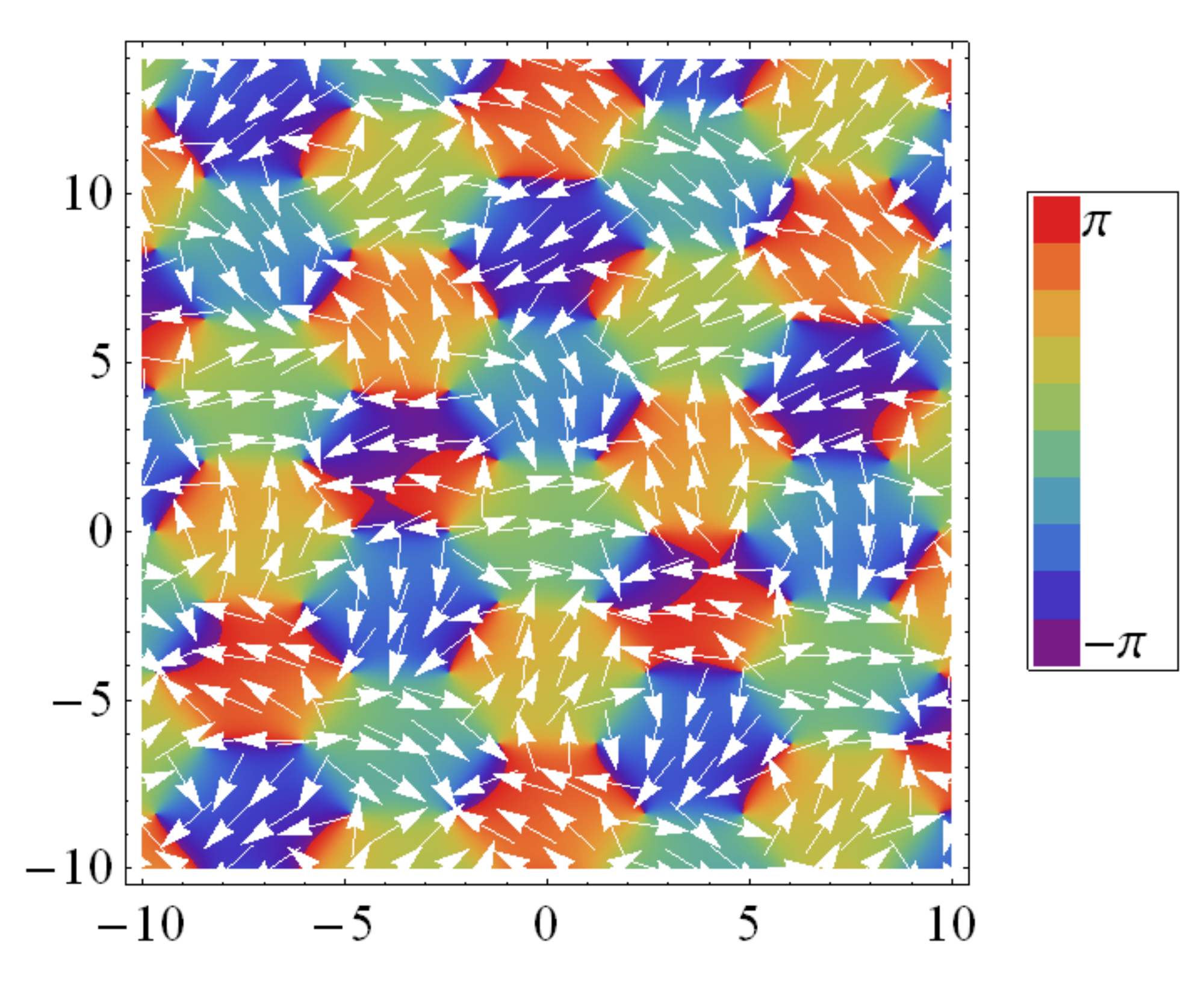}}
\end{center}
\caption{Pseudospin texture (represented as white arrows) of $H(\mf{k})$ after the displacement $\mf{\eta}$, in reciprocal space [$k_x$ and $k_y$ are in units of $1/a$]. The background color is proportional to the azimuthal angle $\phi(\mf{k})$. (a) Displacement vector $\boldsymbol{\eta}=\boldsymbol{\delta}_3/5$ leading to a periodic pseudospin texture with a unit cell five times larger than the BZ. The area covered by the dash lines shows the tripled pseudospin periodicity of the un-displaced honeycomb lattice.  (b) Displacement vector $\boldsymbol{\eta}=(1,-1)/5$ leading to a non-periodic pseudospin texture.} \label{stretch2}
\end{figure}

By displacing the $B$-sites by a constant vector $\boldsymbol{\eta}$, the Bloch Hamiltonian becomes \beq
H(\mf{k})=\left( \begin{array}{cc}0&f_{\boldsymbol{\eta}} (\mf{k})^*\\f_{\boldsymbol{\eta}} (\mf{k})&0 \end{array}\right)
\eeq 
with
\beqn\label{stretch}
f_{\boldsymbol{\eta}} (\mf{k})&=&e^{-i\mf{k}\cdot \mf{\eta}} f(\mf{k})\nn \\
&=&-e^{-i\mf{k}\cdot (\boldsymbol{\delta}_3+\boldsymbol{\eta})}\bigl( 1+e^{-i\mf{k}\cdot \mf{a}_1}+e^{-i\mf{k}\cdot \mf{a}_2}  \bigr).
\label{feta}
\eeqn
By assuming the hopping parameters to be constant, the energy spectrum $\pm |f_{\boldsymbol{\eta}} (\mf{k})|=\pm |f(\mf{k})|$ is unaltered. However, the Bloch Hamiltonian as well as the cell-periodic Bloch states do get modified. In Eq. (\ref{stretch}), while the second factor always has the periodicity of the reciprocal lattice, the first factor has this periodicity only if there exist reciprocal lattice vectors $\mf{G}$ such that $\mf{G}\cdot (\mf{\delta}_3+\mf{\eta})=0$ modulo $2\pi$. 
For example, when $\boldsymbol{\eta}=0$ the tripled \textit{commensurate} structure discussed so far is spanned by the extended basis vectors $\mf{A}_1^*=2\mf{a}_1^*+\mf{a}_2^*$ and $\mf{A}_2^*=\mf{a}_1^*+ 2\mf{a}_2^*$, with the area of the new elementary cell which is three times that of the original BZ, see the dashed line in Fig. \ref{stretch2}(a). 

We now consider two illustrative displacements shown in Fig. \ref{stretch1}, which lead to Bloch Hamiltonians with either a commensurate or an incommensurate pseudospin structure. In the first case, we take $\boldsymbol{\eta}=\f{p}{q}\boldsymbol{\delta}_3$, with $p,q$ positive integers and $p<q$. This leads to a new commensurate periodicity with $\mf{A}^*=m_1 \mf{a}_1^*+m_2 \mf{a}_2^*$ and $m_1,m_2$ being the smallest integer solution to $m_1+m_2=3 q/(p+q)$. In Fig. \ref{stretch2}(a), we show such an example with $p=1, q=5$, leading to $\mf{A}_1^*=2\mf{a}_1^*+ 3 \mf{a}_2^*$ and $\mf{A}_2^*=3\mf{a}_1^*+ 2 \mf{a}_2^*$. In the second example, we take the displacement to be away from the ``principal axis" $\boldsymbol{\delta}_3$ with $\boldsymbol{\eta}=\epsilon (1,-1)$, for small $\epsilon>0$. We then find that it leads to incommensurability with no periodicity in the pseudospin texture, see Fig. \ref{stretch2}(b). This is not surprising because the reciprocal lattice vectors $\mf{a}_{1,2}^*$ contain an irrational component which forbids integer solution to the periodicity equation.

We note that this kind of non-periodicity related to incommensurability also occurs in X-ray diffraction of a crystal. Even for a crystal with standard translational order (not a quasi-crystal), if the position of atoms within the unit cell are not commensurate with the unit cell, the diffraction pattern (taking the intensity of Bragg peaks into account) is not periodic due to the geometric structure factor. Using the sudden limit procedure, the highly unusual pseudo-spin textures we discuss here should be observables.

\subsubsection{Tuning the Berry curvature}
\begin{figure}[ht]
\begin{center}
\subfigure[]{\includegraphics[width=4.25cm]{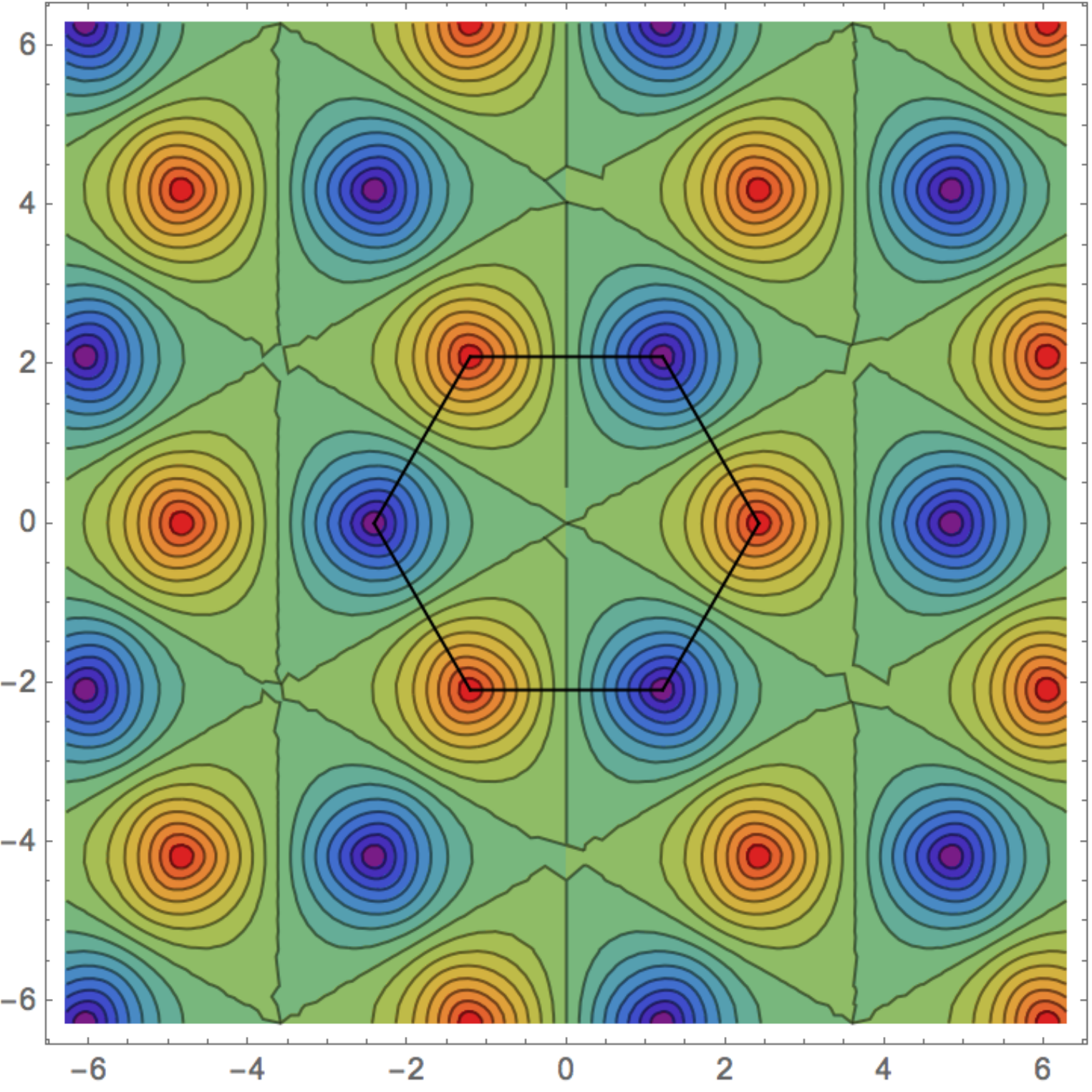}}
\subfigure[]{\includegraphics[width=4.25cm]{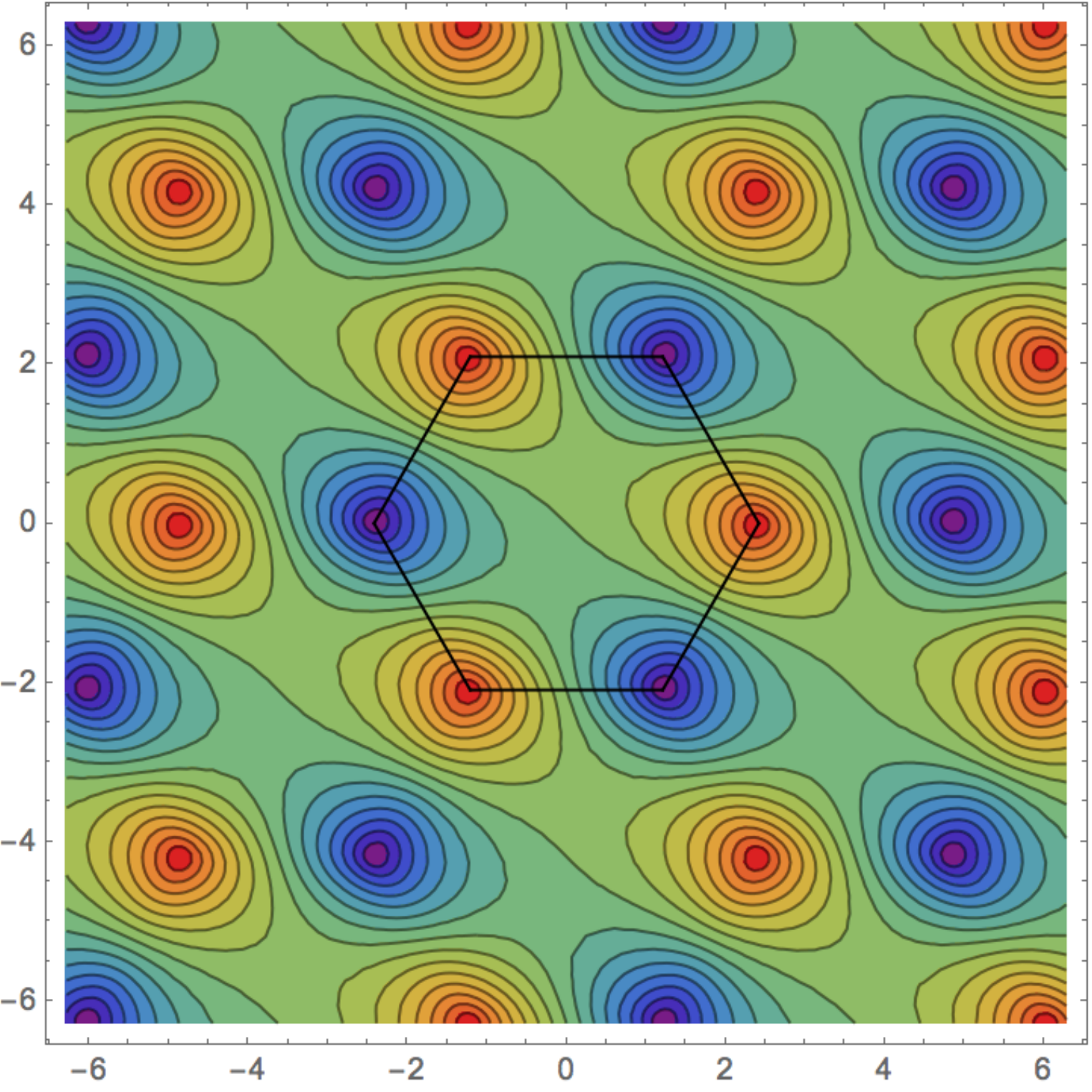}}
\end{center}
\caption{Contour plot of the Berry curvature [in units of $a^2$] for the lower band of gapped graphene with $M=1.5J$ as a function of $(k_x,k_y)$ [in units of $1/a$] for a few BZ (the first BZ is indicated by a black line). Bumps are in red and hollows in blue. (a) is the undeformed honeycomb lattice ($\mf{\eta}=\mf{0}$) and (b) is the deformed honeycomb lattice with $\mf{\eta}=(1,-1)/5$. In case (b), although the pseudo-spin texture is a-periodic (see Fig. \ref{stretch2}(b)), the Berry curvature remains periodic.} \label{bcurv}
\end{figure}

The same toy-model can further illustrate the importance of spatial embedding for the Berry curvature. Here, we consider a honeycomb lattice with staggered on-site energy $\pm M$, a usual model for boron nitride, see e.g. \cite{FPGM2010}. The on-site potential is needed to break inversion symmetry and to create a non-singular Berry curvature \cite{Xiao2010}. It also opens a gap at the Dirac points. The Bloch Hamiltonian is
\beq
H(\mf{k})=\left(\begin{array}{cc}M&f_{\boldsymbol{\eta}}(\mf{k})^*\\ f_{\boldsymbol{\eta}}(\mf{k})&-M\end{array}\right)
\eeq
where $f_{\boldsymbol{\eta}}(\mf{k})$ is given in Eq. (\ref{feta}). Following the calculation in \cite{FPGM2010}, the Berry curvature 
$\Omega_n(\mf{k})=i\langle \partial_{k_x}u_n|\partial_{k_y}u_n\rangle+\textrm{c.c.}$ \cite{Xiao2010} is found to be:
\beqn
\Omega_- (\mf{k})
&=&-\f{M}{2 E_{+}(\mf{k})^3}\bigl[ \mf{\delta}^\eta_1\times\mf{\delta}^\eta_2\sin(\mf{k}\cdot (\mf{a}_1-\mf{a}_2))\bigr. \nn \\
&+&\mf{\delta}^\eta_2\times\mf{\delta}^\eta_3\sin(\mf{k}\cdot \mf{a}_2)\ \nn \\
&+&\bigl. \mf{\delta}^\eta_1\times\mf{\delta}^\eta_3\sin(\mf{k}\cdot \mf{a}_1)\bigr].
\eeqn
where $\mf{\delta}_j^\eta = \mf{\delta}_j+\mf{\eta}$, the vector product here means $\mf{\delta}_1\times\mf{\delta}_2=\delta_1^x \delta_2^y - \delta_1^y \delta_2^x$ and the energy spectrum $E_\pm(\mf{k})=\pm \sqrt{M^2+|f_{\boldsymbol{\eta}}(\mf{k})|^2}=\pm \sqrt{M^2+|f(\mf{k})|^2}$. The displacement vector tunes the Berry curvature, while the energy spectrum and the Bloch states $|\psi_{n,\mf{k}}\rangle$ are left invariant. One can easily check that the Berry curvature always has the periodicity of the reciprocal lattice. In the limit where the displacement vanishes, one recovers the Berry curvature of boron nitride computed in \cite{FPGM2010}, see Fig. \ref{bcurv}(a). In the limit where the displacement is such that the two sublattices coincide (i.e. $\mf{\eta}=-\mf{\delta}_3$), the Berry curvature is actually the same as that obtained in ``basis I'' as defined in \cite{BM2009,FPGM2010,Fruchart2014}. In Fig. \ref{bcurv}(b), we show the Berry curvature for $\mf{\eta}=(1,-1)/5$ corresponding to the pseudo-spin texture of Fig. \ref{stretch2}(b).

\section{Conclusions}
\label{seccon}
We studied \s interferometry realized with saddle points (the so-called M points) of the energy bandstructure in the honeycomb tight-binding model. The resulting interferometer is quite different from those realized with the low-energy double Dirac cones Hamiltonians studied in our previous work. A knowledge of the full Bloch Hamiltonian is required, with the pseudospin structure generally displaying a periodicity different from that of the reciprocal lattice. This results in a sequence of out-of-phase oscillation pattern in the double M point interferometer, which is a direct consequence of the tripled Brillouin zone periodicity. The applied force $F$ is the only parameter of the problem, and we studied the interferometry problem covering the full force range, from the adiabatic $(F\ll1)$ to the sudden limit $(F\gg1)$. The insensitivity of the out-of-phase oscillation across the full regime explains the geometric origin of the \s phase, which is here due to the lattice effects. We provide physical explanations of the out-of-phase oscillation by the untwisted adiabatic impulse model, as well as by the adiabatic perturbation theory. In the sudden limit, the problem simplifies to merely computing the overlap between the initial and final cell-periodic Bloch states. This is another interesting observable as it distinguishes from quantities usually derived from the adiabatic assumption, such as the energy spectrum or the Berry curvature that are periodic in the Brillouin zone. In this regime, we propose to experimentally measure the pseudospin structure with arbitrary periodicity, and even with no periodicity. This is achieved by modifying the intracell positions while keeping the Bravais lattice and hopping amplitude fixed.

\acknowledgments
We thank I. Bloch, U. Schneider and M. Schleier-Smith for sharing ideas of realizing the \s interferometer with a BEC in a honeycomb lattice. L.-K. Lim would also like to thank A. Lazarides, R. Moessner and S. Sondhi for useful discussions. J.-N. Fuchs thanks R. Mosseri and D. Gratias for discussing the periodicity or not of the X-ray diffraction pattern of a crystal.

\appendix
\section{Uniform electric field and Peierls substitution: importance of the Bloch Hamiltonian}
\label{ps}
In this appendix, we recall results obtained by Zak \cite{Zak1989} and others about the way to treat Bloch oscillations in the presence of a constant and uniform force. The important point is that the presence of a force imposes to work with the Bloch Hamiltonian and the cell-periodic Bloch states, both of which depend on the position operator and therefore do not necessarily have the periodicity of the reciprocal lattice. This clarifies the issue about different basis conventions used in the literature in order to write Bloch-like Hamiltonians in the case of graphene, for example, see \cite{BM2009,FPGM2010,Fruchart2014,Milovanovic2015}.

Consider a quantum particle in a periodic potential. In the absence of a force, the Hamiltonian is called $\hat{H}$ (typically, we have a tight-binding Hamiltonian for a non-Bravais lattice in mind). A constant and uniform force $\mf{F}$ is introduced using a time-independent scalar gauge such that the Hamiltonian becomes:
\beq
\hat{H}_F=\hat{H}-\mf{F}\cdot \hat{\mf{r}}
\eeq
However, in order to study Bloch oscillations and to preserve translational invariance of the Hamiltonian, it is more convenient to work in a time-dependent vectorial gauge. We therefore perform a gauge transformation to obtain the transformed Hamiltonian:
\beq
\hat{H}_F(t)=e^{-it\mf{F}\cdot \hat{\mf{r}}} \left(\hat{H}_F-i\partial_t\right) e^{it \mf{F}\cdot \hat{\mf{r}}}=\hat{H}(\mf{F}t),
\eeq
where $\hat{H}(\mf{k})\equiv e^{-i\mf{k}\cdot \hat{\mf{r}}}\hat{H}e^{i\mf{k}\cdot \hat{\mf{r}}}$. The latter has the important property that it does not have the periodicity of the reciprocal lattice when there are several sites per unit cell. The reason is that it involves the position operator $\hat{\mf{r}}$. Indeed $\hat{H}(\mf{k}+\mf{G})=e^{-i\mf{G}\cdot \hat{\boldsymbol{r}}}\hat{H}e^{i\mf{G}\cdot \hat{\boldsymbol{r}}}=e^{-i\mf{G}\cdot \hat{\boldsymbol{\delta}}}\hat{H}e^{i\mf{G}\cdot \hat{\boldsymbol{\delta}}}$, where the position operator $\hat{\mf{r}}$ is split into the sum of a Bravais lattice $\hat{\mf{R}}$ and an intra-cell position $\hat{\boldsymbol{\delta}}$ operator. For example, in the case of the honeycomb lattice, if the shortest reciprocal lattice vector $\mf{G}$ is such that $\mf{G}\cdot (\boldsymbol{\delta}_A-\boldsymbol{\delta}_B)\neq 0$ modulo $2\pi$ -- where $\boldsymbol{\delta}_A$ and $\boldsymbol{\delta}_B$ are the intra-cell positions of the $A$ and $B$ sites --, then $\hat{H}(\mf{k})$ does not have the periodicity of the reciprocal lattice. In the case of the honeycomb lattice and with our choice of position origin, $\boldsymbol{\delta}_A=0$ and $\boldsymbol{\delta}_B=\boldsymbol{\delta}_3$.

The unitary operator $e^{-it \mf{F}\cdot \hat{\mf{r}}}$ performs a gauge transformation as can be seen from $\hat{H}_F(t)=e^{-it \mf{F}\cdot \hat{\mf{r}}}(\hat{H}-\mf{F}\cdot \hat{\mf{r}}-i\partial_t)e^{it \mf{F}\cdot \hat{\mf{r}}}$ and $\hat{H}-\mf{F}\cdot \hat{\mf{r}}=e^{it \mf{F}\cdot \hat{\mf{r}}}(\hat{H}_F(t)-i\partial_t)e^{-it \mf{F}\cdot \hat{\mf{r}}}$ relating the Hamiltonian $\hat{H}_F(t)$ in the vectorial gauge to the Hamiltonian $\hat{H}_F$ in the time-independent scalar gauge.

The key point of the above derivation is that, because the vector potential $\mf{A}=-\mf{F}t$ is merely proportional to the identity when the force is homogeneous, the gauge transformation $e^{-it \mf{F}\cdot \hat{\mf{r}}}$ is the exponential of the position operator $\hat{\mf{r}}$. Therefore $\hat{H}_F(t)$ is actually equal to $\hat{H}(\mf{k})$, obtained for the system {\it in the absence of a force}, upon replacing the parameter $\mf{k}$ by $\mf{F}t$. This kind of Peierls's substitution is actually exact here:
\beq
\hat{H}_F(t)=\hat{H}(\mf{k}\to \mf{F}t)
\eeq

At this point, it is convenient to project $\hat{H}(\mf{k})$ onto the $N_b\times N_b$ $\mf{k}$-subspace due to the conservation of crystal momentum and in order to obtain the Bloch Hamiltonian $H(\mf{k})=P(\mf{k})\hat{H}(\mf{k})P(\mf{k})$, where $N_b$ is the number of bands (the total size of the Hilbert space being $N\times N_b$ where $N$ is the number of unit cells) and $P(\mf{k})=\sum_n |u_{n,\mf{k}}\rangle \langle u_{n,\mf{k}}|$ is the projector. This procedure is detailed in Appendix B of \cite{FPGM2010}. One should keep in mind the difference between the three Hamiltonians: $\hat{H}$, $\hat{H}(\mf{k})$ (both of dimension $NN_b\times NN_b$) and $H(\mf{k})$ (of dimension $N_b\times N_b$).

The Schr\"odinger equation for the time evolution in the presence of a force is therefore:
\beq
i\frac{d}{dt}|\psi(t)\rangle=H(\mf{F}t)|\psi(t)\rangle
\eeq
An adiabatic basis is provided by the cell-periodic Bloch states $|u_{n}(\mf{k})\rangle$ -- that satisfy $H(\mf{k})|u_{n}(\mf{k})\rangle=E_{n}(\mf{k})|u_{n}(\mf{k})\rangle$ where $E_{n}(\mf{k})$ is the band energy spectrum and $n$ is the band index -- so that
\beq
H_F(t)|u_{n}(\mf{F}t)\rangle =H(\mf{F}t)|u_{n}(\mf{F}t)\rangle = E_{n}(\mf{F}t)|u_{n}(\mf{F}t)\rangle
\eeq
The cell-periodic Bloch have the periodicity of the Bravais lattice in real space $\langle \mf{r}+\mf{R}|u_{n}(\mf{k})\rangle=\langle \mf{r}|u_{n}(\mf{k})\rangle$, but do not have the periodicity of the reciprocal lattice and satisfy $|u_{n}(\mf{k}+\mf{G})\rangle=e^{-i \mf{G}\cdot \hat{\mf{r}}}|u_{n}(\mf{k})\rangle=e^{-i \mf{G}\cdot \hat{\boldsymbol{\delta}}}|u_{n}(\mf{k})\rangle$ instead. In particular with a Hamiltonian $H_F(t)=H(\mf{k}\to \mf{F}t)$ that depends smoothly on time as a parameter, it is possible to properly define geometric quantities \cite{Berry1984}. This is due to the fact that $|u_{n}(\mf{k})\rangle$ evolves smoothly as a function of $\mf{k}$ in a given band (unlike the Bloch state $|\psi_{n\mf{k}}\rangle=e^{i\mf{k}\cdot \hat{\mf{r}}}|u_{n\mf{k}}\rangle$ that evolves very rapidly). The key quantity is the overlap matrix $S_{n,n'}(\mf{k},\mf{k'})=\langle u_{n}(\mf{k})|u_{n'}(\mf{k}')\rangle$ which is diagonal in band indices but not with $\mf{k}$ indices, unlike $\langle \psi_{n\mf{k}}|\psi_{n'\mf{k}'}\rangle=\delta_{n,n'}\delta(\mf{k}-\mf{k}')$. The reason behind that is that $|\psi_{n\mf{k}}\rangle$ and $|\psi_{n'\mf{k}'}\rangle$ are eigenvectors of the same Hamiltonian $\hat{H}$, whereas $|u_{n}(\mf{k})\rangle$ and $|u_{n'}(\mf{k}')\rangle$ are eigenvectors of two different Hamiltonians $H(\mf{k})$ and $H(\mf{k}')$ when $\mf{k}\neq \mf{k}'$. In other words, for Bloch states $|\psi_{n\mf{k}}\rangle$, $\mf{k}$ acts as a quantum number, whereas for cell-periodic Bloch states $|u_{n}(\mf{k})\rangle$, it is merely a parameter. In both cases, the band index always acts as a quantum number.

If the initial state is $|\psi(t=0)\rangle = |\psi_{n,\mf{k}_0}\rangle = e^{i \mf{k}_0\cdot \hat{\mf{r}}} |u_{n,\mf{k}_0}\rangle$, it is actually more convenient to work with a slightly different adiabatic basis \cite{Zak1989}
\beq
H(\mf{F}t) \left[e^{i \mf{k}_0\cdot \hat{\mf{r}}}|u_{n}(\mf{k}(t))\rangle\right] = E_{n}(\mf{k}(t)) \left[e^{i \mf{k}_0\cdot \hat{\mf{r}}}|u_{n}(\mf{k}(t))\rangle\right]
\eeq
with $\mf{k}(t)=\mf{k}_0+\mf{F}t$, such that the initial state is one of the adiabatic basis states (that at $t=0$). Note that $e^{i \mf{k}_0\cdot \hat{\mf{r}}}|u_{n,\mf{k}(t)}\rangle$ and $|\psi_{n,\mf{k}(t)}\rangle$ are distinct and related by a gauge transformation $e^{it \mf{F}\cdot \hat{\mf{r}}}$, see \cite{Zak1989}.

This explains the key role played by the Bloch Hamiltonian $H(\mf{k})$ (rather than the tight-binding Hamiltonian $\hat{H}$) and the cell-periodic Bloch states $|u_{n}(\mf{k})\rangle$ (rather than the Bloch states $|\psi_{n\mf{k}}\rangle$) in the description of Bloch oscillations. It also shows that there is no such thing as a ``choice of basis'' in writing the $\mf{k}$-dependent Hamiltonian to describe the dynamics of a Bloch electron in the presence of a force \cite{BM2009,FPGM2010,Fruchart2014,Milovanovic2015}.

\section{Adiabatic perturbation theory}
\label{app:apt}
\subsection{Single M point avoided crossing}
\begin{figure}[ht]
\begin{center}
\includegraphics[width=7cm]{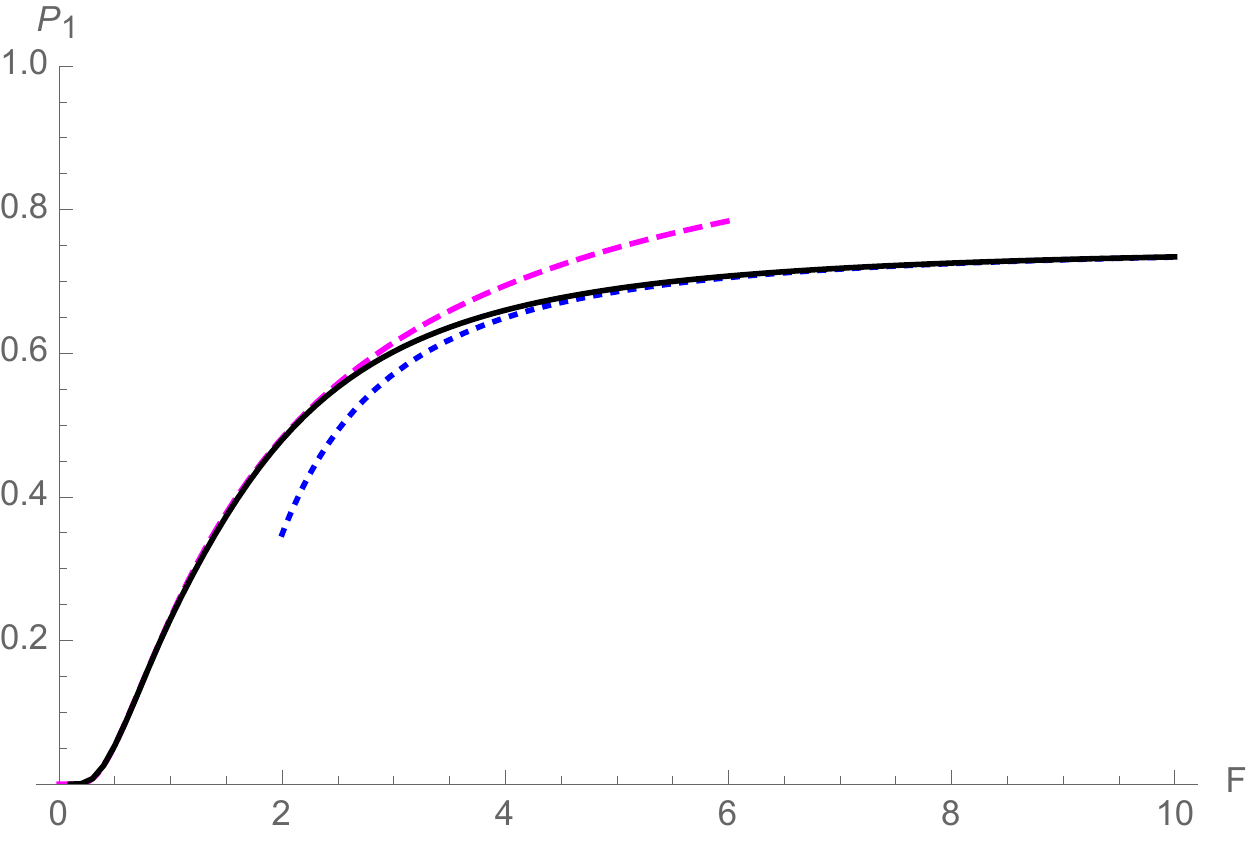}
\end{center}
\caption{Probability $P_1$ for a single M point crossing as a function of the force $F$ [in units of $J/a$]. Full (black) line is the numerical solution, dashed (magenta) line is the adiabatic approximation $P_1\approx e^{-1.459/F}$ and dotted (blue) line is the diabatic perturbation theory at second order $P_1\approx 3/4-(27-14\pi/\sqrt{3})/F^2$.}
\label{singleMpointsudden}
\end{figure}

We start by considering the case of a single M point avoided crossing as described by the Hamiltonian
\beqn
H_1(t)&=&-[2\cos(\frac{Ft}{2}+\frac{\pi}{3})+\cos(Ft+\frac{2\pi}{3})]\sigma_x\nn\\
&+&[2\sin(\frac{Ft}{2}+\frac{\pi}{3})-\sin(Ft+\frac{2\pi}{3})]\sigma_y
\label{h1}
\eeqn
that is obtained from $H_a(t)$ after shifting the initial and final times to  $Ft_i=-2\pi/3$ and $Ft_f=2\pi/3$ so that the avoided crossing is now at $t=0$. The Dykhne method \cite{Dykhne1960} in the adiabatic limit gives \beq
P_1\approx e^{-2\textrm{Im}\int_0^{t_c} dt 2E_+}=e^{-\frac{16}{3F}\textrm{Im}E(i\frac{\ln 2}{2},-8)}\approx e^{-1.459/F}
\label{p1apt}
\eeq
where $E_+(t)=\sqrt{5-4\cos \frac{3Ft}{2}}$ is the adiabatic energy spectrum, $t_c=i\frac{2\ln 2}{3F}$ is such that $E_+(t_c)=0$ and $E(\phi,m)=\int_0^\phi d\theta \sqrt{1-m\sin^2 \theta}$ is the elliptic integral of the second kind. See Fig. \ref{singleMpointsudden} for a comparison between the numerically obtained $P_1$ and analytical expressions obtained in the adiabatic or in the diabatic limits. There are two ways to compare this result to a LZ-type formula. 

First, in the vicinity of the M point, when $Ft\to 0$, the Hamiltonian reads $H_1(t)\approx (-1/2+\sqrt{3}Ft)\sigma_x + (\sqrt{3}/2+Ft)\sigma_y$ and the corresponding adiabatic energy spectrum is $E_\pm (t) \approx \pm \sqrt{1+4F^2t^2}$. After a rotation in pseudo-spin space, the Hamiltonian acquires a familiar LZ form $\tilde{H}_1(t)\approx 2Ft\sigma_x + \sigma_y=Vt \sigma_x + W\sigma_y$ from which the corresponding LZ probability can be immediately read as $P_Z=e^{-\pi W^2/V}=e^{-\pi/(2F)}$ \cite{LZ1932}. This is slightly different from the above Dykhne result (\ref{p1apt}), which better matches the numerical solution.

Second, the above adiabatic energy spectrum $E_+(t)\approx \sqrt{1+4F^2t^2}$ is different from the exact spectrum $E_+(t)=\sqrt{5-4\cos \frac{3Ft}{2}}\approx \sqrt{1+\frac{9}{2}F^2t^2}$ even when $Ft \to 0$. If we correct the local LZ Hamiltonian for that by considering $H'_1(t)\approx \frac{3}{\sqrt{2}}Ft\sigma_x + \sigma_y$ instead of $H_1(t)$, we find a different LZ probability $P'_Z=e^{-\pi\sqrt{2}/(3F)}$, which happens to be much closer to both the numerical result and the Dykhne formula. 

If we identify the Dykhne result (\ref{p1apt}) with the standard LZ formula $P_{LZ}=e^{-2\pi \delta}$, we find an adiabaticity parameter $\delta\approx 0.232/F$ close to $1/(3\sqrt{2}F)\approx 0.236/F$ if $P'_Z=e^{-\pi\sqrt{2}/(3F)}$ or to $1/(4F)=0.25/F$ if $P_Z=e^{-\pi/(2F)}$. In all cases, the adiabaticity parameter is qualitatively $\delta \sim F^{-1}$ as anticipated.

\subsection{Double M point interferometer}
At first order of adiabatic perturbation theory $c_- (t)\approx c_- (t_i)=1 \gg |c_+ (t)|$ in Eq. (\ref{cfode}), so that the probability amplitude for the particle to end in the upper band becomes:
\beq
c_+(t_f)\approx -\int_{t_i}^{t_f} dt \langle u_+|\dot{u}_-\rangle e^{i\int_0^t dt' [2E_+(t')+\mathcal{A}_- (t) - \mathcal{A}_+ (t)]}.
\eeq
Following the convention of Refs. \cite{LFM2015,LFM2014}, in the south pole gauge, $i\mathcal{A}_{+,-}(t)=-\langle u_+|\dot{u}_-\rangle=\frac{\dot{\theta}-i\dot{\phi}\sin \theta}{2}e^{-i\phi}=-\frac{i}{2}\dot{\phi}e^{-i\phi}$ and $\mathcal{A}_- (t) - \mathcal{A}_+ (t)=\langle u_-|i\partial_t|u_-\rangle-\langle u_+|i\partial_t|u_+\rangle=\dot{\phi}(1-\cos \theta)=\dot{\phi}$. We also define $\alpha(t)=\int_0^t dt' 2E_+(t')$ such that $\dot{\alpha}=2E_+$. Note that $\alpha(t)=\frac{8}{F}E(\frac{3Ft}{4},\frac{8}{9})$ is a monotonically increasing function of $t$ (in a rough approximation $\alpha(t) \sim 4.25 t$). Therefore, and not paying attention to the irrelevant overall phase of $c_+(t_f)$, we obtain
\beq
c_+(t_f)\sim \int_{t_i}^{t_f} dt \frac{\dot{\phi}}{2} e^{i\alpha(t)}=\int_{\alpha_i}^{\alpha_f} d\alpha g(\alpha) e^{i\alpha}
\eeq
where we called $g(\alpha)=\frac{\dot{\phi}}{2\dot{\alpha}}$. Note that all phases apart from the dynamical one have disappeared from $c_+(t_f)$ and the main role will be played by the sign of the angular velocity $\dot{\phi}$ at the two crossing points.

\subsubsection{Case (a)}
\begin{figure}[ht]
\begin{center}
\subfigure[]{\label{apta}\includegraphics[width=6.5cm]{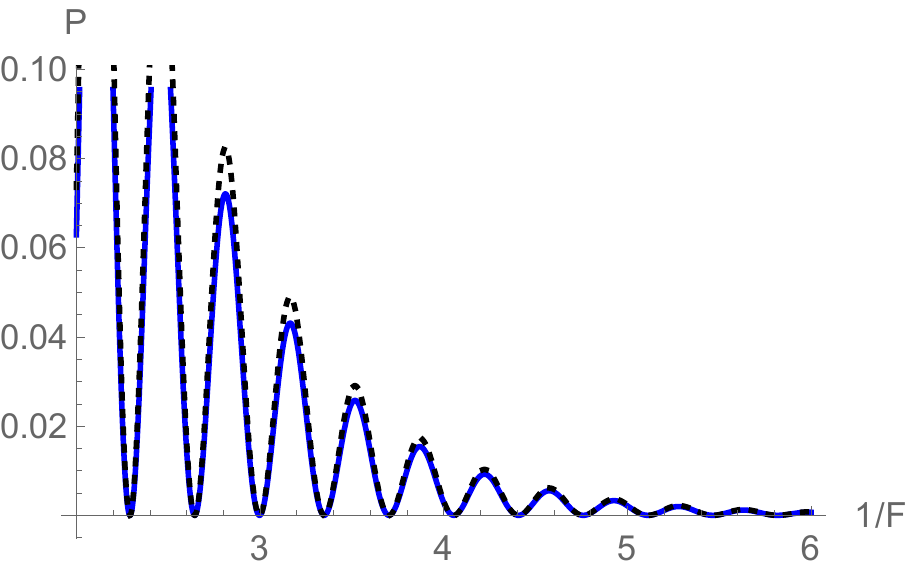}}
\subfigure[]{\label{aptb}\includegraphics[width=6.5cm]{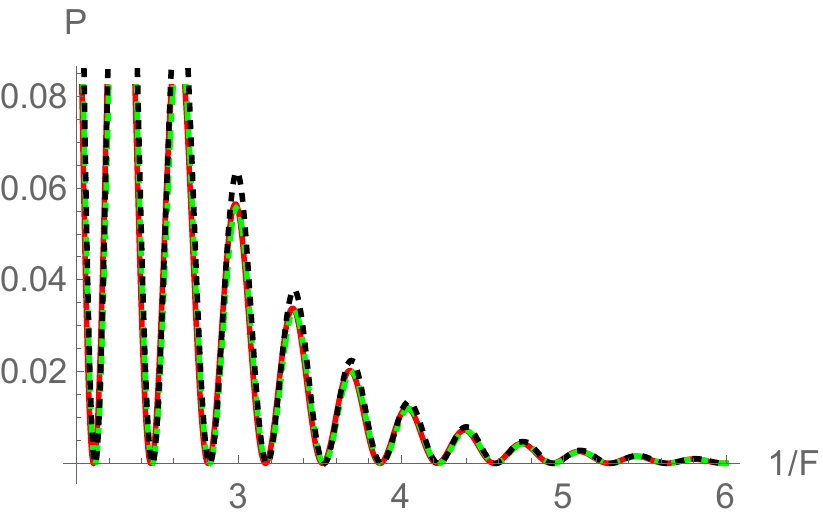}}
\end{center}
\caption{Probability $P=|c_+(t_f)|^2$ for a double M point interferometer computed in first order adiabatic perturbation theory as a function of $F^{-1}$ [with the force $F$ in units of $J/a$] between 2 and 6. Comparison between different approximations for computing the integral giving the amplitude $c_+(t_f)$. (a) is the case of $H_a$. Dashed line (black) is the numerical evaluation of the integral and full line (blue) is the approximate analytical result using the residue theorem, Eq. (\ref{Paapt}). (b) is the case of $H_b$. Dashed line (black) is the numerical evaluation of the integral and full line (red) is the approximate analytical result using the residue theorem, Eq. (\ref{Pbapt}). Dot-dashed line (green) is Eq. (\ref{Pbapt}) except for the phase shift $\textrm{arg} \alpha_2\approx 0.082$ being replaced by $0$.}
\label{apt}
\end{figure}

We have $g(\alpha)=F\frac{\cos\frac{3Ft}{2}-1}{4E_+^3}$ as $\tan \phi=(\sin Ft-2\sin\frac{Ft}{2})/(\cos Ft +2\cos\frac{Ft}{2})$ and $\dot{\phi}=F(\cos\frac{3Ft}{2}-1)/E_+(t)^2$. This function has poles in $\alpha$ when $E_+=0$ corresponding to $\frac{3}{2}Ft=\pi (2n+1)\pm i \ln 2$, with $n\in \mathbb{Z}$. Four of these poles ($\frac{3}{2}Ft=\pm \pi\pm i \ln 2$) correspond to the two avoided crossings at $t_1=-\frac{2\pi}{3F}$ and $t_2=\frac{2\pi}{3F}$. In order to use the residue theorem to compute the above integral, we extend the integration to $\alpha \in \mathbb{R}$ but only retain the contribution of the above four poles (actually when closing the contour in the upper half-plane we only retain the contribution of the two poles $\frac{3}{2}Ft=\pm \pi +i \ln 2$ i.e. $t_1^c=t_1+i\frac{2\ln2}{3F}$ and $t_2^c=t_2+i\frac{2\ln 2}{3F}$). Then
\beq
c_+(t_f)\sim 2\pi(\textrm{Res}(g,\alpha_1)e^{i \alpha_1}+\textrm{Res}(g,\alpha_2)e^{i \alpha_2})
\eeq
where $\alpha_2=\alpha(t_2^c)$ and $\alpha_1=\alpha(t_1^c)$ such that
\beqn
\textrm{Re}\alpha_2&=&-\textrm{Re}\alpha_1=\frac{4}{3F}\int_0^\pi dz \sqrt{5+4\cos z}\nn\\
&=&\frac{8}{3F}E(-8)\approx \frac{8.91}{F} \label{B5}\\
\textrm{Im}\alpha_2&=&\textrm{Im}\alpha_1=\frac{4}{3F}\int_0^{\ln2} dx \sqrt{5-4\cosh x}\nn\\
&=&\frac{8}{3F}\textrm{Im}E(i \frac{\ln 2}{2},-8)\approx \frac{0.729}{F}
\label{B6}
\eeqn
Close to $t_1^c$, we have $\alpha-\alpha_1\approx 2\sqrt{2iF}(t-t_1^c)^{3/2}$ so that $E_+(t)^3\approx \frac{27}{8}iF(\alpha-\alpha_1)$, which shows that the branch point in $t$ is actually a simple pole in $\alpha$. Similarly, close to $t_2^c$, we have $\alpha-\alpha_2\approx 2\sqrt{2iF}(t-t_2^c)^{3/2}$ so that $E_+(t)^3\approx \frac{27}{8}iF(\alpha-\alpha_2)$. In the end, $\textrm{Res}(g,\alpha_1)=\textrm{Res}(g,\alpha_2)=i/6$ and
\beqn
P_a=(\frac{\pi}{3})^2 4e^{-2 \textrm{Im} \alpha_2}\cos^2(\textrm{Re}\alpha_2)
\label{Paapt}
\eeqn
Evacuating the familiar $\pi/3\to 1$ problem \cite{BM1972}, this probability has the \s form $4P_1(1-P_1)\sin^2((\varphi_\textrm{dyn}+\Delta \varphi)/2+\varphi_\textrm{S})\approx 4P_1\sin^2((\varphi_\textrm{dyn}+\Delta \varphi)/2)$ with the Stokes phase $\varphi_\textrm{S}\to 0$ in the adiabatic limit, the single tunneling probability $P_1 = e^{-2 \textrm{Im} \alpha_2}\approx e^{-1.459/F}\ll 1$, the dynamical phase $\varphi_\textrm{dyn}=\textrm{Re} \alpha_2 - \textrm{Re} \alpha_1 = \int_{t_1}^{t_2} dt 2 E_+(t)=2\textrm{Re}\alpha_2$ and the phase shift $\Delta \varphi=\pi$. To check the validity of our approximate contour calculation, we computed the integral
\beqn
c_+(t_f)&\sim& \int_{t_i}^{t_f}dt F\frac{\cos\frac{3Ft}{2}-1}{2E_+(t)^2}e^{i2\int_0^t dt' E_+(t')} \\
&=&\int_{-4\pi/3}^{4\pi/3}d\tau \frac{\cos\frac{3\tau}{2}-1}{2(5+4\cos\frac{3\tau}{2})}e^{i2F^{-1}\int_0^\tau d\tau' \sqrt{5+4\cos \frac{3\tau'}{2}}}\nn
\eeqn
numerically as a function of $F^{-1}\sim \delta$, using that $\int_0^\tau d\tau' \sqrt{5+4\cos \frac{3\tau'}{2}}=4 E(\frac{3\tau}{4},\frac{8}{9})$.
We find a very good agreement between the numerically exact result (still at first order in adiabatic perturbation theory) $|c_+(t_f)|^2$ and the approximate analytical result (\ref{Paapt}) when $F\ll 1$ (see Fig. \ref{apt}(a)).

\subsubsection{Case (b)}
The only difference with case (a) is the presence of a -sign function in front of $\sigma_y$ in the time-dependent Hamiltonian. Note that $\textrm{sign}(t)=t/|t|=e^{i\textrm{arg} t}$ when $t$ becomes complex. The calculation is very similar to the previous case, except that $\dot{\phi}$ is replaced by $-\textrm{sign}(t)\dot{\phi}$. Indeed $\phi(t)=-\textrm{sign}(t) \phi_a(t)$, where $\phi_a$ refers to the previous case, and $\phi_a(0)=0$ so that $\dot{\phi}(t)=-2\delta(t)\phi_a(0)-\textrm{sign}(t) \dot{\phi}_a(t)=-\textrm{sign}(t) \dot{\phi}_a(t)$. Therefore
\beq
c_+(t_f)\sim \int_{\alpha_i}^{\alpha_f} d\alpha F\frac{1-\cos\frac{3Ft}{2}}{4E_+^3}e^{i\textrm{arg} \alpha} e^{i\alpha}
\eeq
so that
\beq
c_+(t_f)\sim 2\pi(\frac{i}{6}e^{i\textrm{arg} \alpha_1}e^{i \alpha_1}+\frac{i}{6}e^{i\textrm{arg} \alpha_2}e^{i \alpha_2})
\eeq
with $e^{i\textrm{arg} \alpha_1}=-e^{-i\textrm{arg} \alpha_2}=\frac{-6.7+i0.5}{\sqrt{6.7^2+0.5^2}}$. As a rough approximation $e^{i\textrm{arg} \alpha_1}\approx -1$ and $e^{i\textrm{arg} \alpha_2}\approx 1$. In the end, we find
\beq
P_b=(\frac{\pi}{3})^2 4e^{-2 \textrm{Im} \alpha_2}\sin^2(\textrm{Re}\alpha_2+\textrm{arg}\alpha_2)
\label{Pbapt}
\eeq
Again, apart from the $\pi/3\to 1$ problem \cite{BM1972}, we recover a \s like probability $4P_1(1-P_1)\sin^2((\varphi_\textrm{dyn}+\Delta \varphi)/2+\varphi_\textrm{S})$ with the Stokes phase $\varphi_\textrm{S}\to 0$ in the adiabatic limit, the single tunneling probability $P_1=e^{-2 \textrm{Im} \alpha_2}=e^{-1.459/F}\ll 1$, the dynamical phase $\varphi_\textrm{dyn}=\textrm{Re} \alpha_2 - \textrm{Re} \alpha_1 = \int_{t_1}^{t_2} dt 2 E_+(t)=2\textrm{Re}\alpha_2$ and the phase shift $\Delta \varphi=2\textrm{arg} \alpha_2\approx 0.163$, which is small but non-zero, whereas we expect to find $\Delta \varphi=0$. One reason for the latter result could be that the gap at the avoided crossings is not small but is a substantial fraction of the bandwidth ($\Delta=2$ compared to $W=6$). Therefore $\textrm{Im}\alpha_2$ is small by a factor of 10 compared to $\textrm{Re}\alpha_2$ but not completely negligible. Hence $\textrm{arg} \alpha_2\sim \tan \textrm{arg} \alpha_2 = \textrm{Im}\alpha_2/\textrm{Re}\alpha_2$. Here also we compared the approximate analytical result with a numerical calculation for the integral
\beq
c_+(t_f)\sim \int_{-4\pi/3}^{4\pi/3}d\tau \textrm{sign}(\tau)\frac{\cos\frac{3\tau}{2}-1}{2(5+4\cos\frac{3\tau}{2})}e^{i\frac{2}{F} E(3\tau/4,8/9)}
\eeq
This $|c_+(t_f)|^2$ agrees very well with $P_b\approx (\frac{\pi}{3})^2 4e^{-2 \textrm{Im} \alpha_2}\sin^2(\textrm{Re}\alpha_2)$ (see Fig. \ref{apt}(b)). In other words, it seems that the phase shift is exactly $\Delta \varphi=0$ rather than $2\textrm{arg} \alpha_2$.

As a conclusion, we find that a good analytical approximation valid also away from the adiabatic limit is given by the following \s-like formula:
\beq
P_{a/b}\approx 4P_1(1-P_1)\sin^2(\frac{\varphi_\textrm{dyn}+\Delta \varphi_{a/b}}{2}+\varphi_\textrm{S})
\label{sfapt}
\eeq
with $P_1=e^{-2\textrm{Im}\alpha_2}$, $\varphi_\textrm{dyn}=2\textrm{Re}\alpha_2$, $\varphi_\textrm{S}(\delta=\textrm{Im}\alpha_2/\pi)$
where the Stokes phase is $\varphi_\textrm{S}(\delta)=\frac{\pi}{4}+\delta(\ln \delta -1)+\textrm{Arg } \Gamma(1-i \delta)$ \cite{SAN2010} and $\Delta \varphi_a=\pi$ for case (a) or $\Delta \varphi_b=0$ for case (b). For the accuracy of such a formula, see Figure \ref{apt_main} in which it is compared to the numerical solution.

\section{Diabatic perturbation theory}
\label{app:dpt}
\subsection{Single M point avoided crossing}
The single M point avoided crossing Hamiltonian is:
\beqn
H_1(t)&=&-[2\cos(\frac{Ft}{2}+\frac{\pi}{3})+\cos(Ft+\frac{2\pi}{3})]\sigma_x\nn\\
&+&[2\sin(\frac{Ft}{2}+\frac{\pi}{3})-\sin(Ft+\frac{2\pi}{3})]\sigma_y
\eeqn
It can be rewritten as $H_1(t)=E_+(t)(\cos\phi(t) \sigma_x+\sin\phi(t) \sigma_y)$ with $E_\pm(t)=\pm\sqrt{5-4\cos\frac{3Ft}{2}}$ and $\tan \phi(t)=\frac{-2\sin(\frac{Ft}{2}+\frac{\pi}{3})+\sin(Ft+\frac{2\pi}{3})}{2\cos(\frac{Ft}{2}+\frac{\pi}{3})+\cos(Ft+\frac{2\pi}{3})}$. We first define an {\it adiabatic basis} $\{|u_+(t)\rangle,|u_-(t)\rangle\}$ by $H_1(t)|u_\pm(t)\rangle = \pm E_+(t)|u_\pm(t)\rangle$. Then we take as a {\it diabatic basis} $|u_- (t_i)\rangle=|\sigma_x +\rangle$ and $|u_+ (t_i)\rangle=|\sigma_x -\rangle$ as $\phi_i=\pi$ and expand the state in this basis $|\Psi(t)\rangle = A(t)|u_- (t_i)\rangle+B(t)|u_+ (t_i)\rangle$. We then compute $P_1=|\langle u_+(t_f)|\Psi(t_f)\rangle|^2=|A(t_f)\frac{3-i\sqrt{3}}{4}+B(t_f)\frac{1+i\sqrt{3}}{4}|^2$, where we used that $\phi_f=\pi/3$. Note that both $A(t_f)$ and $B(t_f)$ are required to compute the transition probability. The time-evolution in the diabatic basis is given by:
\beqn
\dot{A}&=&-iE_+ \cos\phi A + E_+\sin \phi B\nn \\
\dot{B}&=& iE_+\cos \phi B -E_+ \sin\phi A
\label{diabeqt}
\eeqn
We then use perturbation theory at second order in $1/F$. We find $A(t_f)=1+i\frac{3\sqrt{3}}{2F}-\frac{27}{2F^2}$ and $B(t_f)=-\frac{9}{2F}+\frac{i}{F^2}\frac{27\sqrt{3}-56\pi}{6}$ so that $P_1\approx 3/4-(27-14\pi/\sqrt{3})/F^2\to 3/4$ when $F\to \infty$ as expected. This result is plotted in Fig. \ref{singleMpointsudden} and compared with the numerical solution and with the adiabatic approximation.

From diabatic perturbation theory at first order in $1/F$, it is also possible to compute the phase $\varphi_1$ acquired upon being reflected at a single M point crossing. This phase is the equivalent of the Stokes phase $\varphi_\textrm{S}$ acquired upon being reflected at a infinite linear avoided crossing (LZ type). On the one hand, diabatic perturbation theory gives $\langle u_-(t_f)|\Psi(t_f)\rangle=A(t_f)\langle u_- (t_f)|u_- (t_i) \rangle + B(t_f)\langle u_- (t_f)|u_+ (t_i) \rangle=A(t_f)\frac{1-i\sqrt{3}}{4}+B(t_f)\frac{1+i\sqrt{3}}{4}=\frac{1}{2}e^{-i\pi/3}e^{i 6\sqrt{3}/F}$. On the other hand, from the N-matrix approach (see \cite{LFM2012}), we know that $\langle u_-(t_f)|\Psi(t_f)\rangle=\sqrt{1-P_1}e^{i\varphi_{cg}}e^{i\varphi_1}e^{-i\int_{t_i}^{t_f}dt E_-(t)}$, where $P_1=3/4$, $\varphi_1$ depends on $F$, the phase of the complex gap $\varphi_{cg}$ is independent of the force (it gives rise to the geometrical phase $\Delta \varphi$ in the double M point interferometer) and the dynamical phase is $-\int_{t_i}^{t_f}dt E_- (t)=\frac{8E(-8)}{3F}$. Equating the two equations, and identifying the $F$ dependent phase, we find
\beq
\varphi_1=\frac{6\sqrt{3}-8E(-8)}{3F}\approx \frac{1.48}{F} \to 0 
\label{dptvarphi1}
\eeq
when $F\to \infty$. This behavior is very different from that of the Stokes phase that reaches $\frac{\pi}{4}$ in the sudden limit.

\begin{figure}[ht]
\begin{center}
\includegraphics[width=7cm]{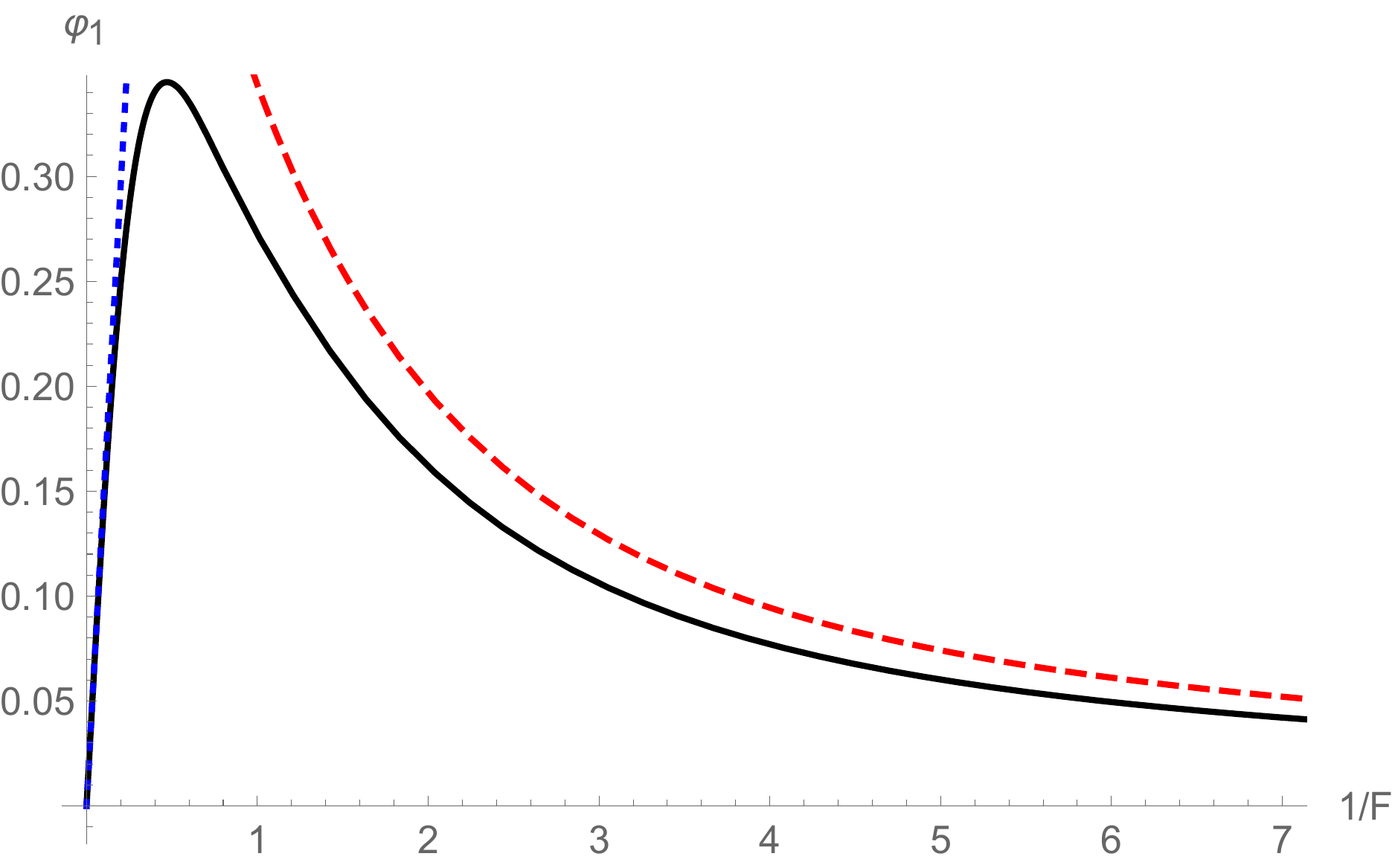}
\end{center}
\caption{Numerically computed $\varphi_1$ phase as a function of $1/F$ [force in units of $J/a$] (full black line). The dotted (blue) line is the diabatic perturbation theory result $1.48/F$ of Eq. (\ref{dptvarphi1}) and the dashed (red) line is the Stokes phase $\varphi_\textrm{S}(\delta)=\pi/4+\delta(\ln \delta -1)+\textrm{Arg }\Gamma(1-i\delta)$ with an adiabaticity parameter $\delta\approx 0.232/F$.}
\label{varphi1}
\end{figure}

We have computed the phase $\varphi_1$ numerically for the whole range of forces (see Fig. \ref{varphi1}). 
In the adiabatic limit, it behaves as the Stokes phase $\varphi_\textrm{S}(\delta)=\pi/4+\delta(\ln \delta -1)+\textrm{Arg }\Gamma(1-i\delta)$ with an adiabaticity parameter $\delta\approx 0.232/F$, reaches a maximum $\sim 0.33$ at $1/F\sim 0.47$ and then vanishes as $1.48/F$ in the sudden limit, in agreement with diabatic perturbation theory. This behavior can be qualitatively understood with the following heuristic argument. On the one hand, in the adiabatic limit, when the total duration of the double M point crossing $t_\textrm{tot}=t_f-t_i \sim 1/F$ is large compared to the tunneling time $t_\textrm{tun}\sim 1/\Delta \sim 1$, where $\Delta$ is the gap, the accumulated phase $\varphi_1$ is the same as that for the infinite linear avoided crossing, i.e. $\varphi_1 \sim \varphi_\textrm{S}(\delta\gg 1)$. On the other hand, when the total duration is short compared to the tunneling time (sudden limit), the accumulated phase is only a fraction $t_\textrm{tot}/t_\textrm{tun}\sim \Delta/F$ of the phase that could have been accumulated $\varphi_\textrm{S}(\delta\ll 1)\approx \pi/4$ and therefore $\varphi_1\sim \varphi_\textrm{S} t_\textrm{tot}/t_\textrm{tun} \sim\Delta/F$. This is a kind of truncated LZ process.

\subsection{Double M point interferometer}
\begin{figure}[ht]
\begin{center}
\includegraphics[width=7cm]{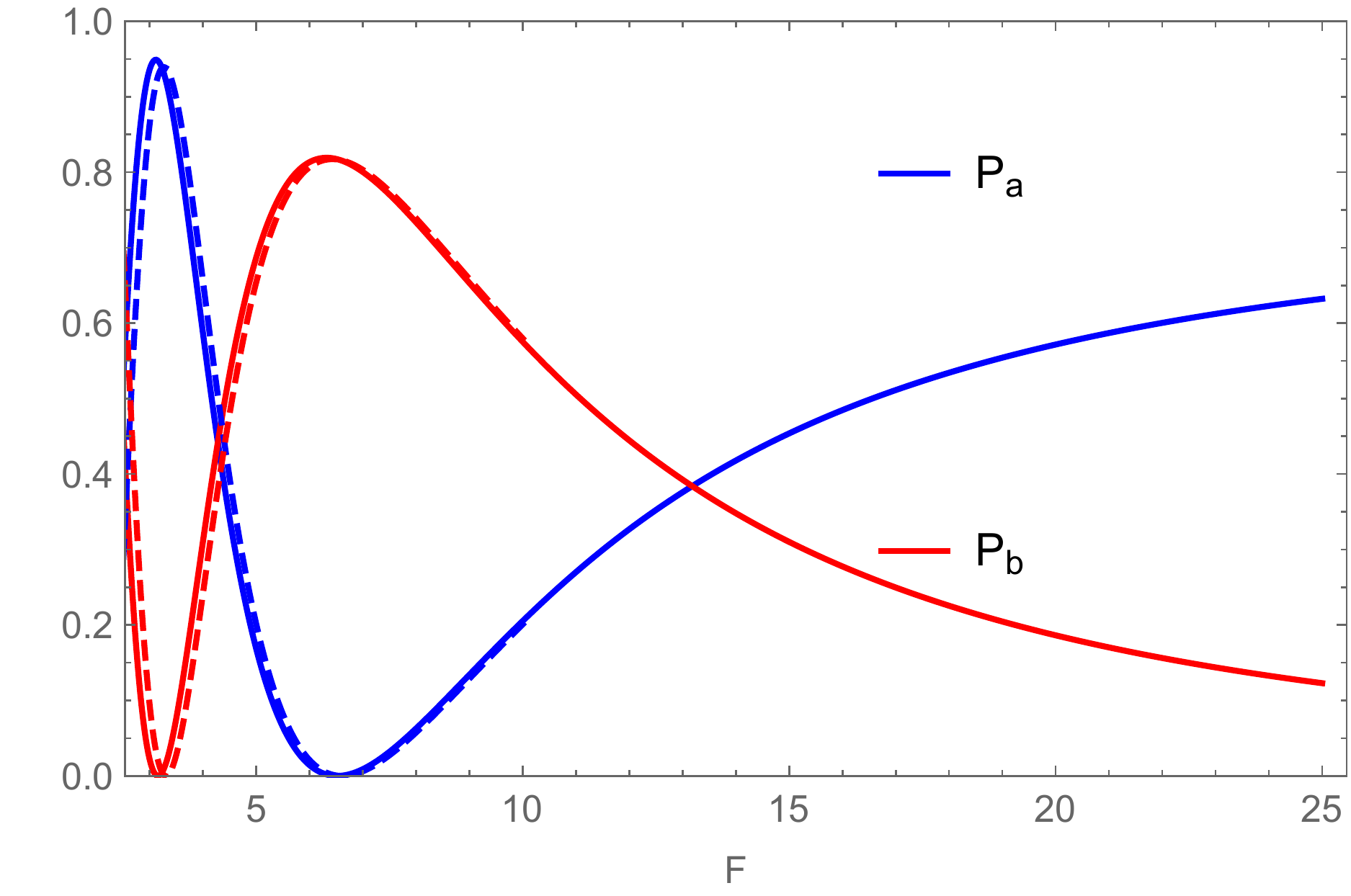}
\end{center}
\caption{Interband transition probability as a function of the force $F$ between $3$ and $25$ [in units of $J/a$]. Comparison between the numerical solutions ($P_a$ is the blue line and $P_b$ the red line) for the double M point interferometer and $4P_1(1-P_1)\sin^2(\varphi_1+(\varphi_\textrm{dyn}+\Delta \varphi)/2)$ with $\varphi_1= 1.48/F$, $\Delta \varphi_a=\pi$ (in blue dashed line) and $\Delta \varphi_b=0$ (in red dashed line) and $\varphi_\textrm{dyn}/2=\textrm{Re} \alpha_2=8E(8/9)/F$. The agreement is very good.}
\label{doubleMpointsudden}
\end{figure}
We already mentioned that $4P_1(1-P_1)$ (with $P_1$ obtained from a numerical solution of the single M point problem) matches $P_a+P_b$ (obtained from a numerical solution of the double M point problem) extremely well, see Figure \ref{papluspb}. This is also true when $F\gg 1$ and means that even in this limit (in which saturation occurs), there is a two-path interferometer of the \s type. It is therefore tempting to guess that
\beq
P_{a/b}= 4P_1(1-P_1)\sin^2(\frac{\varphi_\textrm{dyn}+\Delta \varphi_{a/b}}{2}+\varphi_1)
\label{guess}
\eeq
with $\Delta \varphi_a=\pi$ and $\Delta \varphi_b=0$ should agree with the numerical result very well also in the diabatic limit. Here, we set $\varphi_1= 1.48/F$ and compare with the numerics in the sudden limit, see Fig. \ref{doubleMpointsudden}.

The $4P_1(1-P_1)\sin^2(...)$ structure of the final probability $P_{a,b}$ explains both the saturation $P_a\to 3/4$ and $P_b\to 0$ when $P_1\to 3/4$ in the large limit. Indeed, when $F\gg 1$, both the dynamical phase and the $\varphi_1$ phase vanish as $1/F$. The phase shift $\Delta \varphi_a=\pi$ or $\Delta \varphi_b=0$ is of geometrical origin and does not depend on the force. Therefore $\sin^2(...)\to 1$ for the (a) case and $\to 0$ for the (b) case when $F\to \infty$. And  $4P_1(1-P_1)\to 3/4$ when $P_1\to 3/4$. Also, from the $4P_1(1-P_1)\sin^2(...)$ structure, we can understand that $P_{a,b}$ shows oscillations overshooting the saturation limit of $3/4$ in an intermediate force range, as $4P_1(1-P_1)\sim 1$ when $P_1\sim 1/2$.

Finally, we mention that Eq. (\ref{guess}) with both $P_1$ and $\varphi_1$ numerically computed reproduce the interband transition probability for a double M point avoided crossing not only in the sudden (see Fig. \ref{doubleMpointsudden}) and adiabatic (see Fig. \ref{apt_main}(b)) limits but for any force.


\begin{thebibliography}{99}

\bibitem{SAN2010} S. N. Shevchenko, S. Ashhab, and F. Nori, Phys. Rep. \textbf{492}, 1 (2010).

\bibitem{Stuck1932}E. C. G. St\"{u}ckelberg, Helv. Phys. Acta {\bf 5}, 369 (1932).

\bibitem{Tarruell2012}L. Tarruell , D. Greif, T. Uehlinger, G. Jotzu, T. Esslinger, Nature \textbf{483}, 302 (2012).

\bibitem{LFM2012}L.-K. Lim, J.-N. Fuchs and G. Montambaux, Phys. Rev. Lett. \textbf{108}, 175303 (2012).

\bibitem{LZ1932} L. Landau, Phys. Z. Sowjetunion {\bf 2}, 46 (1932); C. Zener, Proc. R. Soc. Lond. A \textbf{137}, 696 (1932); E. Majorana, Nuovo Cimento {\bf 9}, 43 (1932). See also C. Wittig, J.Phys. Chem. B \textbf{109}, 8428 (2005).

\bibitem{LFM2014}L.-K. Lim, J.-N. Fuchs and G. Montambaux, Phys. Rev Lett. \textbf{112}, 155302 (2014).

\bibitem{LFM2015}L.-K. Lim, J.-N. Fuchs and G. Montambaux, Phys. Rev. A \textbf{91}, 042119 (2015).

\bibitem{Montambaux2009}G. Montambaux, F. Pi\'echon, J.-N. Fuchs, and M.O. Goerbig, Phys. Rev. B \textbf{80}, 153412 (2009); Eur. Phys. J. B \textbf{72}, 509 (2009).

\bibitem{FLM2012}J.-N. Fuchs, L.-K. Lim and G. Montambaux, Phys. Rev. A \textbf{86}, 063613 (2012). See also E. Shimshoni and Y. Gefen, Ann. Phys. \textbf{210}, 16 (1991); K.-A. Suominen, Opt. Commun. \textbf{93}, 126 (1992).

\bibitem{BlochSchneider}I. Bloch, private communication and seminar at Coll\`ege de France, Paris in May 2014, 
; U. Schneider, oral presentation at the ``Advanced Workshop on Landau-Zener Interferometry and Quantum Control in Condensed Matter'', Izmir, Turkey (October 2014).

\bibitem{Chen2011}Z. Chen and B. Wu, Phys. Rev. Lett. \textbf{107}, 065301 (2011).

\bibitem{Blount} E. I. Blount, in \textit{Solid State Physics}, edited by F. Seitz and D. Turnbull (Academic Press, Sandiego, 1962), Vol. 13, pp. 305-373.

\bibitem{Xiao2010} D. Xiao, M.-C. Cheng, and Q. Niu, Rev. Mod. Phys. {\bf 82}, 1959 (2010).

\bibitem{FPGM2010}J.-N. Fuchs, F. Pi\'echon, M. O. Goerbig and G. Montambaux, Eur. Phys. J. B \textbf{77}, 351 (2010).

\bibitem{Li2015}T. Li, L. Duca, M. Reitter, F. Grusdt, E. Demler, M. Endres, M. Schleier-Smith, I. Bloch and U. Schneider, arXiv:1509.02185 (2015).

\bibitem{Wallace1947}P. R. Wallace, Phys. Rev. \textbf{71}, 622 (1947).

\bibitem{BM2009}C. Bena and G. Montambaux, New J. Phys. \textbf{11}, 095003 (2009).

\bibitem{Haldane2014}F. D. M. Haldane, arXiv:1401.0529; see also talks available on his webpage http://wwwphy.princeton.edu/~haldane/research.html .

\bibitem{Berry1984}M.V. Berry, Proc. Roy. Soc. Lond. A \textbf{392}, 45 (1984).

\bibitem{Wu2012} Y.-L. Wu, B. Andrei Bernevig and N. Regnault, Phys. Rev. B \textbf{85}, 075116 (2012).

\bibitem{Jackson2014} T. S. Jackson, G. M\"{o}ller and R. Roy, arXiv: 1408.0843.

\bibitem{Fruchart2014}M. Fruchart, D. Carpentier and K. Gawedzki, Europhys. Lett. \textbf{106}, 60002 (2014).

\bibitem{Milovanovic2015}E. Dobardzic, M. Dimitrijevic, and M. V. Milovanovic, Phys. Rev. B \textbf{91}, 125424 (2015).

\bibitem{Houston1940}W.V. Houston, Phys. Rev. \textbf{57}, 184 (1940).

\bibitem{Berry1990}M.V. Berry, Proc. R. Soc. A \textbf{430}, 405-411 (1990).

\bibitem{Messiah}A. Messiah, {\it Quantum mechanics}, Dunod, Paris 1995, 2nd edition, chapter XVII.

\bibitem{Blochnotes} Such a saturation of the interband transition probability across a single avoided crossing was first mentioned to us in May 2014 by I. Bloch.

\bibitem{AshcroftMermin}N.W. Ashcroft, N.D. Mermin, \textit{Solid State Physics} (Brooks/Cole editors, 1976).

\bibitem{Duca2015}L. Duca, T. Li, M. Reitter, I. Bloch, M. Schleier-Smith, U. Schneider, Science \textbf{347}, 288 (2015).

\bibitem{Esslinger2014}G. Jotzu, M. Messer, R. Desbuquois, M. Lebrat, T. Uehlinger, D. Greif, T. Esslinger, Nature \textbf{515}, 237-240 (2014).

\bibitem{Sengstock2015}N. Fl\"aschner, B. S. Rem, M. Tarnowski, D. Vogel, D.-S. L\"uhmann, K. Sengstock, C. Weitenberg, arXiv:1509.05763 (2015).

\bibitem{pmmnote}What is measured is actually $P^{--}$, while the quantity we compute is $P^{+-}$. In a two-band model, as considered in the present work, the two are simply related as $P^{--}=1-P^{+-}$. This is no longer the case when more than two bands are involved.

\bibitem{Berry1989}M.V. Berry ``The quantum phase, five years after'' in \textit{Geometric Phases in Physics}, eds: A. Shapere, F. Wilczek (World Scientific, 1989), pages 7-28.

\bibitem{Zak1989}J. Zak, Phys. Rev. Lett. \textbf{62}, 2747 (1989).

\bibitem{Dykhne1960}A. M. Dykhne, Sov. Phys. JETP \textbf{14}, 941 (1962); \textbf{11}, 411 (1960).

\bibitem{BM1972} M. V. Berry and K. E. Mount, Rep. Prog. Phys. \textbf{35}, 315 (1972), in particular, Sec. 2.3.

\end{thebibliography}
\end{document}